\documentclass{jfm}
\usepackage{natbib}
\usepackage{graphicx}
\usepackage{amsmath}
\usepackage{color}
\usepackage{bm}
\usepackage{pgfplots}
\usepackage{tikz-3dplot}
\usepackage{ar}
\usepackage{hyperref}

\newcommand{\vect}[1]{\boldsymbol{#1}}
\newcommand{\aver}[1]{ \left\langle {#1} \right \rangle }

\title[Turbulent flow around a rectangular cylinder]{Structure of turbulence \\ in the flow around a rectangular cylinder}

\author[A.~Chiarini, D.~Gatti, A.~Cimarelli and M.~Quadrio]
{Alessandro Chiarini$^1$, Davide Gatti$^2$, Andrea Cimarelli$^3$, Maurizio Quadrio$^1$}

\affiliation{
$^1$ Dipartimento di Scienze e Tecnologie Aerospaziali, Politecnico di Milano,
via La Masa 34, 20156 Milano, Italy \\ [\affilskip]
$^2$ Institute for Fluid Mechanics, Karlsruhe Institute of Technology, Kaiserstr. 10, 76131 Karlsruhe, Germany \\ [\affilskip]
$^3$ DIEF, University of Modena and Reggio Emilia, 41125 Modena, Italy
}

\begin{document}

\maketitle

\begin{abstract}
The separating and reattaching turbulent flow past a rectangular cylinder is studied to describe how small and large scales contribute to the sustaining mechanism of the velocity fluctuations. The work is based on the Anisotropic Generalised Kolmogorov Equations (AGKE), exact budget equations for the second-order structure function tensor in the space of scales and in the physical space. Scale-space energy fluxes show that forward and reverse energy transfers simultaneously occur in the flow, with interesting modelling implications.  

Over the longitudinal cylinder side, the Kelvin-Helmholtz instability of the leading-edge shear layer generates large spanwise rolls, which get stretched into hairpin-like vortices and eventually break down into smaller streamwise vortices. Independent sources of velocity fluctuations act at different scales. The flow dynamics is dominated by pressure-strain: the flow impingement on the cylinder surface in the reattachment zone produces spanwise velocity fluctuations very close to the wall and, at larger wall distances, reorients them to feed streamwise-aligned vortices.

In the near wake, large von K\'arm\'an-like vortices are shed from the trailing edge and coexist with smaller turbulent structures, each with its own independent production mechanism. At the trailing edge, the sudden disappearance of the wall changes the structure of turbulence: streamwise vortices progressively vanish, while spanwise structures close to the wall are suddenly turned into vertical fluctuations by the pressure strain. 

\end{abstract}

%%%%%%%%%%%%%%%%%%%%%
\section{Introduction}

The flow past bluff bodies with sharp corners is of fundamental importance and occurs in several applications. In civil engineering, for example, structural elements such as pylons, high-rise buildings and decks often feature sharp corners \citep{tamura-miyagi-kitagishi-1998}. In addition to the classic von K\'arm\'an-like vortex street typical of bluff bodies, the flow past bodies with sharp corners presents a separation at the leading-edge (LE) corner: the shear layer detaches, becomes unstable and possibly reattaches if the body is sufficiently long.

The cylinder with rectangular cross-section is the prototype of such bodies. The flow past rectangular cylinders depends on the aspect ratio $\AR \equiv L/D$ (where $L$ and $D$ are the streamwise and cross-stream sizes of the cylinder). For small aspect ratios, i.e. $\AR < 2$, the flow cannot reattach, whereas for intermediate $\AR$, i.e. $2 \le \AR \le 3$, the reattachment is intermittent. For larger $\AR$, the flow reattaches permanently, generating a large recirculating region over the cylinder side, and separates again at the trailing edge (TE). In this case, vortex shedding occurs from both LE and TE  \citep{okajima-1982,hourigan-thompson-tan-2001,chiarini-quadrio-auteri-2022}. The value $\AR=5$ defines the Benchmark for the Aerodynamics of the 5:1 Rectangular Cylinder (BARC), the geometry considered in the present work. The BARC (see \url{https://www.aniv-iawe.org/barc-home/}), launched at the VI International Colloquium on Bluff Body Aerodynamics and Applications, is meant to characterise the main features of the turbulent flow and to set the standards for simulations and experiments. Several studies have recently considered this benchmark, with results that can differ already for the mean flow field, showing how challenging is the correct description of the fundamental features of the BARC flow \citep{bruno-salvetti-ricciardelli-2014}.

The recirculating region over the longitudinal cylinder side is important in the dynamics of the flow, and has been the subject of several studies. Along the sides of a blunt flat plate with sharp corners, \cite{cherry-1984} experimentally identified a low-frequency motion throughout the recirculating region accompanied by a weak flapping of the shear layer, consisting in a shedding of pseudo-periodic train of vortical structures followed by a quiescent phase. In their experiments, \cite{kiya-sasaki-1983} found that the large-scale unsteadiness is accompanied by an enlargement and shrinkage of the recirculating region and by a flapping motion of the shear layer near the separating line. Later, \cite{kiya-sasaki-1985} observed that the shrinkage rate is larger than the enlargement rate, and that the strength of the shedding of large-scale vortices depends on the phase of the low-frequeny unsteadiness. They also proposed a mathematical model for the unsteady flow in the reattachment zone. The picture was later numerically confirmed by \cite{tafti-1991}. The recirculating region also defines the three-dimensional pattern of the flow, described by \cite{sasaki-kiya-1991} for a wide range of Reynolds numbers. They observed that the separated shear layer rolls up to form hairpin-like structures whose arrangement depends on the Reynolds number. A similar scenario is described by \cite{chaurasia-thompson-2011} and \cite{huang-etal-2017}, who studied the three-dimensional instability of the flow over a long, sharp rectangular plate. They found that vortices shed from the LE are elliptically unstable to three-dimensional perturbations, and originate hairpin-like structures. 

\cite{cimarelli-leonforte-angeli-2018} were first to perform a Direct Numerical Simulation (DNS) of the BARC, at a value of the Reynolds number such that the flow is turbulent. They found that the recirculating region over the cylinder side is populated by small-scale motions, namely quasi-streamwise vortices and streamwise velocity streaks induced by hairpin-like structures, and observed spanwise vortices in the reverse flow region. Moreover, a self-sustaining cycle was identified that involves both the small- and large-scale motions, and links the velocity fluctuations generated over the cylinder side with those in the wake. Later, \cite{chiarini-quadrio-2021} studied via DNS the single-point budget of the Reynolds stresses of the same flow at the same $Re$, and located where production, redistribution and dissipation of each component of the Reynolds stress tensor are most relevant. Energy is drained from the mean flow to feed the streamwise fluctuations mainly along the LE shear layer and in the core of the recirculating region. For the vertical component, energy moves from the mean field to the fluctuating field within the recirculating region and along the centreline of the wake, but the opposite occurs over the shear layer, where production is negative. Pressure-strain was found to partially reorient the streamwise fluctuations towards the cross-stream ones almost everywhere, except close to the cylinder side and along the centreline of the wake. 

The two studies mentioned above observed certain turbulent structures in different regions of the flow, but how these structures contribute to flow statistics was not described; thus, a complete scale-space characterisation of the flow is still lacking. Moreover, the evolution of the flow from a non-equilibrium boundary layer over the horizontal wall of the cylinder towards a free shear flow in the wake warrants a detailed description, which should disentangle the highly multiscale nature of the flow. In this work, these points are accomplished by leveraging the Anisotropic Generalised Kolmogorov Equations (AGKE) \citep{hill-2001}, used by \cite{gatti-etal-2020} to extend the analysis of turbulence made possible by the Generalised Kolmogorov Equation (GKE). The GKE \citep{danaila-etal-2001,marati-casciola-piva-2004} is an exact budget equation for the second-order structure function, whereas the AGKE are a set of exact budget equations for each component of the second-order structure function tensor. As such, the AGKE separately address each component of the Reynolds stresses in the compound space of scales and positions. Unlike the GKE, the AGKE feature a pressure-strain term that describes redistribution in scale and physical spaces, and that is important in the BARC flow.

Hence, in this work we employ the AGKE to study the structure of turbulence in the BARC flow, starting from the DNS database produced by \cite{chiarini-quadrio-2021}. The specific goals are: (i) to provide an exhaustive scale-space characterisation of the flow, by identifying the statistically significant structures in the various regions; (ii) to describe the role and the dynamical significance of these structures in terms of production and redistribution of Reynolds stresses and energy transfers; (iii) to characterise the near-wake region, where structures generated over the cylinder side interact with the large-scale motions associated with the von K\'arm\'an-like vortex street. In the preliminary
\S\ref{sec:methods} the DNS database is briefly described, and the main features of the BARC flow are recalled. This Section also summarises the AGKE tailored to the present flow. The AGKE are then used to describe the flow over the cylinder side in \S\ref{sec:side}, and in the near wake in \S\ref{sec:wake}. Finally, a concluding discussion is presented in \S\ref{sec:conclusions}.

%%%%%%%%%%%%%%%%%%%%%%%%%%%%%%%%%%%%%%%%%%%%%%%%%%%%%%%%%%%%%%%%%%%%%%%%
\section{Prerequisites}
\label{sec:methods}
%%%%%%%%%%%%%%%%%%%%%%%%%%%%%%%%%%%%%%%%%%%%%%%%%%%%%%%%%%%%%%%%%%%%%%%%
 
%-------------------------
\subsection{The DNS database}

\begin{figure}
\centering
	\tdplotsetmaincoords{73}{145}	
	\begin{tikzpicture}[scale=0.3,tdplot_main_coords]
	
	\tikzset{myptr/.style={decoration={markings,mark=at position 1 with {\arrow[scale=3,>=stealth]{>}}},postaction={decorate}}}

        \coordinate (Aup) at ( -6, -8,  0.5);
        \coordinate (Bup) at ( 2, -8,  0.5);
        \coordinate (Cup) at ( 2,  8,  0.5);
        \coordinate (Dup) at ( -6,  8,  0.5);
        \coordinate (Alo) at ( -6, -8, -0.5);
        \coordinate (Blo) at ( 2, -8, -0.5);
        \coordinate (Clo) at ( 2,  8, -0.5);
        \coordinate (Dlo) at ( -6,  8, -0.5);

        \coordinate (A1) at ( 2,  9.5, -0.5);
        \coordinate (A2) at (-6,  9.5, -0.5);
        \coordinate (A3) at ( 2,    8, -0.5);
        \coordinate (A4) at (-6,    8, -0.5);
        \draw[<->] (A1)--(A2);
        \draw[] (A1)--(A3);
        \draw[] (A2)--(A4);
        \node at (-0.5,12.2,-0.5) {$L=5D$};
        
        \draw[dotted] (2,8,-0,5) -- (2,8,-6);
        
        \coordinate (B1) at ( -6.5, 8, 0.5);
        \coordinate (B2) at ( -6.5, 8, -0.5);
        \coordinate (B3) at ( -6,   8, 0.5);
        \coordinate (B4) at ( -6,   8, -0.5);
        \draw[<->] (B1)--(B2);
        \draw[] (B1)--(B3);
        \draw[] (B2)--(B4);
        \node at (-7.7,8,0) {$D$};

        \draw[very thick,black] (Aup) -- (Bup);
        \draw[very thick,black] (Bup) -- (Cup);
        \draw[very thick,black] (Cup) -- (Dup);
        \draw[very thick,black] (Dup) -- (Aup);

        \draw[very thick,black,dashed] (Alo) -- (Blo);
        \draw[very thick,black] (Blo) -- (Clo);
        \draw[very thick,black] (Clo) -- (Dlo);
        \draw[very thick,black,dashed] (Dlo) -- (Alo);

        \draw[very thick,black,dashed] (Aup) -- (Alo);
        \draw[very thick,black] (Bup) -- (Blo);
        \draw[very thick,black] (Cup) -- (Clo);
        \draw[very thick,black] (Dup) -- (Dlo);

        \coordinate (AAup) at (-19, -8,  0.5);
        \coordinate (BBup) at ( 12, -8,  0.5);
        \coordinate (CCup) at ( 12,  8,  0.5);
        \coordinate (DDup) at (-19,  8,  0.5);
        \coordinate (AAlo) at (-19, -8, -0.5);
        \coordinate (BBlo) at ( 12, -8, -0.5);
        \coordinate (CClo) at ( 12,  8, -0.5);
        \coordinate (DDlo) at (-19,  8, -0.5);

        \coordinate (AAupp) at (  -6, -8,  6);
        \coordinate (BBupp) at (  2, -8,  6);
        \coordinate (CCupp) at (  2,  8,  6);
        \coordinate (DDupp) at (  -6,  8,  6);
        \coordinate (AAloo) at (  2, -8, -6);
        \coordinate (BBloo) at (  2, -8, -6);
        \coordinate (CCloo) at (  2,  8, -6);
        \coordinate (DDloo) at (  -6,  8, -6);

  %      \draw[dashed] (Aup) -- (AAup);
  %      \draw[dashed] (Bup) -- (BBup);
  %      \draw[dashed] (Cup) -- (CCup);
  %      \draw[dashed] (Dup) -- (DDup);

  %      \draw[dashed] (Alo) -- (AAlo);
  %      \draw[dashed] (Blo) -- (BBlo);
  %      \draw[dashed] (Clo) -- (CClo);
  %      \draw[dashed] (Dlo) -- (DDlo);

        \draw[dotted] (AAlo) -- (AAup);
        \draw[] (BBlo) -- (BBup);
        \draw[] (CClo) -- (CCup);
        \draw[] (DDlo) -- (DDup);

        \draw[] (AAupp) -- (BBupp);
  %      \draw[dashed] (BBupp) -- (CCupp);
        \draw[] (CCupp) -- (DDupp);
  %      \draw[] (DDupp) -- (AAupp);

        \draw[dotted] (AAloo) -- (BBloo);
  %      \draw[dashed] (BBloo) -- (CCloo);
        \draw[] (CCloo) -- (DDloo);
  %      \draw[dashed] (DDloo) -- (AAloo);

  %      \draw[dashed] (AAloo) -- (Alo);
   %     \draw[dashed] (BBloo) -- (Blo);
    %    \draw[dashed] (CCloo) -- (Clo);
     %   \draw[dashed] (DDloo) -- (Dlo);
   %     \draw[dashed] (AAupp) -- (Aup);
   %     \draw[dashed] (BBupp) -- (Bup);
   %     \draw[dashed] (CCupp) -- (Cup);
   %     \draw[dashed] (DDupp) -- (Dup);

        \coordinate (OAup) at (-19, -8,  6);
        \coordinate (OBup) at ( 12, -8,  6);
        \coordinate (OCup) at ( 12,  8,  6);
        \coordinate (ODup) at (-19,  8,  6);
        \coordinate (OAlo) at (-19, -8, -6);
        \coordinate (OBlo) at ( 12, -8, -6);
        \coordinate (OClo) at ( 12,  8, -6);
        \coordinate (ODlo) at (-19,  8, -6);

        \draw[] (OAup) -- (AAupp);
        %\draw[dashed] (AAupp) -- (DDupp);
        \draw[] (DDupp) -- (ODup);
        \draw[] (ODup)  -- (OAup);

        \draw[dotted] (OAlo) -- (AAloo);
        %\draw[dashed] (AAloo) -- (DDloo);
        \draw[] (DDloo) -- (ODlo);
        \draw[dotted] (ODlo)  -- (OAlo);

        \draw[] (BBupp) -- (OBup);
        \draw[] (OBup)  -- (OCup);
        \draw[] (OCup)  -- (CCupp);
        %\draw[dashed] (CCupp) -- (BBupp);

        \draw[dotted] (OAup) -- (AAup);
        \draw[] (OCup) -- (CCup);
        \draw[] (ODup) -- (DDup);
        \draw[] (OBup) -- (BBup);

      %  \draw[] (BBup) -- (CCup);
      %  \draw[] (BBlo) -- (CClo);
        \draw[] (OBlo) -- (OClo);
        \draw[] (BBlo) -- (OBlo);
        \draw[] (CClo) -- (OClo);

        \draw[dotted] (OBlo) -- (BBloo);
        \draw[] (OClo) -- (CCloo);
       % \draw[dashed] (AAlo) -- (DDlo);
       % \draw[] (AAup) -- (DDup);
        \draw[dotted] (OAlo) -- (AAlo);
        \draw[] (ODlo) -- (DDlo);	

        \coordinate (R1) at ( -20.5, 8, 6);
        \coordinate (R2) at ( -20.5, 8, -6);
        \coordinate (R3) at ( -19, 8, 6);
        \coordinate (R4) at ( -19, 8, -6);
        \draw[<->] (R1)--(R2);
        \draw[] (R1)--(R3);
        \draw[] (R2)--(R4);
        \node at (-22.5,8,0) {$42D$};
        
        \coordinate (P1) at ( 2, 9.5, -6);
        \coordinate (P2) at (-19, 9.5, -6);
        \coordinate (P3) at ( 2, 8, -6);
        \coordinate (P4) at ( -19, 8, -6);
        \draw[<->] (P1)--(P2);
        \draw[] (P1)--(P3);
        \draw[] (P2)--(P4);
        \node at (-5.75,11.7,-6) {$42.5D$};
        
        \coordinate (Q1) at ( 12, 9.5, -6);
        \coordinate (Q2) at (  2, 9.5, -6);
        \coordinate (Q3) at ( 12, 8, -6);
        \coordinate (Q4) at ( 2, 8, -6);
        \draw[<->] (Q1)--(Q2);
        \draw[] (Q1)--(Q3);
        \draw[] (Q2)--(Q4);
        \node at (5.25,11.7,-6) {$20D$};
        
        \coordinate (S1) at ( 13.5,-8, -6);
        \coordinate (S2) at ( 13.5, 8, -6);
        \coordinate (S3) at ( 12,-8, -6);
        \coordinate (S4) at ( 12, 8, -6);
        \draw[<->] (S1)--(S2);
        \draw[] (S1)--(S3);
        \draw[] (S2)--(S4);
        \node at (14.7,0,-6) {$5D$};
        
        \coordinate (F1) at (27,0,0);
        \coordinate (F2) at (21,0,0);
        \draw [->,>=stealth] (F1)--(F2);
        \node at (26,0,1) {$U_{\infty}$};

        \coordinate (O0) at ( 2,0,0); % (21,0,-6);
        \coordinate (O1) at (-3,0,0); % (18,0,-6);
        \coordinate (O2) at ( 2,0,5); % (21,0,-4);
        \coordinate (O3) at ( 2,5,0); %(21,2,-6);

        \node at (  -3.5,0,0) {$x$};
        \node at ( 2,0,5.5) {$y$};
        \node at ( 2,5.5,0) {$z$};
        \draw [->] (O0)--(O1);
        \draw [->] (O0)--(O2);
        \draw [->] (O0)--(O3);

	\end{tikzpicture}	
\caption{Sketch of the geometry, computational domain and reference system. The reference length is the body height $D$.}
\label{fig:sketch}
\end{figure}
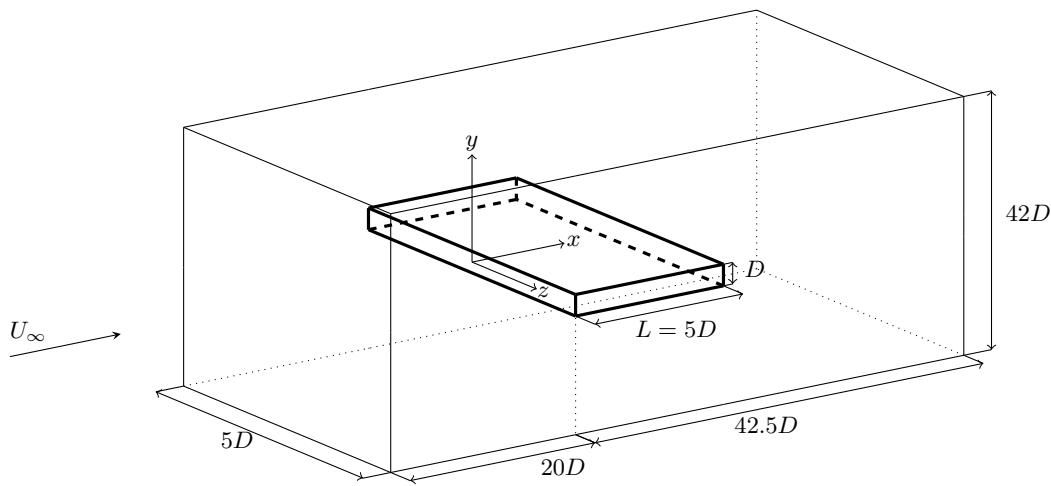

The BARC (Benchmark on the Aerodynamics of a Rectangular 5:1 Cylinder) considers the flow over a spanwise-indefinite rectangular cylinder with length $L$ and thickness $D$, and a 5:1 length-to-thickness ratio. The present work is based upon a DNS dataset produced by \cite{chiarini-quadrio-2021}; its main characteristics are briefly recalled below.

Figure \ref{fig:sketch} describes the geometry and the reference system. A Cartesian coordinate system is placed at the leading edge of the cylinder, with the $x$, $y$ and $z$ axes denoting the streamwise, vertical and spanwise directions (the alternative notation $x_1$, $x_2$, $x_3$ is also used). The body extends for $0 \le x \le 5D$ and $y=0$ corresponds to its symmetry plane. The computational domain extends for $-20 D \le x \le 42.5D $ in the streamwise direction, for $-21D \le y \le 21D $ in the vertical direction and $-2.5D \le z \le 2.5D$ in the spanwise direction. The incoming velocity is uniform and aligned with the $x$ axis, i.e. $(U_\infty,0,0)$. 
The flow is governed by the incompressible Navier--Stokes equations for velocity $\vect{u}=(u,v,w)$ and pressure $p$. Unperturbed flow is enforced at the inlet and at the far field at $y=\pm 21D$, periodic conditions are set at the spanwise boundaries to account for spanwise homogeneity, and a convective condition $\partial \vect{u} / \partial t = U_\infty \partial \vect{u}/\partial x$ is used at the outlet. No-slip and no-penetration conditions are applied at the cylinder surface. The Reynolds number based on the unperturbed velocity, cylinder thickness and kinematic viscosity $\nu$, is $Re=U_\infty D / \nu = 3000$. Unless otherwise noted, all quantities are made dimensionless with $D$ and $U_\infty$; hereinafter capital letters indicate mean fields and small letters denote fluctuations around them.

The Navier--Stokes equations are solved using a DNS code introduced by \cite{luchini-2016}, which employs second-order finite differences on a staggered grid in the three directions. The cylinder is described via an implicit, second-order accurate immersed-boundary method \citep{luchini-2013,luchini-2016}. The computational domain is discretised with $N_x=1776$, $N_y=942$ and $N_z=150$ points in the three directions. In the spanwise direction their distribution is uniform, whereas in the streamwise and vertical directions the resolution becomes finer near the body and is maximum close to the LE and TE corners, where grid spacing is $\Delta x  = \Delta y \approx 0.0015$. 

The momentum equation is advanced in time using a third-order Runge-Kutta scheme; the Poisson equation for pressure is solved by an iterative SOR algorithm. The time step $\Delta t$ varies so as to fulfill the condition of unitary Courant-Frederic-Levy number; its average value is $\Delta t \approx 0.0013$. Overall, the total averaging time is $1819 D/U_\infty$, and the database consists in $1819$ snapshots of the entire flow field saved every time unit. 

%---------------------------------------------------
\subsection{The BARC flow}
\begin{figure}
\centering
\includegraphics[width=1\textwidth]{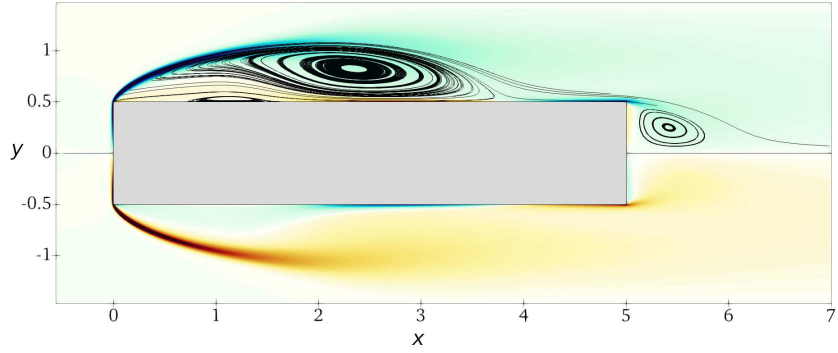}
\caption{Mean streamlines (top half only) plotted over a colour contour of the mean spanwise vorticity $\Omega_z$ (blue/red indicates negative/positive values in the range $-20 \le \Omega_z \le 20$).}
\label{fig:mean}
\end{figure}

To provide a qualitative illustration of the flow, figure \ref{fig:mean} plots the mean streamlines overlaid to a colour map of the mean spanwise vorticity $\Omega_z$. The flow separates at the LE corner and reattaches over the cylinder longitudinal side, before eventually separating again at the TE. Three recirculating regions exist. A large recirculating region is identified by the shear layer separating from the LE corner and reattaching at $x \approx 3.95$, and is hereafter referred to as the primary vortex. A further, thin recirculating region is located below the primary vortex: the reverse boundary layer induced by the primary vortex separates due to the adverse pressure gradient \citep{simpson-1989}, and originates a smaller counter-rotating recirculating region, referred to as the secondary vortex. The third recirculating region is delimited by the shear layer separating from the TE corner, and is referred to as the wake vortex. The symmetry of the plot in figure \ref{fig:mean} with respect to $y=0$ demonstrates the adequacy of the statistical sample: in fact, using one half of the sample yields perfectly overlapping results.

\begin{figure}
\centering
\includegraphics[width=1\textwidth]{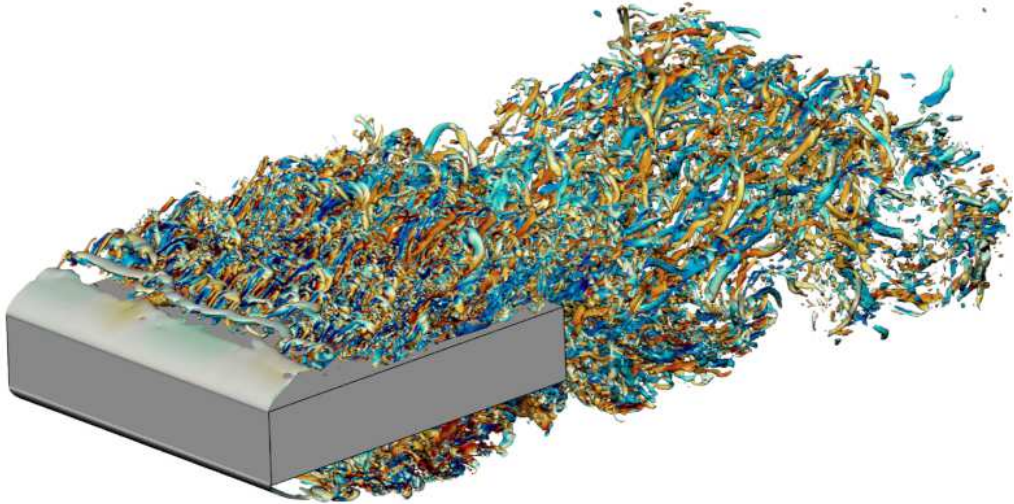}
\caption{Instantaneous snapshot of the BARC flow: perspective view of the isosurfaces at $\lambda_2 = -5$. Color depicts streamwise vorticity $\omega_x$, with the blue-to-red colormap ranging in $-10 \le \omega_x \le 10$.}
\label{fig:lambda2-vorticity}
\end{figure}

The BARC flow contains a wide range of structures, visualised in figure \ref{fig:lambda2-vorticity} via the $\lambda_2$ criterion \citep{jeong-hussain-1995}. Immediately after the LE, the shear layer remains two-dimensional and laminar, but already at $x \approx 0.5$ a Kelvin--Helmholtz instability generates large spanwise tubes \citep{sasaki-kiya-1991,tenaud-etal-2016,cimarelli-leonforte-angeli-2018,moore-etal-2019}. Further downstream the spanwise tubes, stretched by the mean gradients, develop a modulation in the spanwise direction, and then roll up into hairpin-like vortices. At $x \ge 2.5$ the stretched hairpin vortices break down to small-scale elongated streamwise vortices \citep{tenaud-etal-2016,cimarelli-leonforte-angeli-2018}. At this stage the flow is fully turbulent, and small- and large-scale structures coexist. At the TE the flow separates again, and structures typical of wall turbulence originating from the longitudinal cylinder side coexist with the large structures created by the instability of the separating shear layers. Eventually, moving downstream the flow recovers the features of a classic turbulent wake after a bluff body.

%------------------------------------------------------------
\subsection{The Anisotropic Generalised Kolmogorov Equations}

In the present work, turbulent production, redistribution and transfers throughout scales and physical positions are studied via the Anisotropic Generalised Kolmogorov Equations (AGKE) \citep{hill-2002,gatti-etal-2020}. They are the exact budget equations for the components of the second-order structure function tensor $\aver{\delta u_i \delta u_j}$. The operator $\aver{\cdot}$ denotes ensemble averaging, as well as averaging in homogeneous directions (if present) and in time (if the flow is statistically steady). In the definition of $\aver{\delta u_i \delta u_j}$, $\delta u_i$ is the $i$-th component of the fluctuating velocity increment between two points $\vect{x}_1$ and $\vect{x}_2$ identified by their midpoint $\vect{X} = \left( \vect{x}_1 + \vect{x}_2 \right) / 2$ and by their separation vector $\vect{r}= \vect{x}_1 - \vect{x}_2$, i.e. $\delta u_i = u_i \left( \vect{x}_1,t \right) - u_i \left( \vect{x}_2,t \right)$. In the most general case, $\aver{\delta u_i \delta u_j}$ depends upon the six coordinates of the vectors $\vect{X}$ and $\vect{r}$ and upon the time $t$. This tensor is linked \citep{davidson-etal-2006,agostini-leschziner-2017} to the single-point Reynolds stresses and to the spatial cross-correlation tensor, as:
\begin{equation}
\aver{\delta u_i \delta u_j} \left( \vect{X}, \vect{r}, t \right) = V_{ij} \left( \vect{X}, \vect{r}, t \right) - R_{ij} \left( \vect{X}, \vect{r},t \right) - R_{ji} \left( \vect{X}, \vect{r}, t \right), 
\label{eq:str-func}
\end{equation}
where
\begin{equation}
V_{ij} \left( \vect{X}, \vect{r}, t \right) = \aver{u_i u_j} \left( \vect{X} + \frac{\vect{r}}{2}, t \right) + \aver{u_i u_j} \left( \vect{X} - \frac{\vect{r}}{2}, t \right)
\end{equation}
is the sum of the single-point Reynolds stresses at the $\vect{X} \pm \vect{r}/2$ points and
\begin{equation}
R_{ij} \left( \vect{X}, \vect{r}, t \right) = \aver{ u_i \left( \vect{X} + \frac{\vect{r}}{2}, t \right) u_j \left( \vect{X} - \frac{\vect{r}}{2}, t \right) }
\end{equation}
is the two-points spatial cross-correlation tensor. For large enough $| \vect{r} | $ the correlation function vanishes, and $\aver{\delta u_i \delta u_j} = V_{ij}$.

The relationship \eqref{eq:str-func} between $\aver{\delta u_i \delta u_j}$ and $R_{ij}$ is worth a brief discussion, as it will be often implicitly used throughout the manuscript, whenever correlation or anticorrelation of velocity fluctuations components at two points $\vect{X} \pm \vect{r}/2$ is inferred directly from $\aver{\delta u_i \delta u_j}$. When $\vect{r}$ only involves homogeneous directions, $V_{ij}$ depends only on $\vect{X}$ and $t$, as $\vect{r}$ drops from the list of its independent variables. In this case, \eqref{eq:str-func} reduces to $V_{ij}(\vect{X},t)=2\aver{u_i u_j}(\vect{X},t)$, and the scale dependence of $\aver{\delta u_i \delta u_j}$ is entirely determined by the correlation functions $R_{ij}(\vect{X},\vect{r},t)=R_{ji}(\vect{X},\vect{r},t)$. Therefore, at a given $\vect{X}$ and time $t$, a local maximum/minimum of $\aver{\delta u_i \delta u_j}$ at a certain $\vect{r}$ is always associated with a negative/positive peak of $R_{ij}$. In contrast, when dealing with separations in inhomogeneous directions, $V_{ij}$ depends on both $\vect{X}$ and $\vect{r}$ and, therefore, the scale behaviour of $\aver{\delta u_i \delta u_j}$ is not only determined by the correlation tensor $R_{ij}$ but also by $V_{ij}$ itself. To simplify the interpretation of the results, whenever non-zero separations in non-homogeneous directions are involved, we always verify whether local peaks of the structure functions are actually due to local maxima/minima of the correlation functions. 

The BARC flow is statistically steady, and statistically homogeneous along the $z$ direction. The independent variables reduce to five, i.e. $(r_x,r_y,r_z,X,Y)$, and the AGKE written for the BARC flow become:
\begin{equation}
\begin{gathered}
\underbrace{-\aver{ u_k^* \delta u_j} \delta \left( \frac{\partial U_i}{\partial x_k} \right) - \aver{u_k^* \delta u_i} \delta \left( \frac{ \partial U_j }{ \partial x_k } \right) -
\aver{\delta u_k \delta u_j } \left(\frac{\partial U_i}{ \partial x_k} \right)^* - \aver{ \delta u_k \delta u_i } \left( \frac{\partial U_j}{\partial x_k} \right) ^*}_{\text{production } (P_{ij})} +\\
\underbrace{+\frac{1}{\rho}\aver{\delta{p} \frac{\partial \delta u_i}{\partial X_j}} + \frac{1}{\rho} \aver{\delta p \frac{\partial \delta u_j}{\partial X_i}}}_{\text{pressure strain } (\Pi_{ij})} - 
\underbrace{4 \epsilon_{ij}^* }_{\text{ps.dissipation } (D_{ij})} = 
\underbrace{\frac{\partial \phi_{k,ij}}{\partial r_k} + \frac{\partial \psi_{\ell,ij}}{\partial X_\ell} \, .}_{\text{divergence of fluxes}}
\end{gathered}
\label{eq:xi_gen}
\end{equation}
where $k=1,2,3$, $\ell=1,2$ and the asterisk indicates the average of a quantity over the two points $\vect{X} \pm \vect{r}/2$. The l.h.s. is interpreted as a source term and describes the net production of $\aver{\delta u_i \delta u_j}$ in the space of scales ($\vect{r}$) and in the physical space ($\vect{X}$). It features the production tensor $P_{ij}$, which describes the energy exchange between the mean and fluctuating field, the pseudo-dissipation tensor $D_{ij}$ and the pressure-strain tensor $\Pi_{ij}$, which accounts for redistribution.
When considering a separation vector with null components in the non-homogeneous directions, $D_{ij}(\bm{X},\bm{r})$ does not have scale dependence and, besides a multiplicative factor, it corresponds to the pseudo-dissipation tensor $\epsilon_{ij}$ for the single-point Reynolds stresses.
Overall, the source term equals the divergence of the five-dimensional flux vector $\vect{\Phi}_{ij}=(\vect{\phi}_{ij},\vect{\psi}_{ij})$ at the r.h.s. $\vect{\phi}_{ij}$ and $\vect{\psi}_{ij}$ are the scale and space components of the flux vector that describe the scale-space interactions and are defined as:
\begin{equation}
\phi_{k, ij} = \underbrace{\left \langle \delta U_k \delta u_i \delta u_j \right \rangle}_{\phi_{k,ij}^{\text{mean}}} + \underbrace{\left \langle \delta u_k \delta u_i \delta u_j \right \rangle}_{\phi_{k,ij}^{\text{turb}}} \underbrace{-\ 2 \nu \frac{\partial}{\partial r_k}\left \langle \delta u_i \delta u_j \right \rangle}_{\phi_{k,ij}^{\text{visc}}} \ \ k=1,2,3
\label{eq:phi}
\end{equation} 
and
\begin{equation}
%\begin{gathered}
\psi_{\ell, ij} = 
 \underbrace{  \aver{ U_k^* \delta u_i \delta u_j }}_{\psi_{\ell,ij}^{\text{mean}}} 
 + \underbrace{ \aver{ u_k^* \delta u_i \delta u_j }}_{\psi_{\ell,ij}^{\text{turb}}}
 +\underbrace{ \frac{1}{\rho} \aver{ \delta p \delta u_i } \delta_{kj} 
              +\frac{1}{\rho} \aver{ \delta p \delta u_j } \delta_{ki}}_{\psi_{\ell,ij}^{\text{press}}} 
\ \underbrace{- \frac{\nu}{2}\frac{\partial}{\partial X_k} \aver{ \delta u_i \delta u_j }}_{\psi_{\ell,ij}^{\text{visc}}} \ \ \ell=1,2.
%\end{gathered}
\label{eq:psi}
\end{equation}
The space flux $\vect{\psi}_{ij}$ features the mean, turbulent and pressure transport and the viscous diffusion, like in the budget equations for the Reynolds stresses \citep{pope-2000}, while the scale flux $\vect{\phi}_{ij}$ features all these contributions, but the pressure transport.
The sum of the three diagonal components of $\aver{\delta u_i \delta u_j}$ yields the Generalised Kolmogorov Equation (GKE) for the turbulent kinetic energy \citep{hill-2001,danaila-etal-2001,marati-casciola-piva-2004}.

The AGKE terms for the BARC flow have been computed by post-processing the DNS database described above. The code used for the analysis extends a high-performance software tool written for the GKE and developed by \cite{gatti-etal-2019}, freely available at \url{https://github.com/davecats/gke}. Given the size of the problem, the AGKE terms have been computed in two sub-boxes within the computational domain, both encompassing the full body width. One is defined by $0 \le x \le 5$, $0.5 \le y \le 1.5$ and is used for the analysis of the region over the cylinder side described in \S\ref{sec:side}; the other is defined by $5 \le x < 10$, $-1 \le y \le 1$, and is used for the near-wake analysis described in \S\ref{sec:wake}.

\section{Flow over the cylinder wall}
\label{sec:side}

%----------------------------------------------
\subsection{Large- and small-scale structures}
\label{sec:side-structures}

\begin{figure}
\centering
\includegraphics[width=0.7\columnwidth]{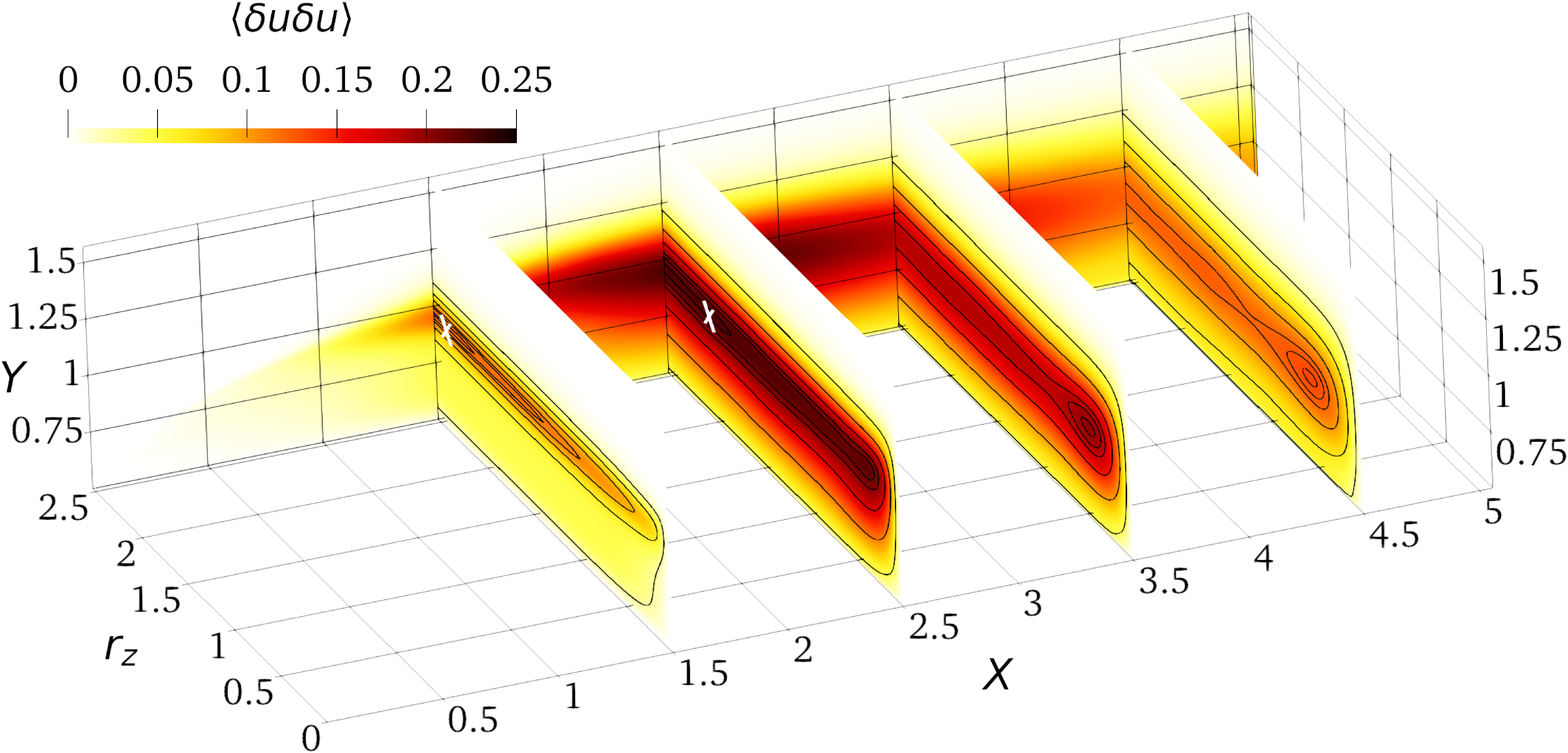}
\includegraphics[width=0.7\columnwidth]{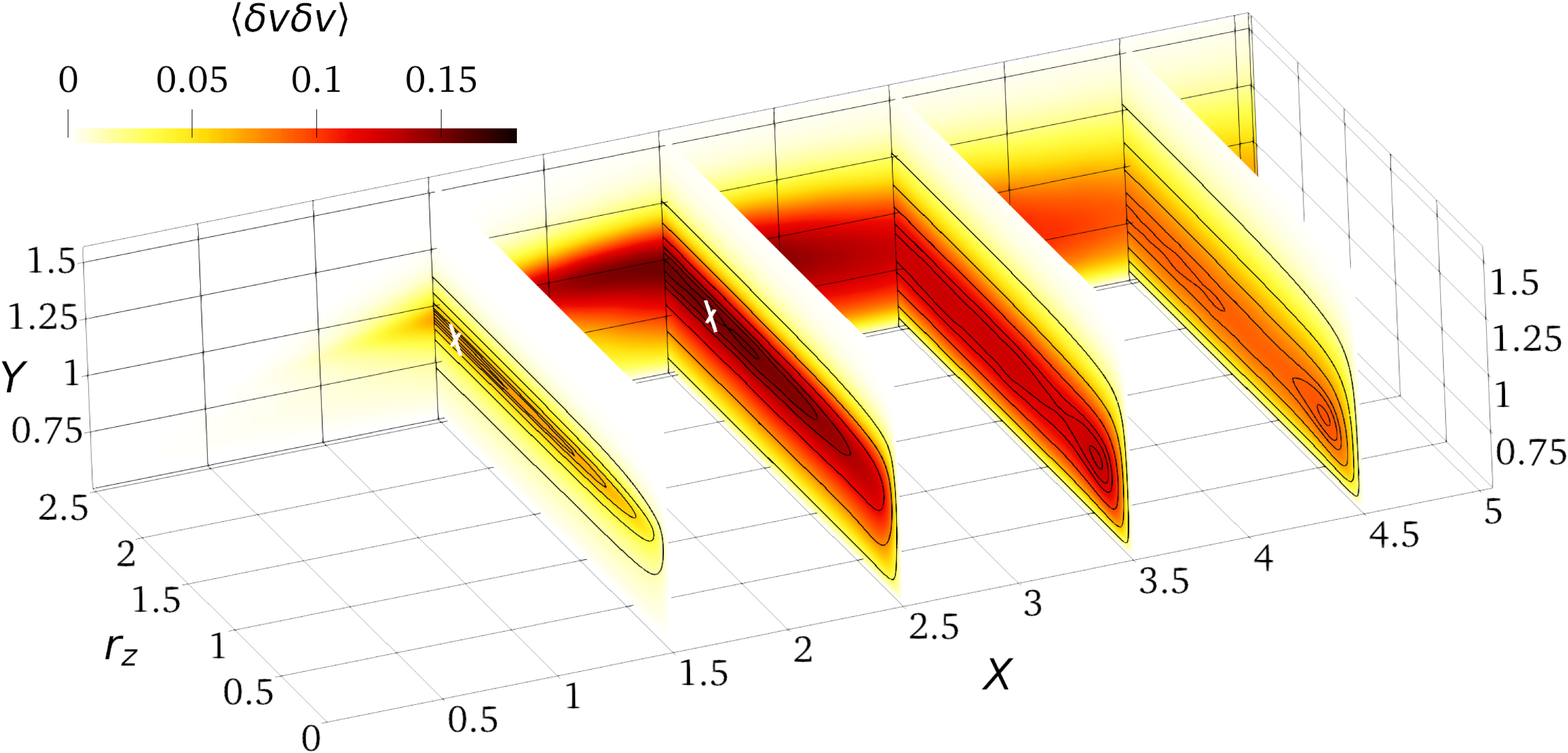}
\includegraphics[width=0.7\columnwidth]{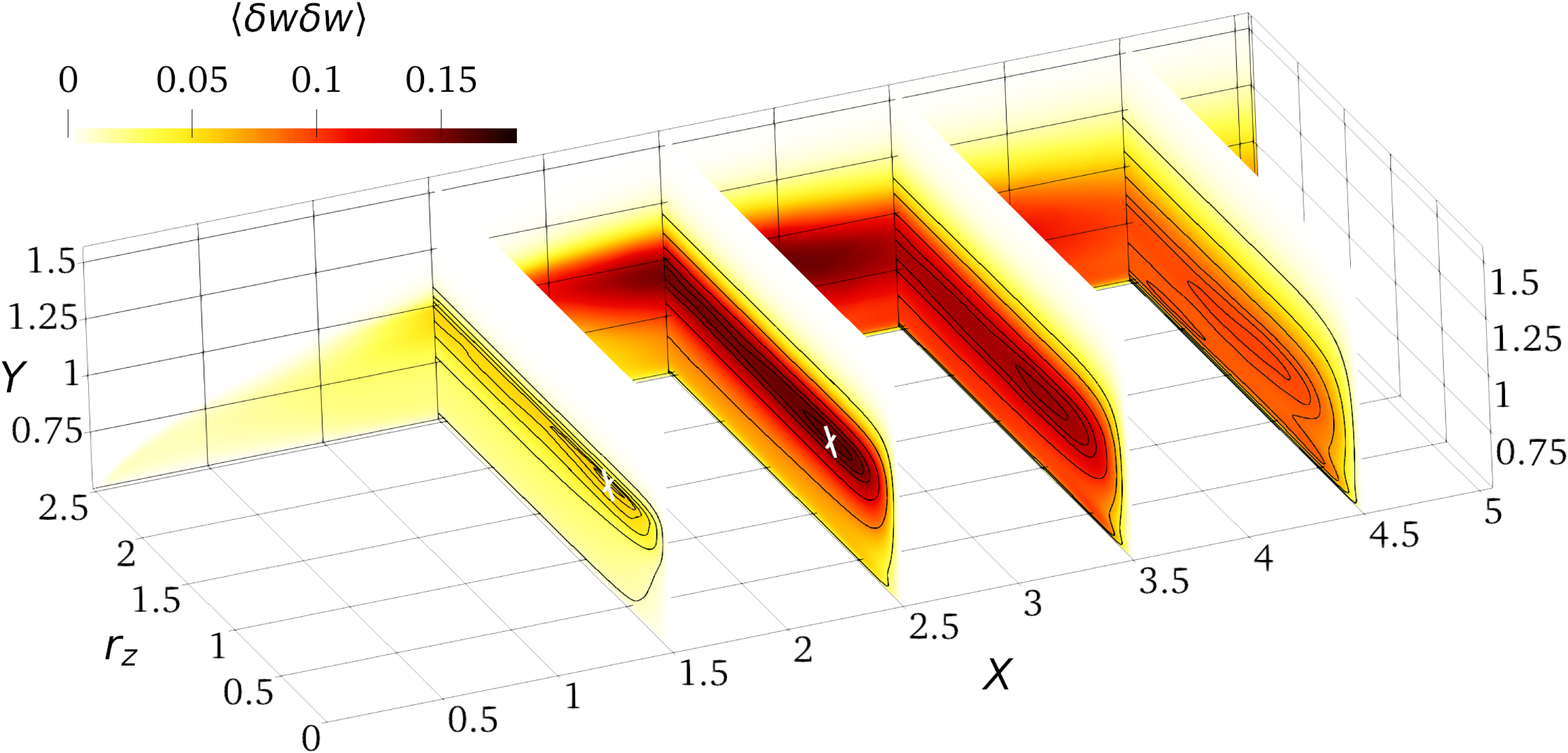}
\includegraphics[width=0.7\columnwidth]{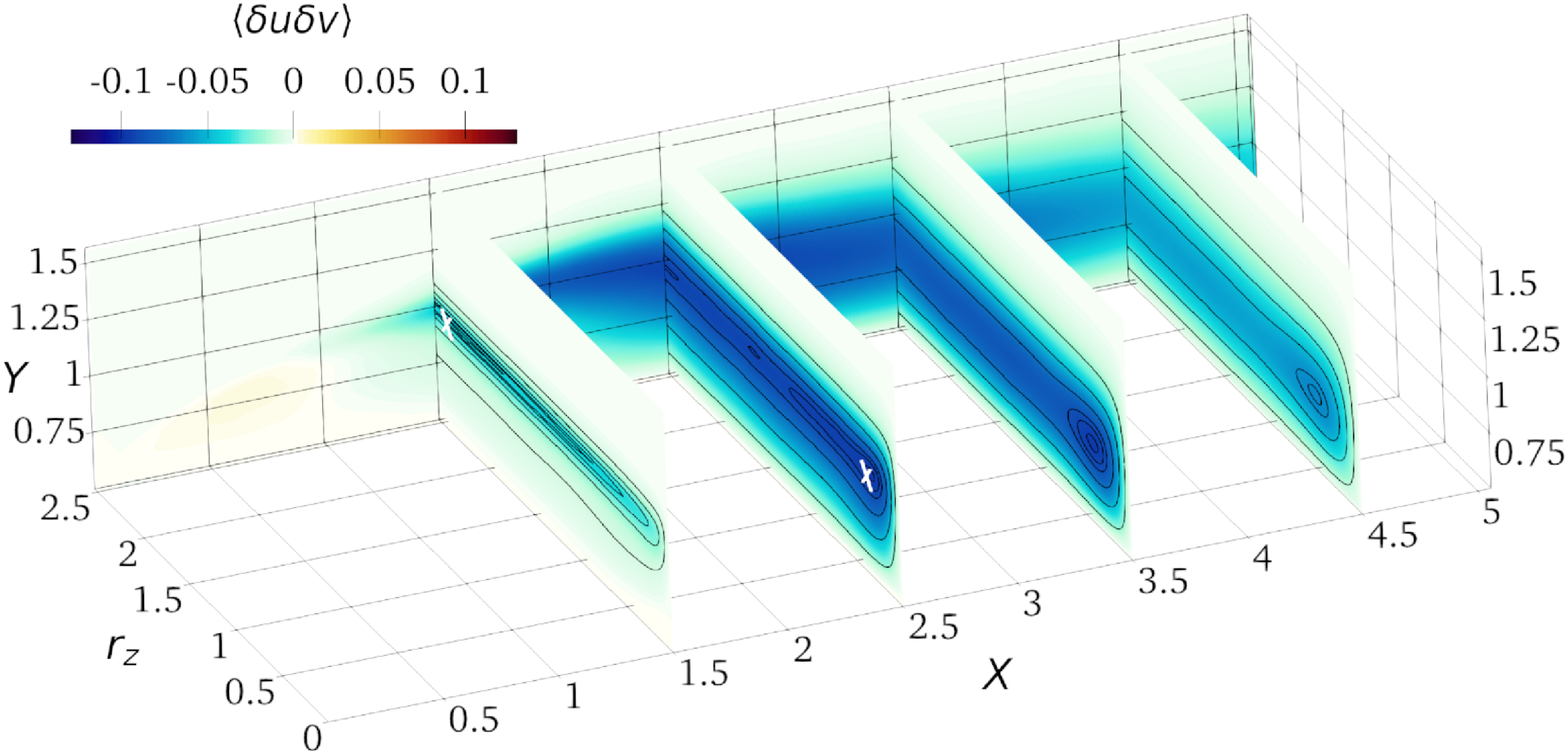}
\caption{Components of the structure function tensor plotted in the space $(X,Y,r_z)$ over the cylinder side. Contours in the planes $X=1.5,2.5,3.5,4.5$ are for the $99\%, 95\%, 90\%, 75\%, 50\%$ and $20\%$ of the in-plane maximum. White crosses are used to identify local maxima when not easily visible.}
\label{fig:side_str-func}
\end{figure}

The flow structures occurring over the cylinder side, already visualised in the snapshot of figure \ref{fig:lambda2-vorticity}, are statistically described through the diagonal components of the structure function tensor and $\aver{\delta u \delta v}$, in the $r_x=r_y=0$ space shown in figure \ref{fig:side_str-func}. 

Close to the LE, two-dimensional structures with spanwise-oriented vorticity are generated through a Kelvin--Helmholtz (KH) instability of the separating shear layer. They are invariant under spanwise translations, hence their contribution to $\aver{\delta u_i \delta u_j}$ is largest at the maximum separation $r_z=L_z/2=2.5$, where $\aver{\delta u_i \delta u_j} \approx V_{ij}$. Indeed, the associated correlation function $R_{ij}$ decreases monotonically with $r_z$, being $R_{ij} = \aver{u_i u_j}$ at $r_z=0$ and $R_{ij} \approx 0$ at $r_z \approx 2.5$. The statistical footprint of these KH rolls does not appear in the map of $\aver{\delta w \delta w}$, since they do not induce $w$ fluctuations. The characteristic scales of the rolls are identified by inspecting the $r_x \neq 0$ and $r_y \neq 0$ spaces.

Further downstream, at about $X \approx 1.3$, under the action of the mean shear the KH rolls become unstable and develop a spanwise modulation with characteristic lengthscale of $r_z \approx 2.4$. The tilted rolls can now induce $w-$fluctuations too and a local maximum in $\aver{\delta w \delta w}$ appears at $r_z \approx 0.5$, indicative of the streamwise scale of the (unmodulated) rolls.
Part of the $u-$ and $v-$fluctuations is reoriented into $w-$fluctuations by the pressure strain (see \S\ref{sec:side-pstrain}) for kinematic reasons. As a result, here the $r_x$ and $r_y$ scale information of the unmodulated KH rolls (see later figure \ref{fig:rx-KH}) is partially transferred to $R_{ww}(r_z)$, and therefore it can be appreciated by this local maximum of $\aver{\delta w \delta w}$ at $r_z \approx 0.5$.
Later on, the spanwise-modulated tubes are further stretched into hairpin-like vortices, with a slightly shrunk spanwise scale of $r_z \approx 1.8$, indicated by the local peaks of $\aver{\delta u \delta u}$, $\aver{\delta v \delta v}$ and $\aver{\delta u \delta v}$. Such hairpin-like structures are visible in the instantaneous picture of figure \ref{fig:lambda2-vorticity}, and have been described by \cite{cimarelli-leonforte-angeli-2018} via spatial correlation functions. These hairpin structures remain visible after $X = 2.5$, but they become progressively weaker and are hardly detectable for $X>3$.

Transition to turbulence occurs at $X \approx 2$, after which small-scale turbulent structures coexist with the large ones created by the instability of the LE shear layer. Figure \ref{fig:lambda2-vorticity} confirms the observation, put forward by \cite{cimarelli-leonforte-angeli-2018} and \cite{chiarini-quadrio-2021}, that the dominant structures here are streamwise-aligned vortical structures. Their spanwise lengthscale is $r_z \approx 0.5$, as deduced from the maxima of the structure functions $\aver{\delta u \delta u}$, $\aver{\delta v \delta v}$ and $\aver{\delta u \delta v}$ at $Y \approx 0.9$ in figure \ref{fig:side_str-func}.
The local peak of $\aver{\delta u \delta u}$ derives from the interaction of the vertical motions induced by these vortices with the local positive $\partial U/\partial y>0$, that produce regions of positive and negative $u-$fluctuations at the lateral sides of the structures, yielding $R_{uu}<0$.
As expected for streamwise-aligned structures, $\aver{\delta w \delta w}$ does not show a localised peak at this scale \citep{gatti-etal-2020}. The magnitude of $\aver{\delta u \delta u}$ and $\aver{\delta v \delta v}$ show that these vortices are mainly organised in streamwise fluctuations. In terms of streamwise evolution, the intensity of the streamwise vortices is maximum at $X \approx 2.5$ and then decreases after the reattachment point, where viscous dissipation becomes stronger \citep[see figure 15 in][]{chiarini-quadrio-2021}.

\begin{figure}
\centering
\includegraphics[trim=0 0 0 400,clip,width=0.49\columnwidth]{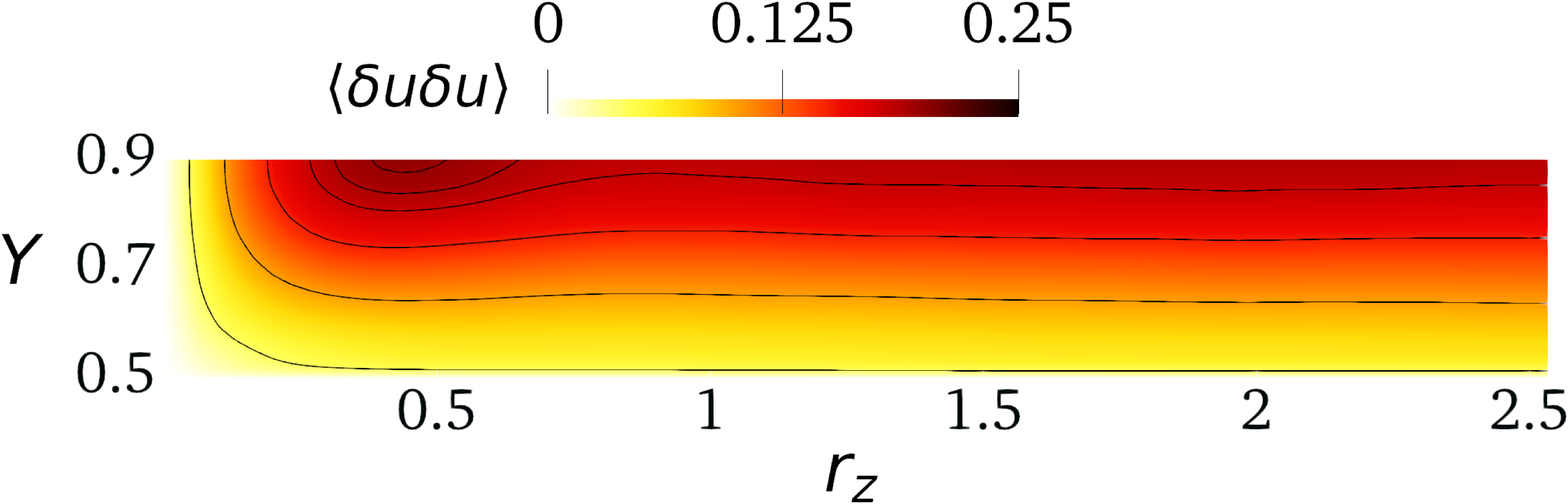}
\includegraphics[trim=0 0 0 400,clip,width=0.49\columnwidth]{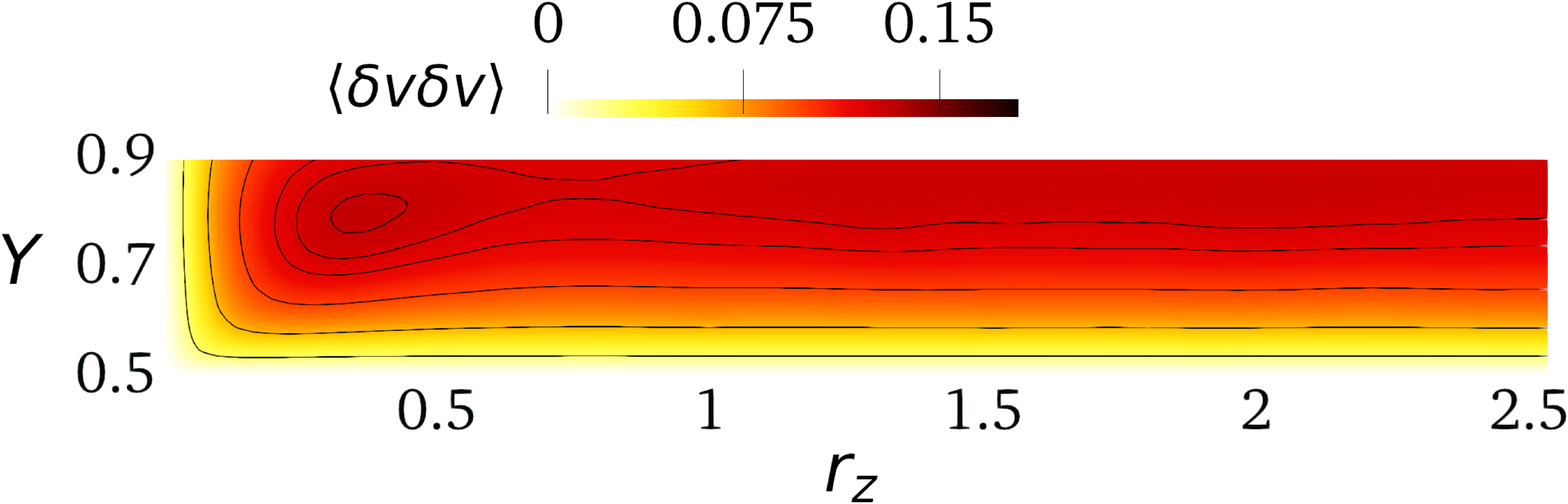}\\
\includegraphics[trim=0 0 0 400,clip,width=0.49\columnwidth]{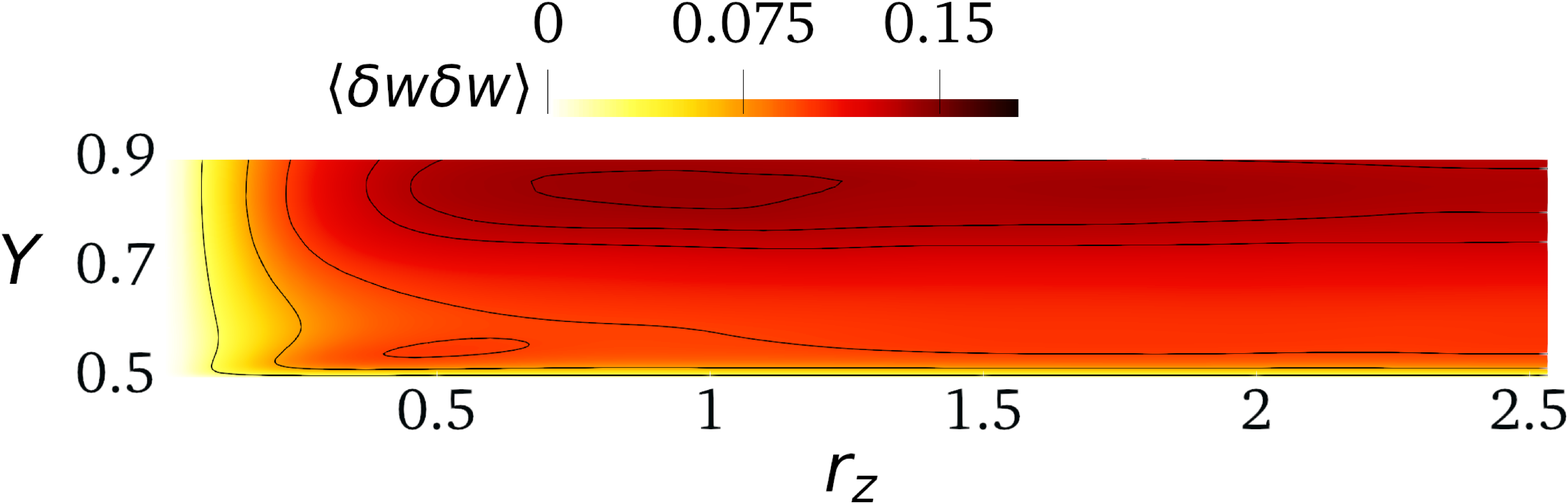}
\includegraphics[trim=0 0 0 400,clip,width=0.49\columnwidth]{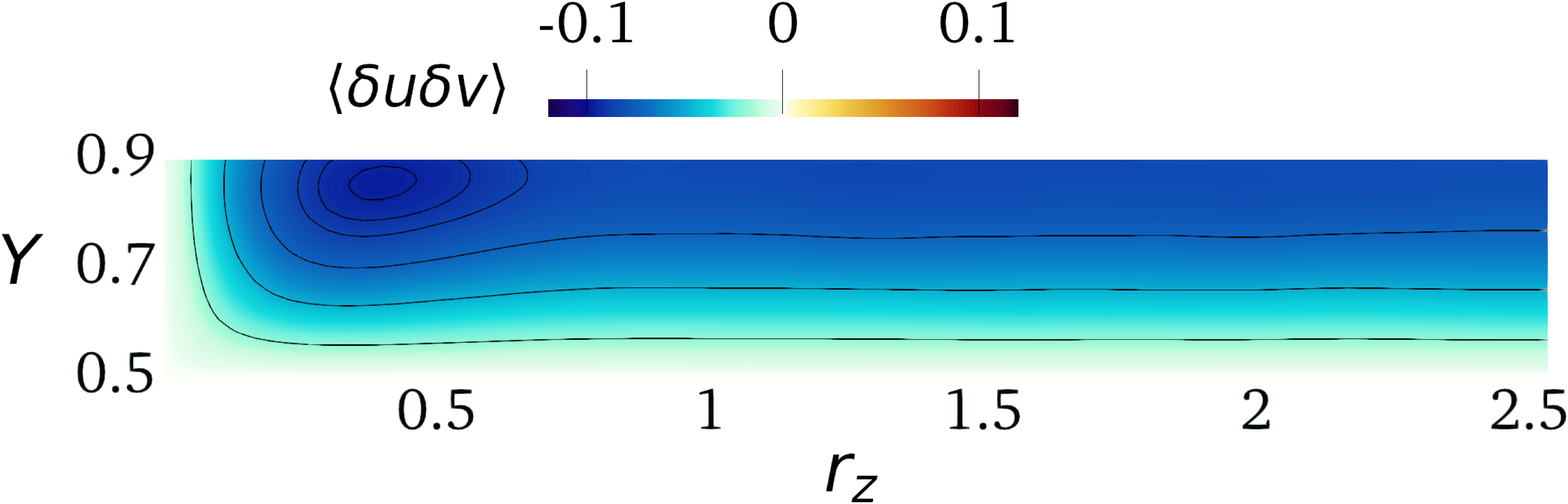}
\caption{Two-dimensional view of the structure functions terms, in the plane $X=3.5$. Panels as in figure \ref{fig:side_str-func}.}
\label{fig:side_str-func-zoom}
\end{figure}

In other regions of the flow, i.e. in the reverse boundary layer close to the wall and downstream of the reattachment, the small-scale fluctuations are preferentially organized in small-scale $w$-structures. Indeed, $\aver{\delta w \delta w}$ is the largest diagonal component for $Y \rightarrow 0.5$ and $ 2 < X < 5$. The characteristic scale of these $w$-structures is $r_z \approx 0.5$, as seen by the local maximum at $Y \approx 0.55$ in figure \ref{fig:side_str-func-zoom} with a zoom of the streamwise location $X=3.5$. This peculiar small-scale organisation of near-wall turbulence will be further described below in \S\ref{sec:side-pstrain}, where it will be linked to flow splatting.

Figures \ref{fig:rx-KH} and \ref{fig:ry-KH} consider non-zero separations in the inhomogenous directions, and identify the characteristic streamwise and vertical scales of the KH rolls close to the LE. Figure \ref{fig:rx-KH} deals with $r_x$ and plots $\aver{\delta v \delta v}$ (left) and $R_{vv}$ (right) in the $(X,r_x)$ space with $r_y=r_z=0$, $Y=0.99$ and $0.5 \le X \le 3$.
\begin{figure}
\centering
\includegraphics[trim=500 0 500 0,clip,width=0.49\textwidth]{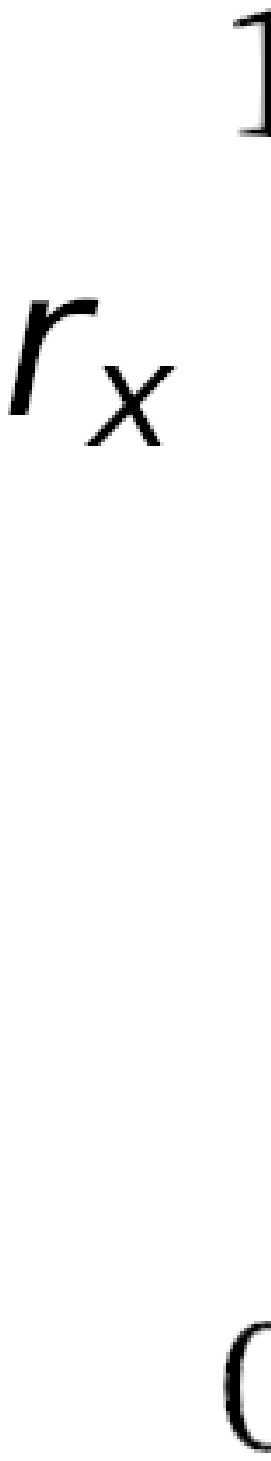}
\includegraphics[trim=500 0 500 0,clip,width=0.49\textwidth]{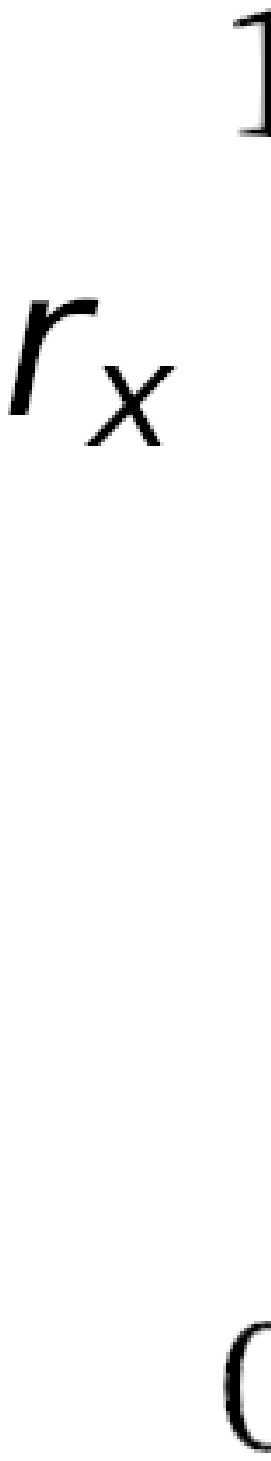}
\caption{$\aver{\delta v \delta v}$ (left) and $R_{vv}$ (right) in the $(X,r_x)$ space with $r_y=r_z=0$, $Y=0.99$ and $0.5 \le X \le 3$.}
\label{fig:rx-KH}
\end{figure}
The characteristic streamwise scale of these rolls is identified by the peak of $\aver{\delta v \delta v}$: for $X>1.3$ it occurs at $r_x \approx 0.3-0.5$, and mildly increases with $X$. The $v-$fluctuations induced by the KH rolls are also observed via the large $R_{vv}<0$ for the same range of $r_x$. For $r_x>0.5$ the correlation function $R_{vv}$ presents alternating positive/negative regions. Similarly to what observed by \cite{thiesset-danaila-antonia-2014} and \cite{alvesportela-papadakis-vassilicos-2017} for the von K\'{a}rm\'{a}n vortices in the circular and square cylinder wakes, this is the statistical trace of the quasi-periodic generation of the KH rolls. The streamwise separation for these peaks is linked to the KH frequency \citep[$\approx 1.2-1.4$; see][]{chiarini-quadrio-2021} by the mean convection velocity $U_c$ of the rolls. Since the first positive peak of $R_{vv}$ has $r_x \approx 0.5-0.7$, it follows that $U_c \approx 0.7-0.8$.

\begin{figure}
\centering
\includegraphics[trim=500 0 500 0,clip,width=0.49\textwidth]{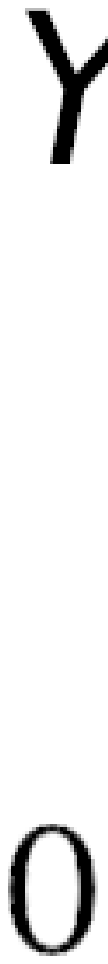}
\caption{$\aver{\delta u \delta u}$ in the $(Y,r_y)$ space with $r_x=r_z=0$, $X=1.5$ and $0.5 \le Y \le 1.55$.}
\label{fig:ry-KH}
\end{figure}
The wall-normal separation $r_y$ is considered in figure \ref{fig:ry-KH}, which plots $\aver{\delta u \delta u}$ in the $(Y,r_y)$ space for $r_x=r_z=0$, $X=1.5$ and $0.5 \le Y \le 1.55$. At $Y \approx 1$ (where the KH rolls are centred) $\aver{\delta u \delta u}$ peaks at $r_y \approx 0.15-0.25$, which identifies their characteristic vertical length scale. The rolls induce $u$ fluctuations at their vertical sides that yield a negative correlation $R_{uu}<0$ at the same range of $r_y$. 
In the $r_x=r_z=0$ space, large values of $\aver{\delta u \delta u}$ fall also along two oblique branches described by $Y=Y_0 \pm r_y/2$. These large values of $\aver{\delta u \delta u}$, however, are not due to $R_{uu}$, but derive from $V_{11}$ (see equation \ref{eq:str-func}). This is consistent with a layer of high velocity fluctuations, i.e. the shear layer separating from the LE, surrounded below and above by two regions where the fluctuations are weaker. Indeed, $V_{11}$, and thus $\aver{\delta u \delta u}$, has large values as long as at least one of the two points $\bm{X} \pm \bm{r}/2$ is within the shear layer. Note that the width of these branches ($r_y \approx 0.2$) is a measure of the vertical width of the shear layer itself at the considered $X$. As shown in the following, this peculiar pattern is also observed for other quantities.

%%%%%%%%%%%%%%%%%%%%%%%%%%%%%%%%%%%%%%%%%%
\subsection{Production}
\label{sec:side_prod}

\begin{figure}
\centering
\includegraphics[width=0.7\columnwidth]{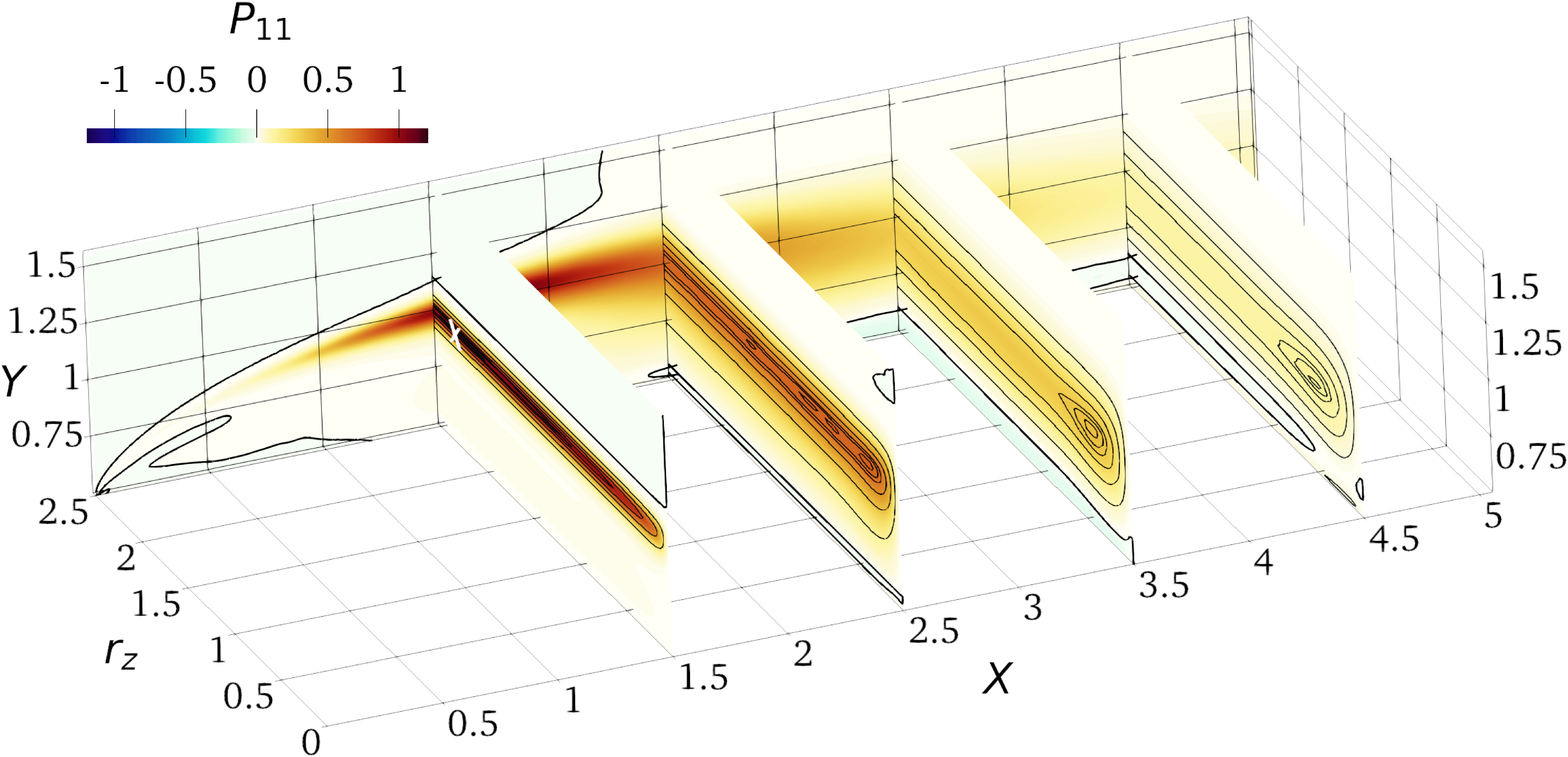}
\includegraphics[width=0.49\columnwidth]{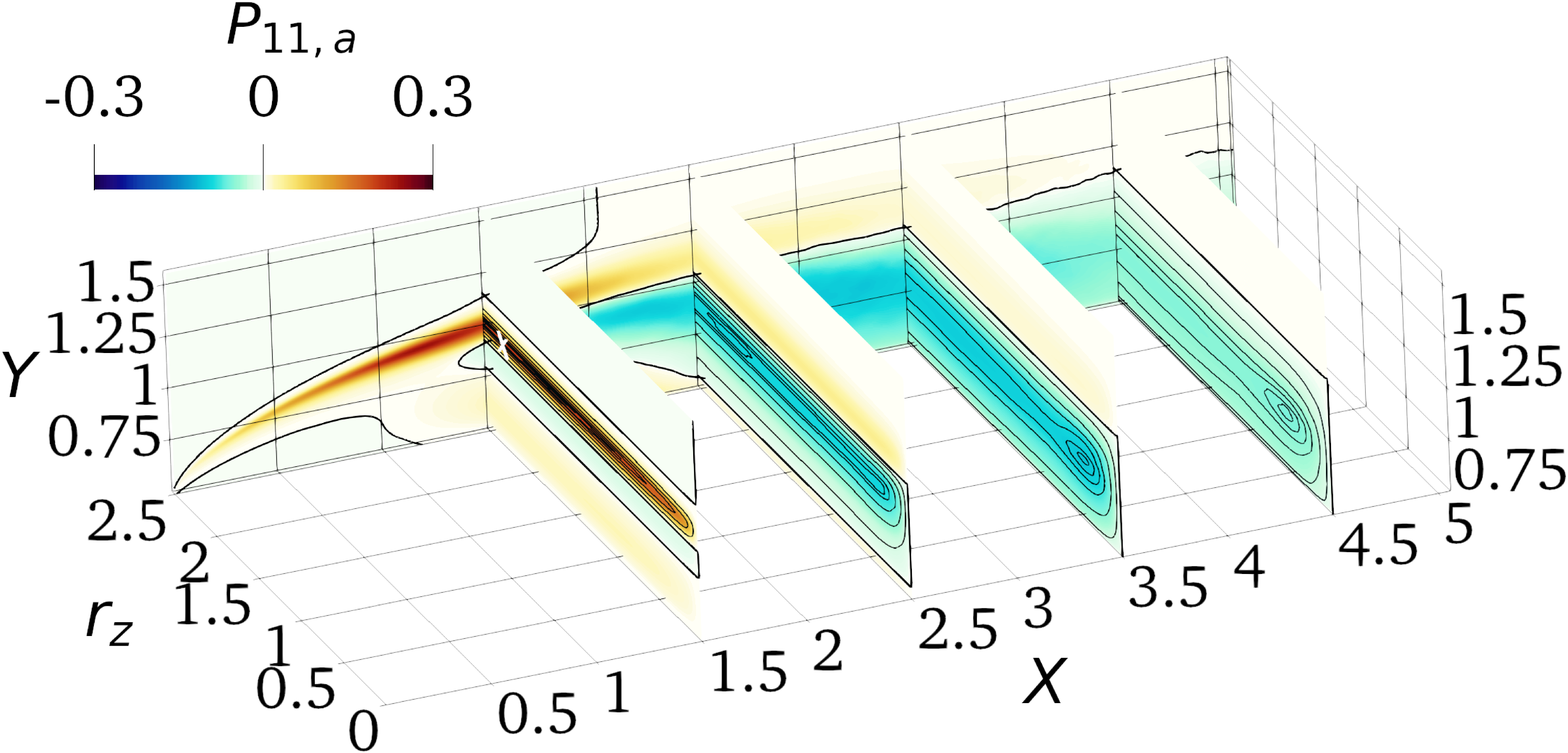}
\includegraphics[width=0.49\columnwidth]{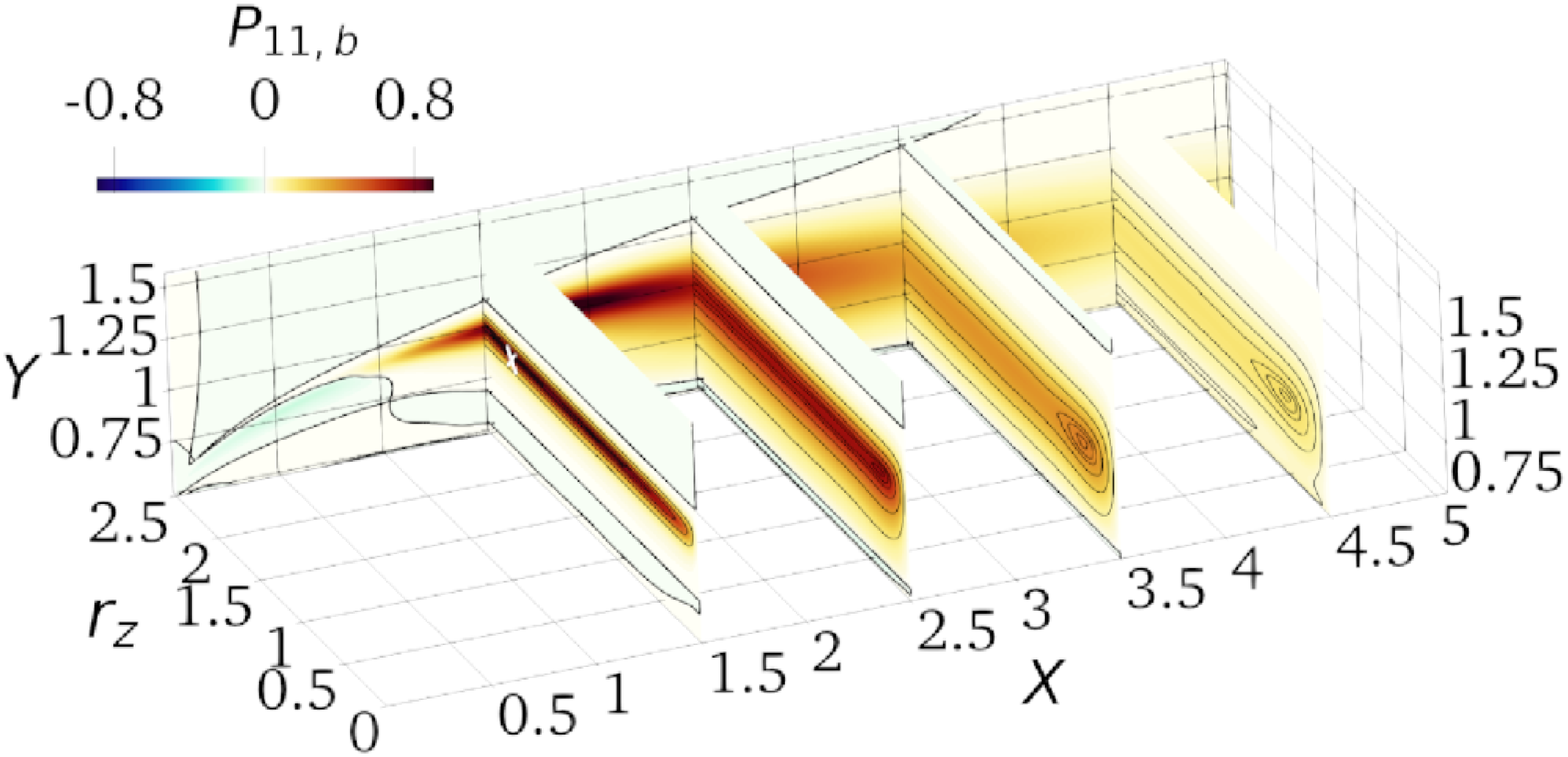}
\includegraphics[width=0.7\columnwidth]{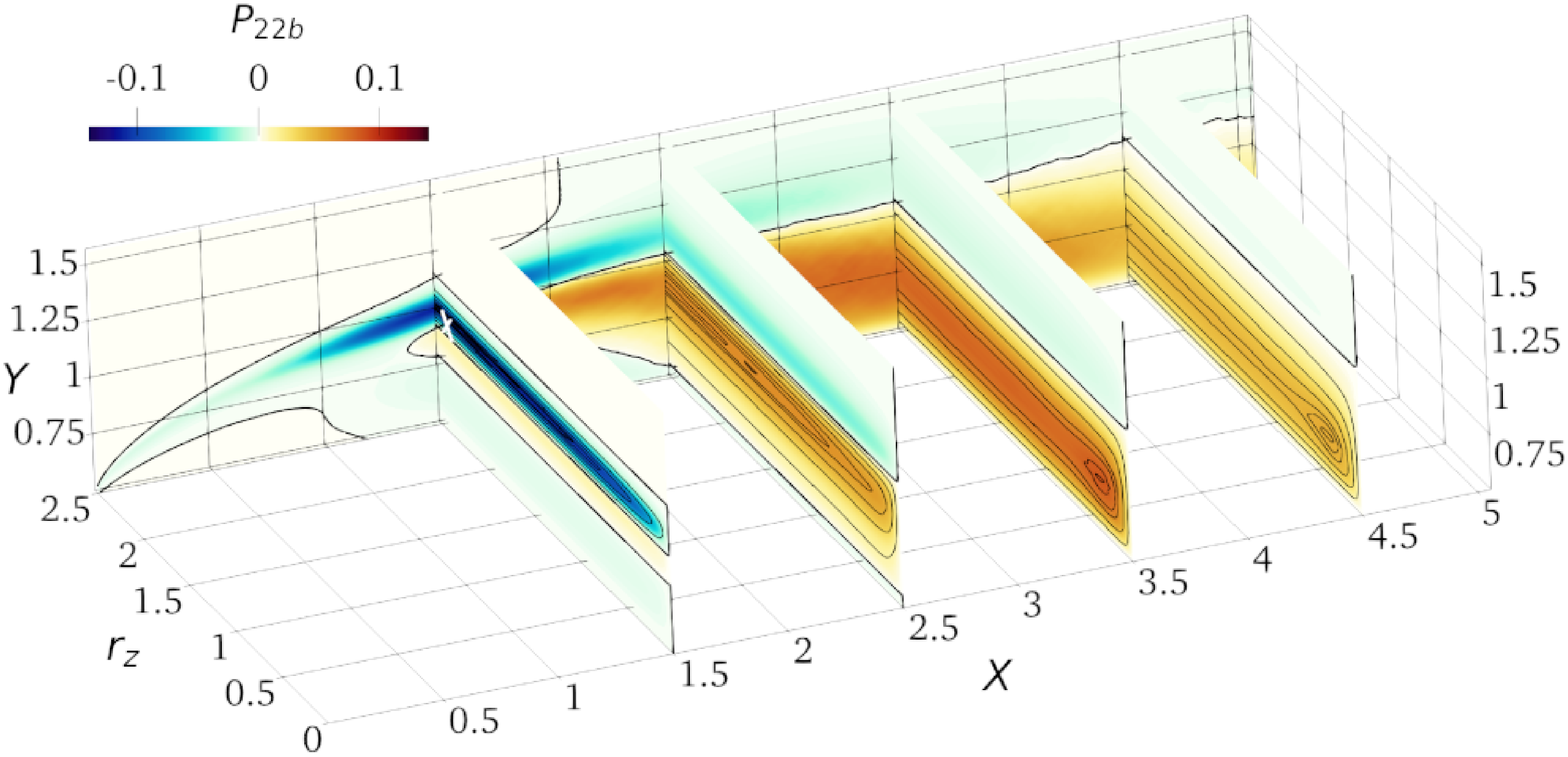}
\caption{Top: $P_{11}$ (with the two contributions $P_{11,a}$ and $P_{11,b}$ plotted separately in the central row). Bottom: $P_{22,b}$. The thick black line marks the zero contour level. White crosses are used to identify local maxima when not easily visible.}
\label{fig:side_production}
\end{figure}

\begin{figure}
\centering
\includegraphics[trim=500 0 500 0,clip,width=0.49\textwidth]{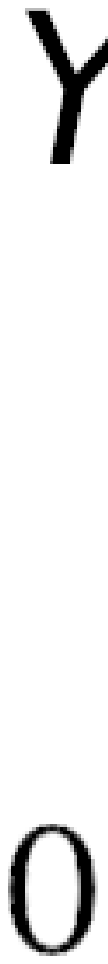}
\includegraphics[trim=500 0 500 0,clip,width=0.49\textwidth]{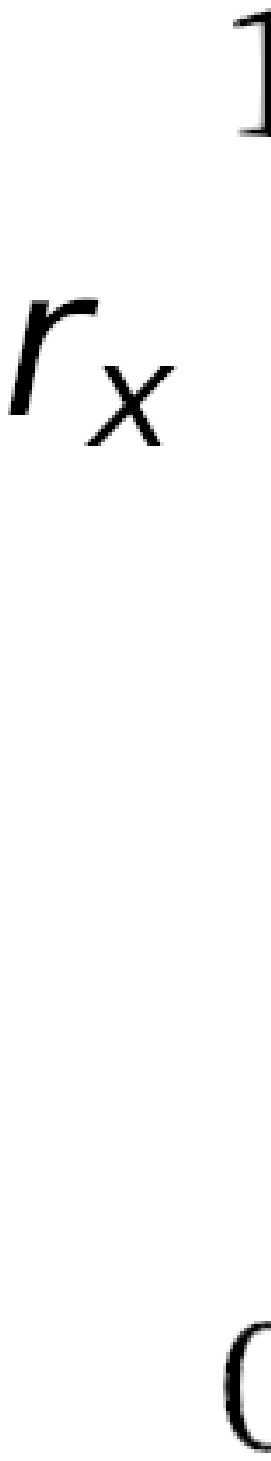}
\caption{Production terms for $\aver{\delta u \delta u}$ and $\aver{\delta v \delta v}$ associated with the KH rolls. Left: $P_{11}$ in the $(Y,r_y)$ space with $r_x=r_z=0$, $X=1.5$ and $0.5 \le Y \le 1.55$. Right: $P_{22}$ in the $(X,r_x)$ space with $r_y=r_z=0$, $Y=0.99$ and $0.5 \le X \le 3$.}
\label{fig:KH-prod}
\end{figure}

Production of the scale Reynolds stresses along the side of the cylinder in the $r_x=r_y=0$ space is considered in figure \ref{fig:side_production}. For the normal components $P_{33}=0$ and the only nonzero production terms in this subspace reduce to:

\begin{equation}
P_{11}= \underbrace{-2\aver{\delta u \delta u} \frac{\partial U}{\partial x} }_{P_{11,a}}  \underbrace{ -2\aver{\delta u \delta v} \frac{\partial U}{\partial y} }_{P_{11,b}}
\label{eq:prod_uu}
\end{equation}
and
\begin{equation}
P_{22}= \underbrace{-2\aver{\delta u \delta v} \frac{\partial V}{\partial x} }_{P_{22,a}}  \underbrace{ -2\aver{\delta v \delta v} \frac{\partial V}{\partial y} }_{P_{22,b}}.
\label{eq:prod_vv}
\end{equation}

The map of $P_{11}$ shows two independent production mechanisms of streamwise velocity fluctuations. The first and most intense is associated with the large-scale KH rolls and thus occurs along the LE free shear layer. It is visualised by the local maximum of $P_{11}$ at $(Y,r_z) \approx (1,2.4)$ for $0.75 \le X \le 2$ and by the maximum at $(Y,r_y) \approx (1.1,0.2)$ in the left panel of figure \ref{fig:KH-prod}, that plots $P_{11}$ in the same $r_x=r_z=0$ plane considered in figure \ref{fig:ry-KH}. The second and weaker peak is related to the small-scale streamwise-aligned vortices and takes place in the core of the primary vortex and downstream the reattachment point; see the peak of $P_{11}$ at $(Y,r_z) \approx (1,0.5)$ for $2.5 \le X \le 5$. %XXX Can this scale be expressed in wall units to see if it matches with canonical flows? If yes, it means that boundary layer is more or less in equilibrium XXX 
Interestingly, unlike in classic turbulent wall-bounded flows, $P_{11}$ is not positive at all positions and scales: in fact, it becomes negative below the first part of the shear layer and in the near-wall region of the aft cylinder side. 

The two contributions $P_{11,a}$ and $P_{11,b}$, shown in the central panels of figure \ref{fig:side_production}, are considered separately to discern the production mechanism. The sign of $P_{11,a}$ is determined by $\partial U/ \partial x$, whereas its scale modulation is prescribed by $\aver{\delta u \delta u}$. In contrast, the sign of $P_{11,b}$ is not enforced by $\partial U / \partial y$ alone, since $\aver{\delta u \delta v}$ is not defined in sign. However, figure \ref{fig:side_str-func} has shown that $\aver{\delta u \delta v}$ is negative everywhere (except below the shear layer very close to the LE), so that positive $P_{11,b}$ generally corresponds to positive $\partial U / \partial y$.

Production along the shear layer is partially due to the large negative $\partial U/ \partial x < 0$ across the layer itself, and is thus seen in the map of $P_{11,a}$, with a positive peak at $(X,Y,r_z) \approx (1.5,1,2.4)$. $P_{11,a}$ is positive also within the secondary vortex, indicating that $\partial U / \partial x<0$ contributes to sustaining streamwise fluctuations, albeit only weakly, also in this flow region. In contrast, within the primary vortex and after the reattachment point, $P_{11,a} <0$ for $0.5 \le Y < 1$. Both large and small structures contribute to production: indeed two negative peaks of $P_{11,a}$ are present, one at $r_z \approx 2$ (associated with the hairpin vortices for $X \le 2.5 $) and the other at $r_z \approx 0.5$ (associated with the streamwise-aligned structures, for larger $X$).

While $P_{11,a}$ dominates in the upstream portion of the LE shear layer, $P_{11,b}$ takes over for $X>1$, where it is positive almost everywhere. The positive production associated with the modulated KH rolls in the shear layer contains both $P_{11,a}$ and $P_{11,b}$, but the latter is larger since the mean shear $\partial U / \partial y >0$ dominates in this region. In contrast, the positive $P_{11}$ associated with the smaller structures in the aft cylinder side is entirely due to $\partial U / \partial y>0$, since  at the corresponding positions and scales $P_{11,b}$ is positive and larger than the negative $P_{11,a}$. $P_{11,b}$ is negative only in the separated shear layer at $X<1$ \citep{cimarelli-etal-2019-negative} and in the near-wall region at $X \le 2$, where the reverse boundary layer separates and creates the secondary vortex.

A further remark concerns the two regions with $P_{11}<0$. The one beneath the shear layer is due to $P_{11,b}$ and has been already addressed by \cite{cimarelli-etal-2019-negative} and \cite{chiarini-quadrio-2021}. The negative region close to the wall, instead, results from the positive $\partial U / \partial x$ across the reattachment and will be further considered in \S\ref{sec:side-pstrain}, since near-wall turbulence here substantially differs from other canonical flows. 

The second production term $P_{22}$ is dominated by $P_{22,b}$, much larger than $P_{22,a}$, so that the production of vertical fluctuations is determined by $\partial V / \partial y$; this is consistent with the observations made by \cite{moore-etal-2019} and \cite{chiarini-quadrio-2021} for the single-point Reynolds stress budget. Hence, $P_{22,b}$ alone is shown in the bottom panel of figure \ref{fig:side_production}. Note that, because of incompressibility, $P_{22,b}$ and $P_{11,a}$ must have opposite sign. The large-scale production mechanism associated with the KH instability is generally a sink for the large-scale vertical fluctuations; in fact, as shown in \S\ref{sec:side-pstrain}, in the LE shear layer the large-scale vertical fluctuations are mainly sustained by the pressure-strain term. The large positive $\partial V / \partial y$ across the LE shear layer, indeed, leads to $P_{22,b}<0$ at all spanwise scales in the $r_x=r_y=0$ space, with a negative peak at $(X,Y,r_z) \approx (1.5,1,2)$, consistently with the scale and location of the modulated KH rolls. Similarly a large negative peak of $P_{22,b}$ is detected at $(X,r_x)=(1.3,0.2)$ in the right panel of figure \ref{fig:KH-prod}, where the same $r_y=r_z=0$ plane as in figure \ref{fig:rx-KH} is considered. Note, however, that when considering $r_x \neq 0$, $P_{22,b}$ becomes slightly positive for $X \ge 1.7$ indicating a weak source of large-scale vertical fluctuations. \cite{cimarelli-etal-2019-negative} and \cite{cimarelli-franciolini-crivellini-2020} observed that the negative $P_{22,b}$ along the LE shear layer is a peculiarity of flows around bodies with sharp corners, and conjectured that this explains, at least partially, the enhanced thickness of the primary vortex compared to flows past bodies with rounded corners; the same results are also observed in \cite{chiarini-quadrio-2021b}. Unlike in the LE shear layer, the small-scale production produces both streamwise and vertical fluctuations. Within the primary vortex and after the reattachment point, indeed, $P_{22,b}>0$ with a distinct peak at the small scale $r_z \approx 0.5$ of the streamwise-aligned vortices.

In closing this section, we remark that the strongly inhomogeneous features observed for $P_{11}$ and $P_{22}$ are challenging for turbulence theories and closures. Indeed, across the relatively small scale made by the length of the rectangular cylinder, we can simultaneously recognise regions where production is driven by flow accelerations/decelerations, $P_{11,a}$ and $P_{22,b}$, and regions where production is on the contrary driven by the more complex interaction of Reynolds shear stresses and mean velocity gradients, $P_{11,b}$. Of particular interest is their behaviour in the reverse flow within the primary vortex (see the near-wall region for $1 \le X \le 2.6$), where 
%a negative production of both streamwise and vertical velocity fluctuations:
 $P_{11,b}<0$ and $P_{22}<0$. This is driven by a weak vertical acceleration $\partial V / \partial y > 0$, leading to $P_{22}<0$, and by a negative correlation of the Reynolds shear stresses $-\aver{\delta u \delta u}$ with the mean streamwise shear $\partial U / \partial y$, thus leading to $P_{11,b}<0$. The latter cannot be described within the classic mixing length hypothesis \citep{cimarelli-etal-2019-negative}, and represents an issue for two-equations eddy viscosity closures.

%%%%%%%%%%%%%%%%%%%%%%%%%%%%%%%%%%%
\subsection{Redistribution}
\label{sec:side-pstrain}

\begin{figure}
\centering
\includegraphics[width=0.7\columnwidth]{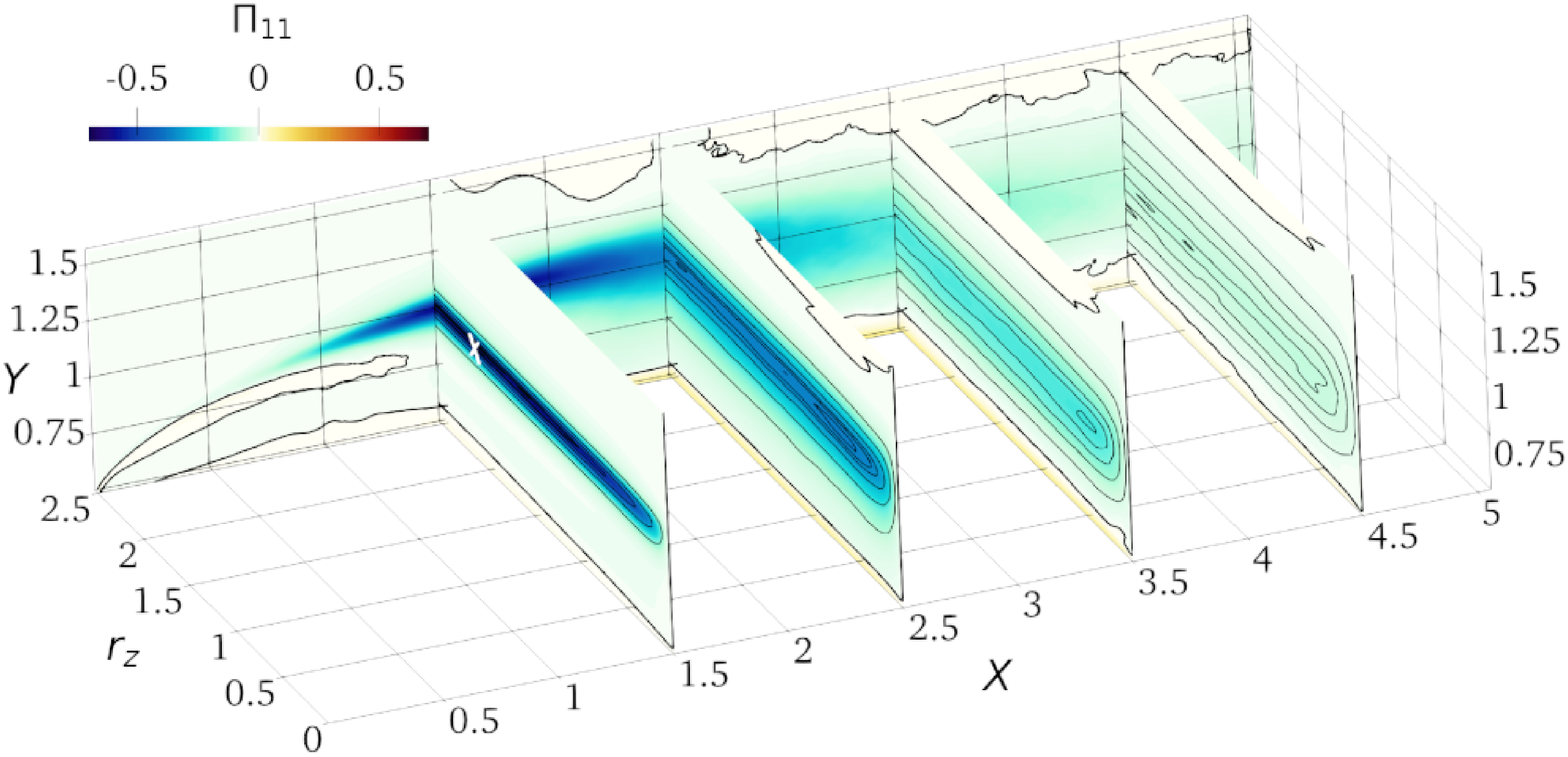}
\includegraphics[width=0.7\columnwidth]{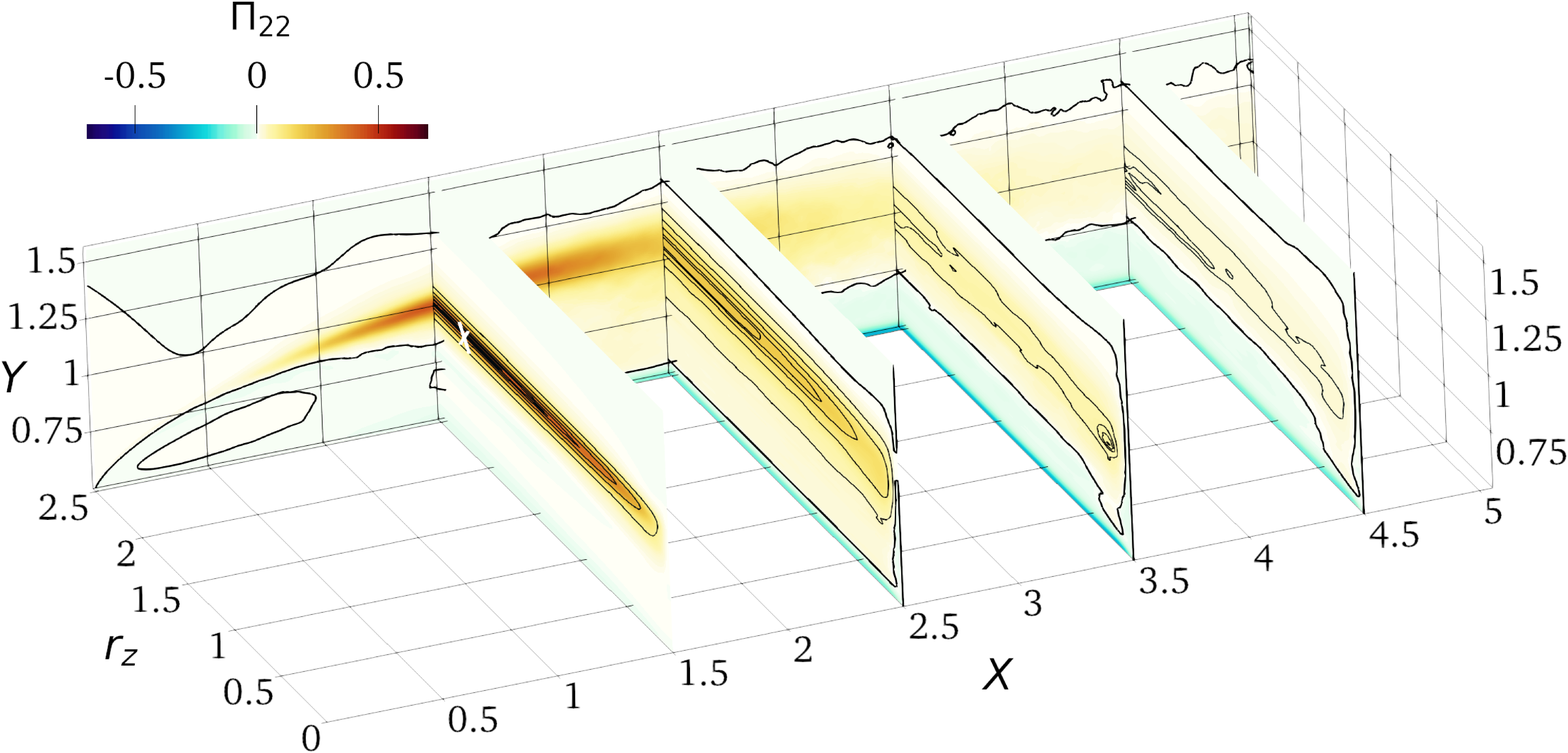}
\includegraphics[width=0.7\columnwidth]{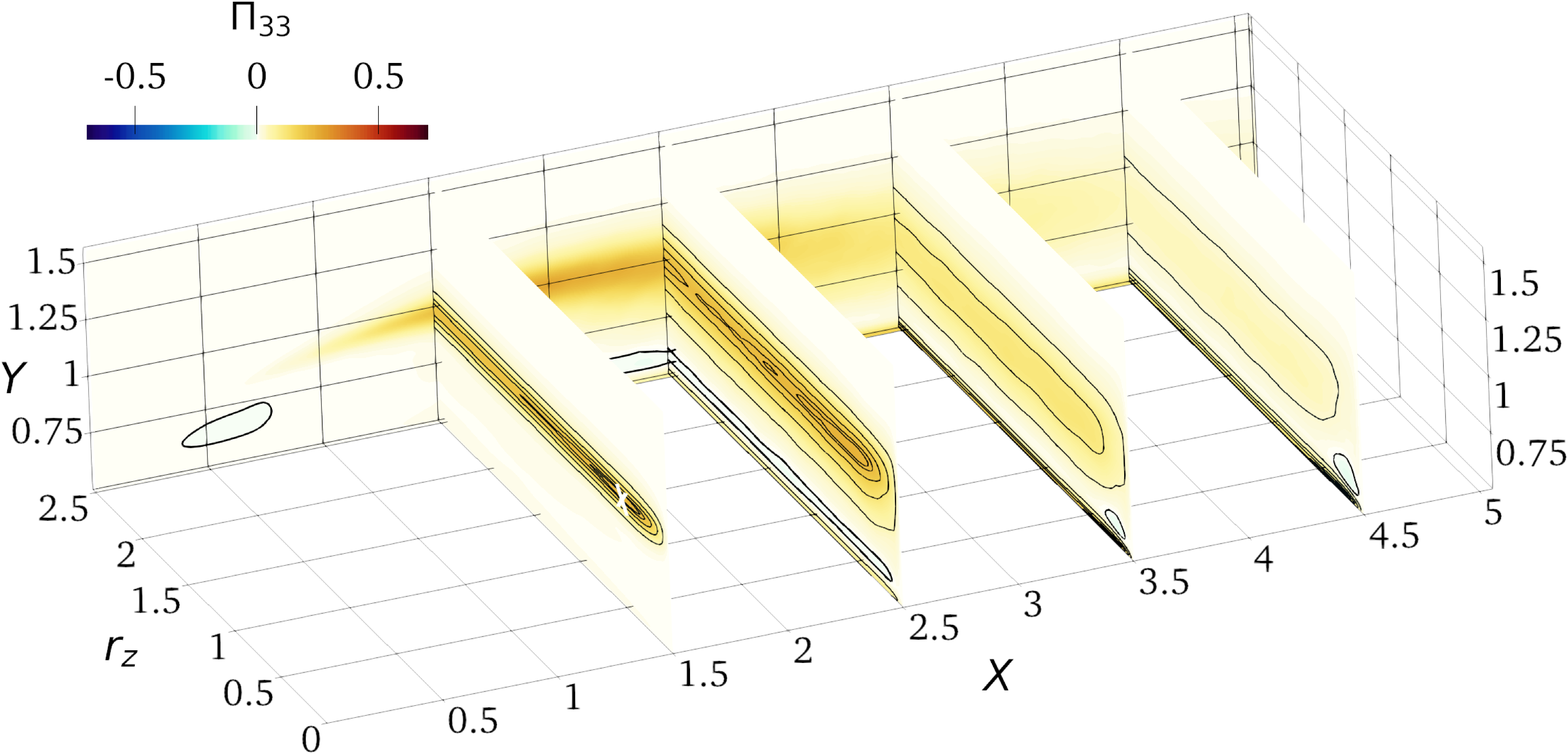}
\caption{Top: $\Pi_{11}$; centre: $\Pi_{22}$; bottom: $\Pi_{33}$. The thick black line marks the zero contour level. White crosses are used to identify local maxima when not easily visible.}
\label{fig:side_pstrain}
\end{figure}

\begin{figure}
\centering
\includegraphics[trim=500 0 500 0,clip,width=0.49\textwidth]{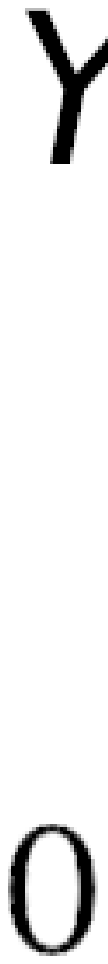}
\includegraphics[trim=500 0 500 0,clip,width=0.49\textwidth]{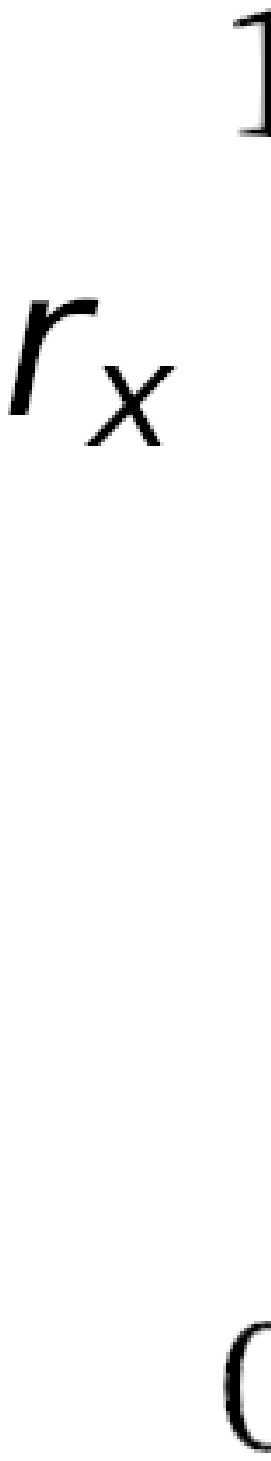}
\caption{As figure \ref{fig:KH-prod}, but for the pressure strain terms.}
\label{fig:KH-pstrain}
\end{figure}

Figure \ref{fig:side_pstrain} plots the three diagonal components of the pressure-strain tensor $\Pi_{ij}$. The streamwise energy $\aver{\delta u \delta u}$ drained from the mean flow by the two production mechanisms discussed above is partially redistributed towards the cross-stream components $\aver{\delta v \delta v}$ and $\aver{\delta w \delta w}$ at all scales and positions, except for the near-wall region where $\Pi_{11}>0$, $\Pi_{33}>0$ and $\Pi_{22}<0$. 

The KH instability is a net source for the large-scale vertical fluctuations despite the negative $P_{22}$. Indeed, here the large-scale vertical fluctuations are sustained by redistribution rather than by direct production, as shown by the maps of $\Pi_{11}$ and $\Pi_{22}$ exhibiting their negative/positive peaks at $(X,Y,r_z) \approx (1.5,1,2.4)$. The same conclusion is arrived at when considering the $r_y=r_z=0$ and $r_x=r_z=0$ around the LE shear layer, where (figure \ref{fig:KH-pstrain}) $\Pi_{11}$ is negative and $\Pi_{22}$ has a large positive peak at $(X,r_x)=(1.3,0.2)$. 

The local peak of $\Pi_{33}$ in the LE shear layer, instead, is found at smaller spanwise scales, i.e. at $(X,Y,r_z) \approx (1.5,1,0.5)$. This indicates that the modulation of the KH rolls is accompanied by a small-scale redistribution of streamwise fluctuations towards spanwise ones: indeed here $\Pi_{11}<0$ and $\Pi_{22}>0$. Note that these scales and positions are consistent with the local peak of $\aver{\delta w \delta w}$ within the LE shear layer discussed above in figure \ref{fig:side_str-func}. Along the LE shear layer the sum $P_{ij} + \Pi_{ij}$ is locally positive for all the three normal components, indicating that the KH instability is a net positive source of $u-$, $v-$ and $w-$fluctuations.

Within the primary vortex and after the reattachment point, pressure-strain activity is concentrated at the small $r_z$ scales associated with the streamwise-aligned vortices; see the local minimum of $\Pi_{11}$ and the local maxima of $\Pi_{22}$ and $\Pi_{33}$ at $r_z \approx 0.5$ for $X \ge 2.5$, $Y \approx 0.8$. This resembles the buffer layer of a turbulent channel flow \citep{gatti-etal-2020}, with streamwise fluctuations partially reoriented into cross-stream ones.

\begin{figure}
\centering
\includegraphics[trim=0 0 0 400,clip,width=0.7\columnwidth]{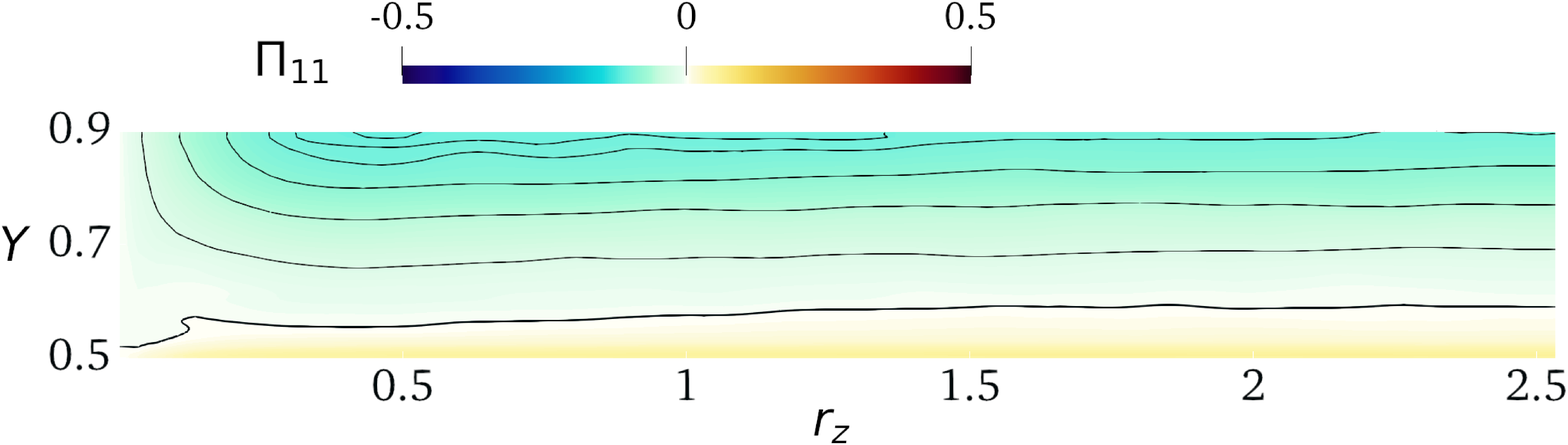}
\includegraphics[trim=0 0 0 400,clip,width=0.7\columnwidth]{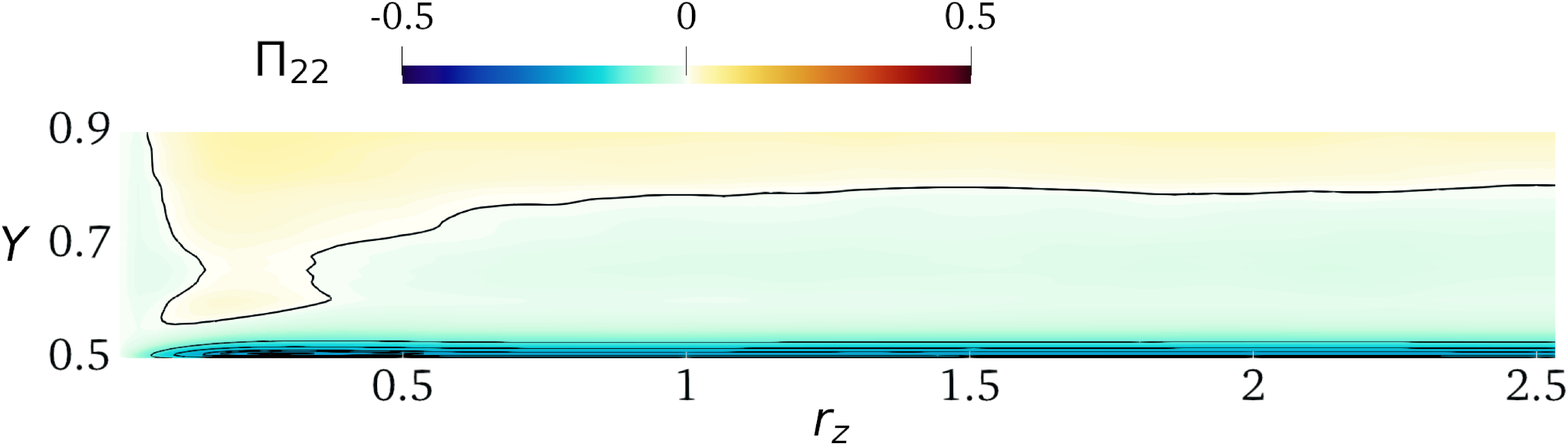}
\includegraphics[trim=0 0 0 400,clip,width=0.7\columnwidth]{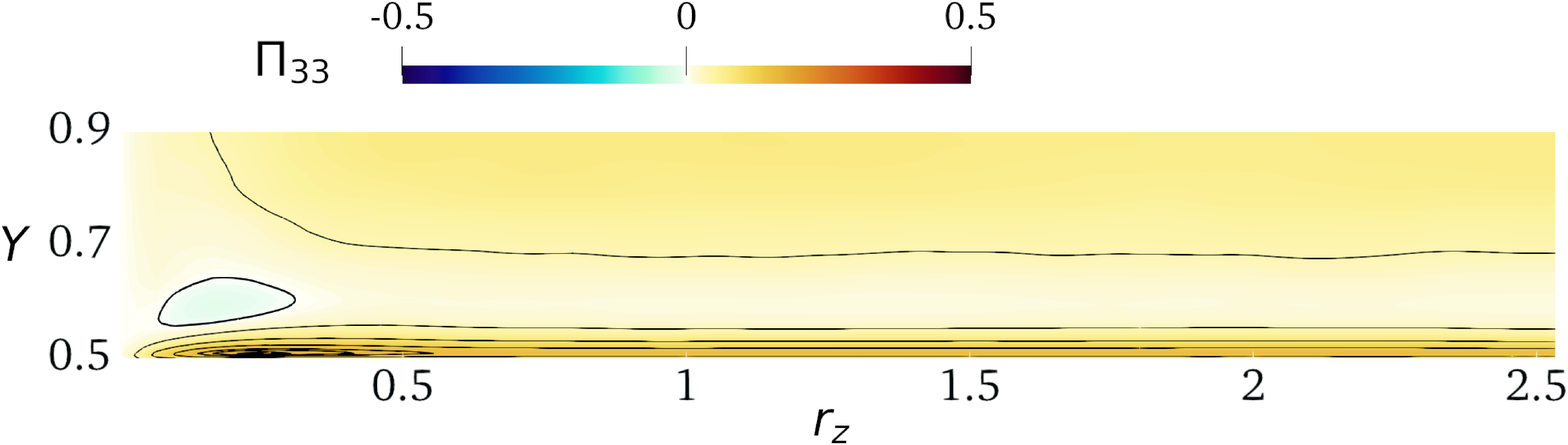}
\caption{Two-dimensional view of the pressure-strain terms, in the plane $X=4$. Panels as in figure \ref{fig:side_pstrain}.}
\label{fig:side_pstrain_zoom}
\end{figure}

The near-wall region in the final part of the cylinder side is different, due to the flow impingement on the wall. A closeup of this region is provided in figure \ref{fig:side_pstrain_zoom}, where the plane $X=4$ is considered. For $X>2.5$ and $Y \rightarrow 0.5$, large negative $\Pi_{22}$ together with positive $\Pi_{11}, \Pi_{33}$ are observed. This is the so-called splatting \citep{mansour-kim-moin-1988}, where vertical velocity fluctuations turn into wall-parallel ones near a solid wall. Interestingly, $\Pi_{33} > \Pi_{11}$ indicates that vertical fluctuations are preferentially redistributed towards spanwise ones. The positive peak of $\Pi_{33}$ occurs at $r_z \approx 0.3$, which agrees with the local maximum of $\aver{\delta w \delta w}$ at $r_z \approx 0.5$, seen in figure \ref{fig:side_str-func-zoom}. Together with the negative $P_{11}$, this explains why in the near-wall region the small-scale fluctuations are predominantly organised into $w-$structures. Moreover, we argue that these $w-$structures are then responsible for the generation of the streamwise vortices populating the flow at slightly larger $Y$. This is visualised by the local positive/negative peaks of $\Pi_{22}$ and $\Pi_{33}$ at $(Y,r_z) \approx (0.6,0.2)$. This type of impingement is sketched in figure \ref{fig:sketch_impinging}:
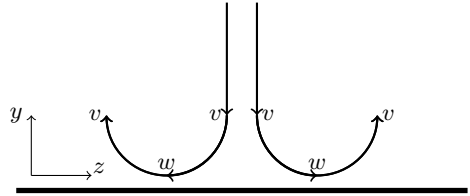
\begin{figure}
\centering
\begin{tikzpicture}[scale=1]	

\coordinate (A) at (-3,0);
\coordinate (B) at (3,0);

\draw[line width=2pt] (A)--(B);

\coordinate (al) at (-0.2,2.5);
\coordinate (bl) at (-0.2,1);

\coordinate (ar) at (0.2,2.5);
\coordinate (br) at (0.2,1);

\node at (-0.35,1) {$v$};
\node at (+0.35,1) {$v$};

\node at (-1,0.35) {$w$};
\node at (+1,0.35) {$w$};

\node at (-1.95,1) {$v$};
\node at (+1.95,1) {$v$};

\draw[thick,->] (al)--(bl);
\draw[thick,->] (-0.2,1) arc (0:-90:0.8);
\draw[thick,->] (-0.2,1) arc (0:-180:0.8);
\draw[thick,->] (-0.2,1) arc (0:-180:0.8);

\draw[thick,->] (ar)--(br);
\draw[thick,->] (0.2,1) arc (180:270:0.8);
\draw[thick,->] (0.2,1) arc (180:360:0.8);
\draw[thick,->] (0.2,1) arc (180:360:0.8);

\coordinate (Oo) at (-2.8,0.2);
\coordinate (Oz) at (-2,0.2);
\coordinate (Oy) at (-2.8,1);

\draw[->] (Oo)--(Oz);
\draw[->] (Oo)--(Oy);

\node at (-3,1) {$y$};
\node at (-1.9,0.3) {$z$};

\end{tikzpicture}
\caption{Sketch of the structures generated by the impingement flow on the cylinder surface in a $z-y$ cross-stream plane.}
\label{fig:sketch_impinging}
\end{figure}
energy is transferred from $\aver{\delta v \delta v}$ mainly towards $\aver{\delta w \delta w}$ in the very near-wall region, generating small-scale $w-$structures. As quantitatively shown by the local peak of $\aver{\delta w \delta w}$ in figure figure \ref{fig:side_str-func-zoom} and qualitatively sketched in figure \ref{fig:sketch_impinging}, these structures induce negative $R_{ww}<0$ for $r_z \approx 0.5$. Then, further from the wall the spanwise fluctuations are reoriented into vertical ones and energy is transferred back from $\aver{\delta w \delta w}$ to $\aver{\delta v \delta v}$ to feed the streamwise-aligned vortices.

In the reverse boundary layer upstream of the reattachment point at $X=3.95$, these streamwise-aligned structures produce local peaks of $\aver{\delta u \delta u}$, $\aver{\delta v \delta v}$ and $\aver{\delta u \delta v}$ for $r_z \approx 0.4$, as seen for example in figure \ref{fig:side_str-func-zoom} at $X=3.5$. While traveling upstream, these structures lose their coherence and eventually disappear when the reverse boundary layer detaches for $X<2$ \citep{cimarelli-leonforte-angeli-2018}. Consistently, the impinging structures described above vanish here: for $X < 2.5$ the local peak of $\aver{\delta w \delta w}$ at $Y \rightarrow 0.5$ (see figure \ref{fig:side_str-func-zoom}) and the negative $\Pi_{33}$ at slightly larger $Y$ disappear. Overall, the spatial organisation of the flow close to the wall differs from the classical near-wall turbulence, as also witnessed by the negative $P_{11}$. Since the impingement generates pairs of streamwise-aligned vortices but only at a certain distance from the wall, neither elongated streamwise vortices nor low-speed streaks are observed in the very near-wall region that, instead, is populated by $w-$structures.

%%%%%%%%%%%%%%%%%%%%%%%%%%%%%%%%%%%%%%%%%
\subsection{Scale-space energy transfers}
\label{side:fluxes}

The description of the structural properties of the flow provided by the AGKE via the separate analysis of each component of $\aver{\delta u_i \delta u_j}$ can be connected with the associated energy transfers by studying the fluxes of scale energy $\aver{\delta q^2}=\aver{\delta u_i \delta u_i}$ \citep{marati-casciola-piva-2004,cimarelli-deangelis-casciola-2013,cimarelli-etal-2016,cimarelli-etal-2021}. (Note that in this Section the subscript $\cdot_{ii}$ implying summation will be omitted for conciseness.)

A complete scale-space characterisation of the energy fluxes requires the analysis of the complete flux vector $\bm{\Phi}=(\bm{\phi},\bm{\psi}$) in the five-dimensional space $(r_x,r_y,r_z,X,Y)$.  However, since the statistical trace of the main turbulent structures is clearly visible for $r_z \neq 0$, here we consider only the $r_y=r_x=0$ subspace, where the flux vector reduces to $(\phi_z,\psi_X,\psi_Y)$: the terms $\partial \phi_x / \partial r_x$ and $\partial \phi_y / \partial r_y$ are moved to the the l.h.s. of equation \ref{eq:xi_gen} and contribute to the source term of the GKE.
The BARC flow is convection-dominated, so that the energy transport in the physical space due to the mean flow, i.e. the convective transfer, overwhelms the other contributions. As a consequence, in the $r_x=r_y=0$ space the fluxes show a minimal scale dependency and closely resemble those observed in the single-point budget equation for the turbulent kinetic energy \citep{chiarini-quadrio-2021b}. To highlight the inter-scale transfer involving turbulent structures, it is useful to remove the relatively trivial effect of the mean convection. Therefore, in equation \eqref{eq:xi_gen} also the divergence of the mean convective flux is moved from the flux vector to the source term and a reduced flux vector and a corresponding extended source term (both indicated with $\hat{\cdot}$) are defined. The components of the reduced flux vector $\hat{\vect{\Phi}}=(\hat{\phi}_z,\hat{\vect{\psi}})$ are: %(here and hereinafter in this Section the subscript $\cdot_{ii}$ implying summation is omitted for conciseness) are:
\begin{equation}
\hat{\phi}_z = \hat{\phi}_z^{turb} + \hat{\phi}_z^{visc}
\end{equation}
and
\begin{equation}
\hat{\vect{\psi}} = \hat{\vect{\psi}}^{turb} + \hat{\vect{\psi}}^{press} + \hat{\vect{\psi}}^{visc}.
\end{equation}
The extended source $\hat{\xi}$, instead, reads:
\begin{equation}
\hat{\xi} = P - D - \frac{\partial \phi_x}{\partial r_x} - \frac{\partial \phi_y}{\partial r_y}
                  - \frac{\partial \psi_x^{mean}}{\partial X}   - \frac{\partial \psi_y^{mean}}{\partial Y};
\end{equation}
recall that $\phi_z^{mean}=0$ as $\vect{U}=(U,V,0)$. Thus the budget equation for $\aver{\delta q^2}$ can be written as:
\begin{equation}
\vect{\nabla} \cdot \hat{\vect{\Phi}}=\frac{\partial \hat{\phi}_z}{\partial r_z} + \frac{\partial \hat{\psi}_j}{\partial X_j} = \hat{\xi}.
\end{equation}
The flux vector describes how $\aver{\delta q^2}$ is transferred in space and across scales. The field lines of $\hat{\vect{\Phi}}$ are used to determine the orientation of the fluxes. Its divergence $\vect{\nabla} \cdot \hat{\vect{\Phi}}$, instead, provides quantitative information about the energetic relevance of the fluxes. When $\vect{\nabla} \cdot \hat{\vect{\Phi}}>0$, i.e. $\hat{\xi}>0$, the fluxes are energised. When $\vect{\nabla} \cdot \hat{\vect{\Phi}}$ is negative, i.e. $\hat{\xi}<0$, the fluxes release energy to locally sustain $\aver{\delta q^2}$. Clearly, as we consider the $r_x=r_y=0$ subspace, the energy transfer across wall-normal and streamwise scales is not considered. Hence, in the following transfers are said direct or inverse only in relation to the spanwise scale.

\begin{figure}
\centering
\includegraphics[width=0.99\textwidth]{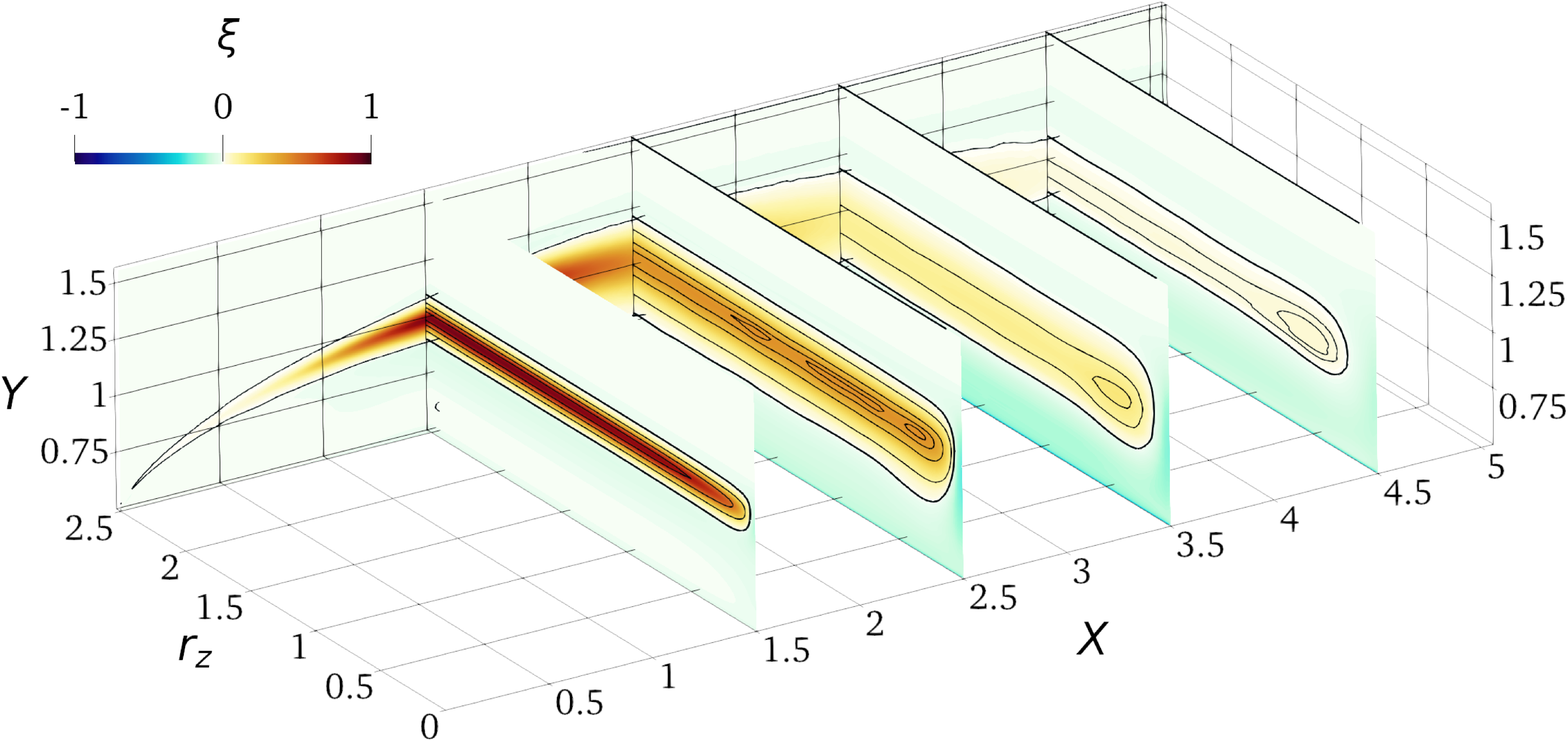}
\includegraphics[width=0.99\textwidth]{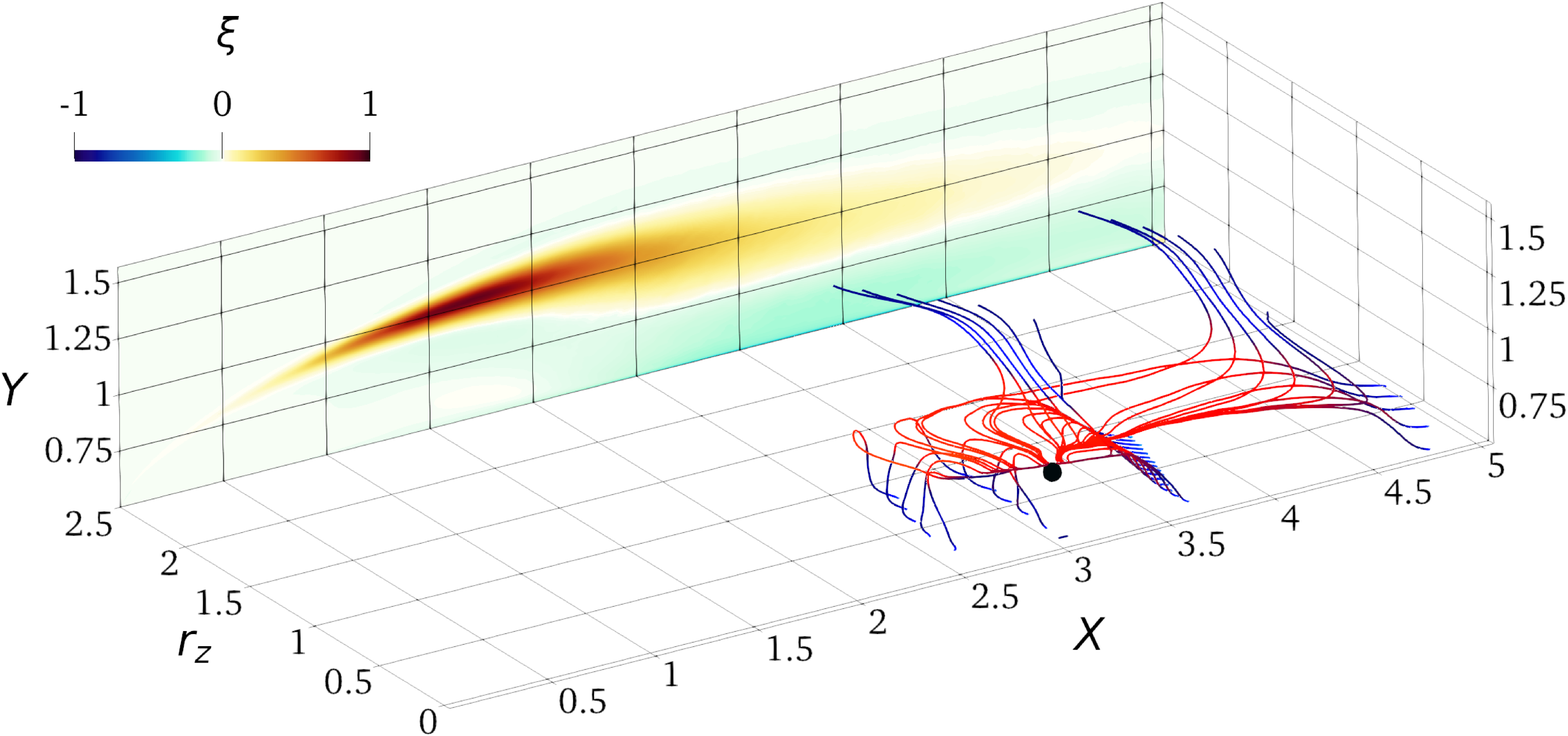}
\includegraphics[width=0.49\textwidth]{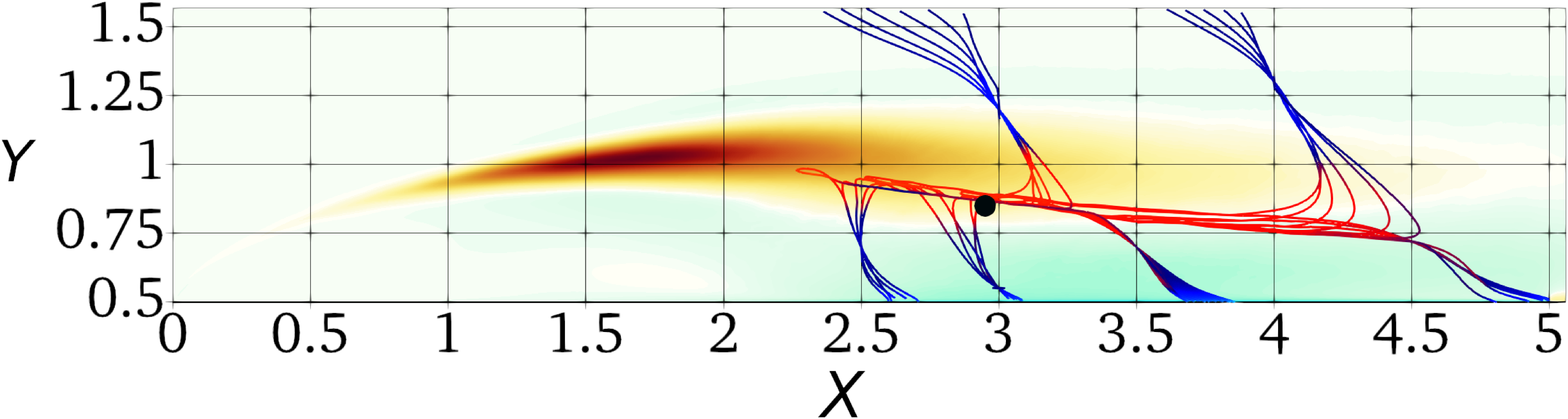}
\includegraphics[width=0.49\textwidth]{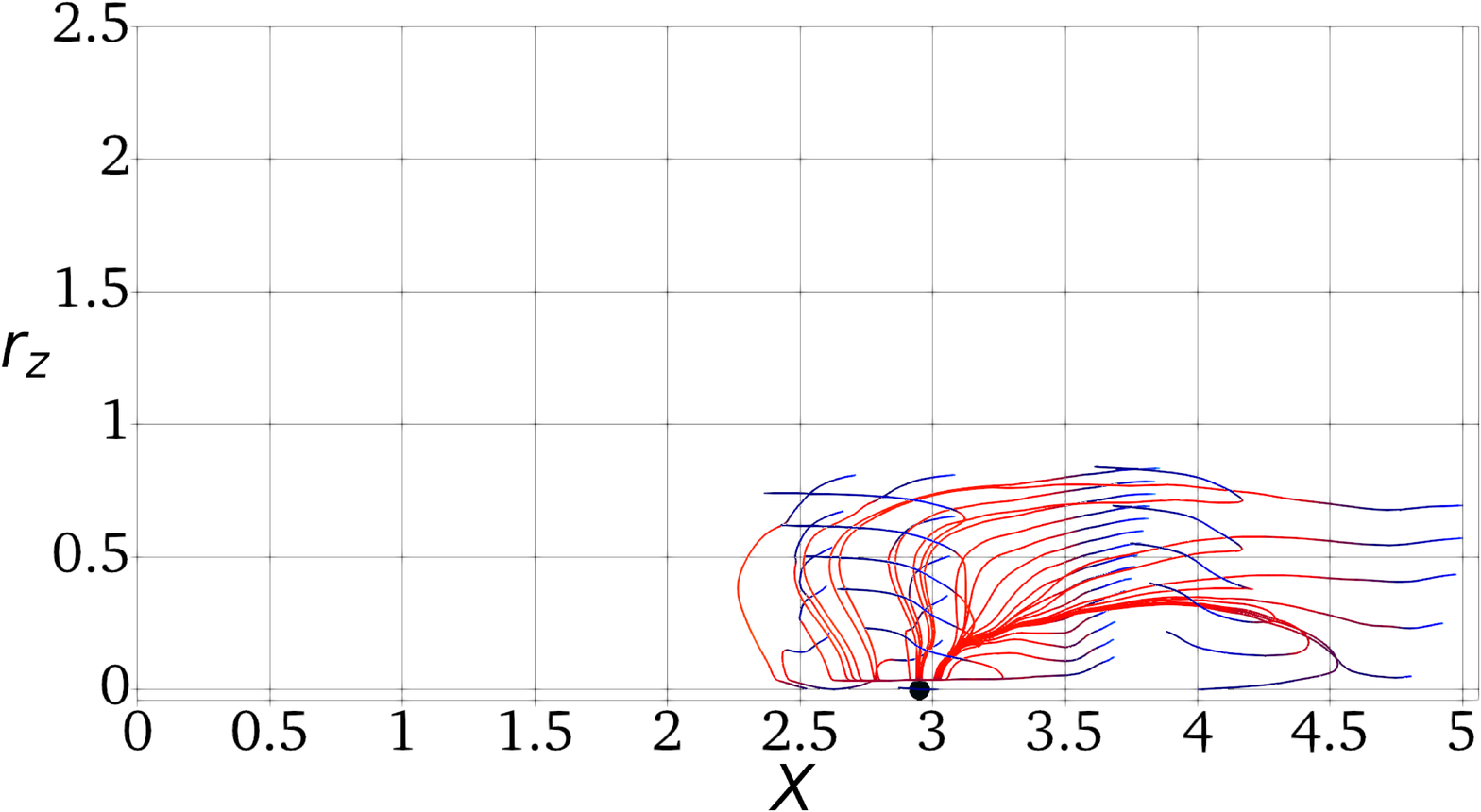}
\caption{Field lines of the reduced fluxes for $\aver{\delta q^2}$ in the $r_x=r_y=0$ space. The top panel is for the source term $\xi$. In the central panel the colourmap on the $r_z=2.5$ plane is for $\xi$ and the lines are coloured with the divergence of the reduced flux vector $\vect{\nabla} \cdot \hat{\vect{\Phi}}$: red/blue indicates positive/negative $\vect{\nabla} \cdot \hat{\vect{\Phi}}$. The lower panels are bottom and side views of the central one.}
\label{fig:side-fluxes}
\end{figure}

Figure \ref{fig:side-fluxes} shows the source term $\xi$ and the field lines of the three-dimensional reduced flux vector $(\hat{\vect{\phi}}_z,\hat{\vect{\psi}}_X,\hat{\vect{\psi}}_Y)$ coloured with its divergence. The source term $\xi$ identifies sink and source regions for $\aver{\delta q^2}$. A wide source area with positive $\xi$ starts from the LE, follows the separating shear layer embedding the core of the primary vortex, and extends downstream the TE. Hereafter this region is referred to as source region. It presents localised peaks at scales and positions associated with both the KH rolls and the streamwise aligned vortices. This means that both the large- and small-scale production dominate local viscous dissipation, leading to a net production of large-scale fluctuations in the shear layer and of small-scale ones in the aft part of the cylinder side. Along the shear layer $\xi>0$ for $r_z > 0.1$; moving downstream this spanwise scale increases and at $X=4.5$ $\xi>0$ for $r_z > 0.25$. Two sink regions, instead, are placed below and above the source region and are referred to as side inner and outer sinks. However, here the negative $\xi$ does not identify a particular scale. As expected at the smallest scales dissipation dominates everywhere, yielding $\xi<0$ for $r_z \rightarrow 0$ at all $(X,Y)$ positions.

The flux lines originate from a singularity point, i.e. a point where the vector flux vanishes and its direction is consequently undefined, located at $(X,Y,r_z) \approx (3,0.82,0)$; their positions and scales are consistent with the statistical footprint of the streamwise-aligned vortices (see figure \ref{fig:side_str-func}). These fluxes remain confined in a small portion of the $(X,Y,r_z)$ space, meaning that the excess of $\aver{\delta q^2}$ produced by the small-scale mechanisms sustains only small-scale fluctuations placed in the aft portion of the cylinder side. Starting from the singularity point, all the lines first lay on the $Y \approx 0.82$ plane (as $\hat{\psi}_Y \approx 0$), where the fluxes gain intensity, and then deviate and release energy both towards larger and smaller scales. 

Three different line families are detected, depending on where energy is released. Lines of the first family pass over the TE towards the wake. Lines of the second family are attracted by the cylinder side and release energy in the side inner sink region for $2.5 \le X \le 5$; they approach the wall with a spiral pattern, thus showing direct and inverse transfers since the fluxes release energy at a certain scale after being energised by both smaller and larger scales \citep{cimarelli-deangelis-casciola-2013,cimarelli-etal-2016}. Lines of the third family deviate towards larger $Y$ where they release energy to sustain the turbulent fluctuations in the range $ 0 \le r_z \le 1$ in the outer sink region. This ascending energy transfer is accompanied by an inverse energy transfer.

The above description of the scale-energy fluxes is far more complex than the Richardson phenomenological description of energy cascade. In particular, the simultaneous presence of forward and reverse transfers is a challenge for turbulence theories, and needs to be accounted for by closures. This is particularly important for Large Eddy Simulation, where a cross-over scale $\ell_{cross}$ needs to be identified separating smaller scales dominated by forward energy transfer and larger scales dominated by reverse transfer. Indeed, when the lengthscale $\Delta$ describing the local grid size (or the filter size) is such that $\Delta < \ell_{cross}$, subgrid motions are dissipative and can be modelled by means of classic eddy viscosity assumptions. On the other hand, when $\Delta > \ell_{cross}$, energy should emerge from the subgrid space, and more sophisticated mixed modelling approaches should be considered \citep{cimarelli-deangelis-2012,cimarelli-deangelis-2014}. A suitable candidate for estimating $\ell_{cross}$ is the smallest scale where the source term $\xi$ is zero. For the spanwise scales considered here, $\ell_{cross,z}$ is such that $\xi<0$ in the $r_x=r_y=0$ subspace for all $(X,Y)$ positions and $r_z < \ell_{cross,z}$. The scale $\ell_{cross,z}$ closely corresponds to the divergence point of the fluxes when projected in the $(Y,r_z)$ plane. By inspecting our data (see the top panel of figure \ref{fig:side-fluxes}) $\ell_{cross,z} \approx 0.1$, which provides a guideline for the selection of the more suitable LES approach as a function of the employed grid.
The proposed $\ell_{cross,z}$ resembles (but is not equal to) the shear length-scale $L_s$ explored by \cite{casciola-etal-2003}.
They applied the GKE (there called K\'{a}rm\'{a}n--Howarth equation) to an homogeneous shear flow and, after integrating the equation on a sphere of radius $r$, defined $L_s$ as the scale $r$ at which the integral of the production term equals the integral of the transport term.

\section{The near wake}
\label{sec:wake}

\begin{figure}
\centering
\includegraphics[trim=0 0 200 0,clip,width=1\columnwidth]{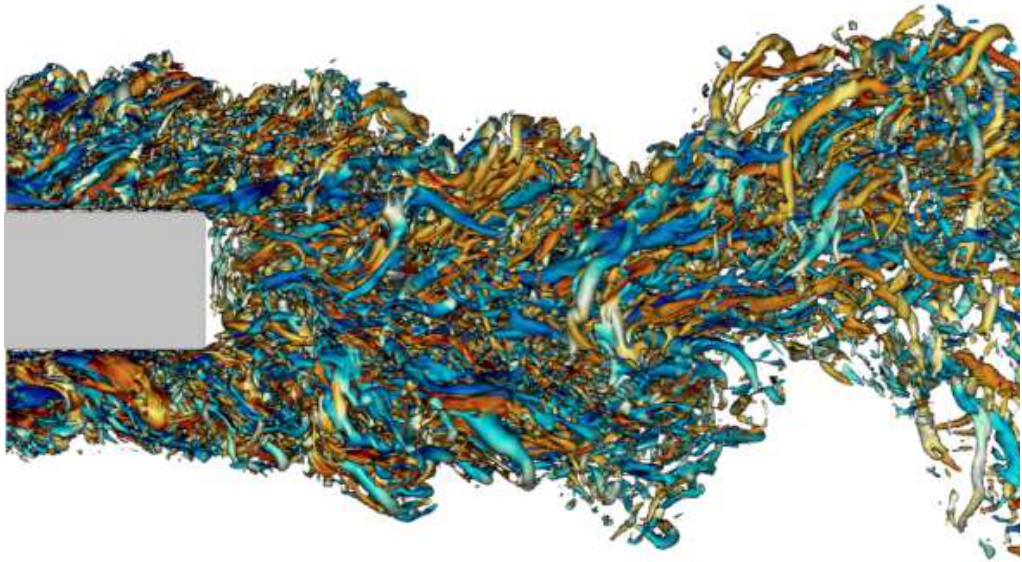}
\caption{Instantaneous snapshot of the BARC flow: lateral view of the isosurfaces at $\lambda_2 = -5$, with zoom into the near-wake region. Colour depicts streamwise vorticity $\omega_x$, with the blue-to-red colormap ranging in $-10 \le \omega_x \le 10$. The vertical size $D$ of the cylinder can be used as reference to identify the vertical and streamwise length scales of the near wake.}
\label{fig:lambda2_zoom_wake}
\end{figure}

The near wake is the wake in the vicinity of TE, before its development into a classic, self-similar turbulent wake \citep{bevilaqua-lykoudis-1978}. It is known \citep{cimarelli-leonforte-angeli-2018,chiarini-quadrio-2021} that the BARC wake becomes self-similar for $X \ge 10$, hence the present analysis is limited to $X<10$. The main feature of the near wake, illustrated in figure \ref{fig:lambda2_zoom_wake}, is the coexistence of the large-scale von K\'arm\'an-like vortices shed from the TE with the small-scale turbulent structures advected from the boundary layer formed over the cylinder side. 

\begin{figure}
\centering
\includegraphics[trim={0 0 0 450},clip,width=1\columnwidth]{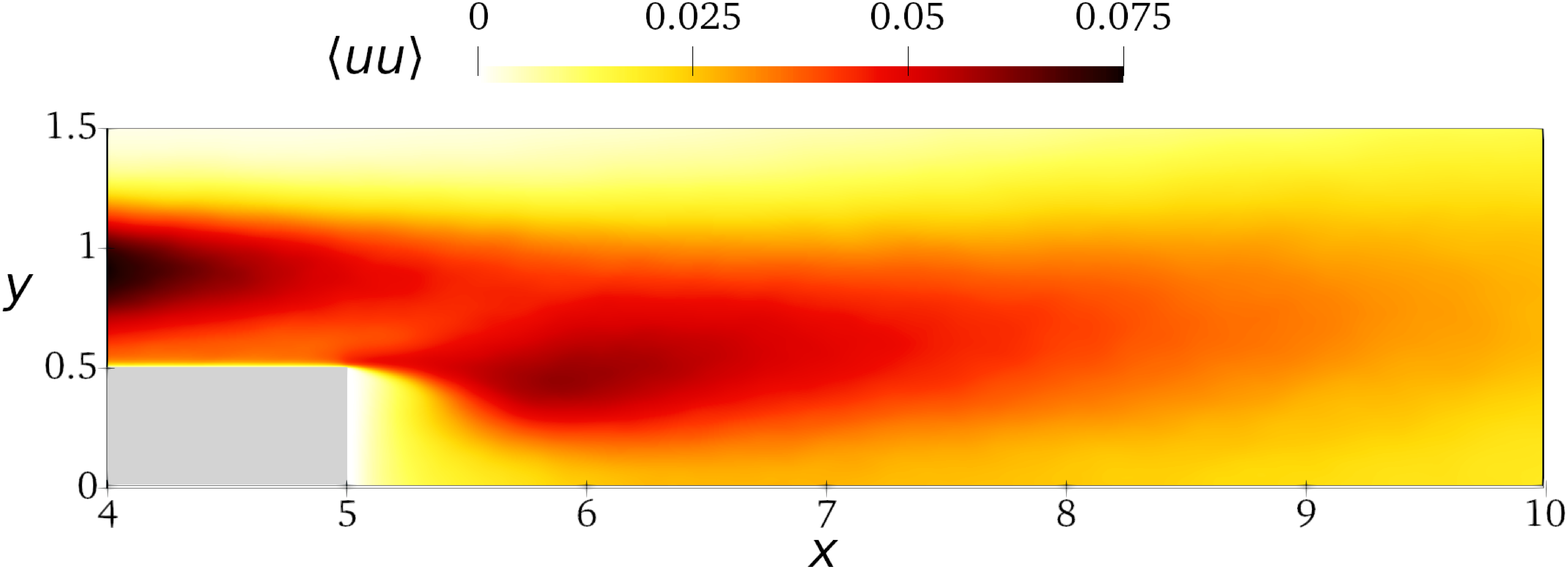}
\includegraphics[trim={0 0 0 450},clip,width=1\columnwidth]{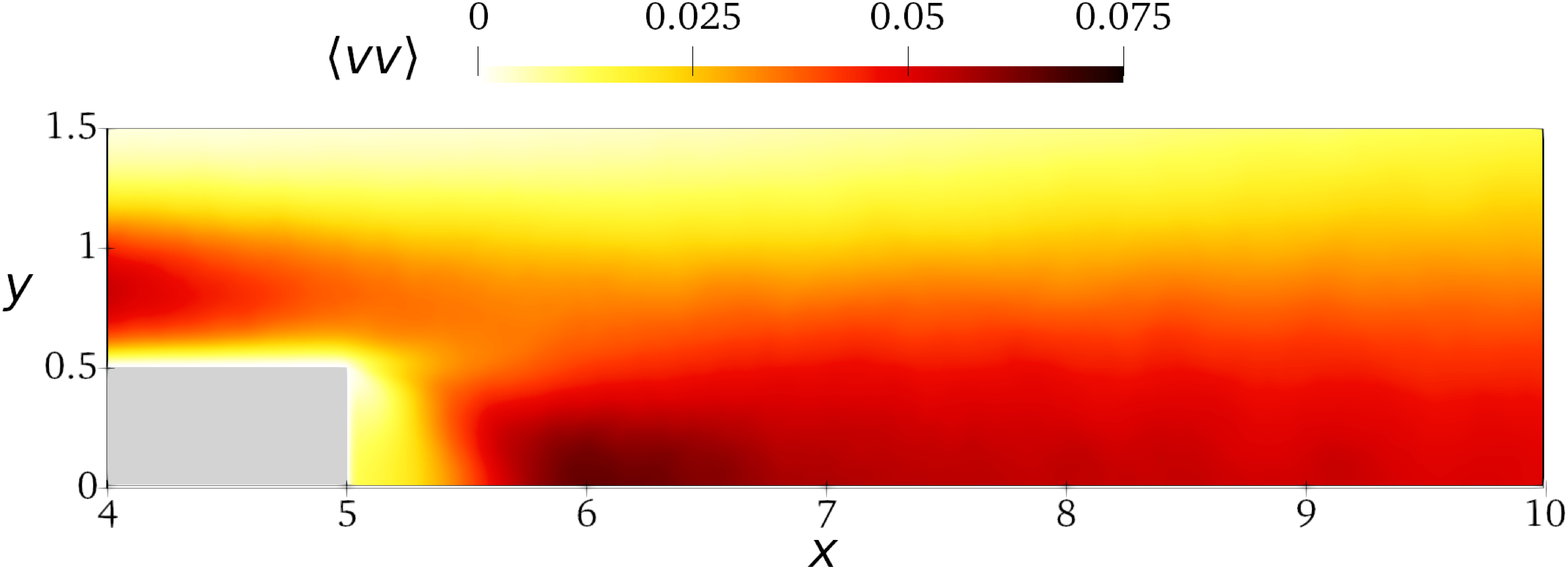}
\caption{Maps of the $\aver{uu}$ (top) and $\aver{vv}$ (bottom) components of the Reynolds stresses in the $x-y$ plane, with focus on the near wake.}
\label{fig:uu_zoom_wake}
\end{figure}

The shed vortices produce past the TE a peak in the maps of $\aver{uu}$ at $Y=0.5$ and $\aver{vv}$ at $Y=0$, as shown in figure \ref{fig:uu_zoom_wake}. 
The vertical position of the peaks suggests large spanwise vortices inducing the strongest $u-$ and $v-$fluctuations at their vertical and lateral sides. The streamwise distance between the TE and the $\aver{vv}$ peak is often used to define the vortex formation length. In the present case, we measure a value of $1.1 D$ that is close to what measured in the wake of a square cylinder by \cite{trias-etal-2015} at $Re=22000$ and \cite{alvesportela-papadakis-vassilicos-2017} at $Re=3900$. It should be noted, however, that the von K\'arm\'an-like vortices for the square cylinder originate from the roll-up of the LE shear layer, since the flow does not reattach over the cylinder side.

%%%%%%%%%%%%%%%%%%%%%%%%%%%%%%%%%%%%%%%%%%%%%%%%%%%%%%%
\subsection{Structures advected from the boundary layer}

\begin{figure}
\centering
\includegraphics[width=0.7\columnwidth]{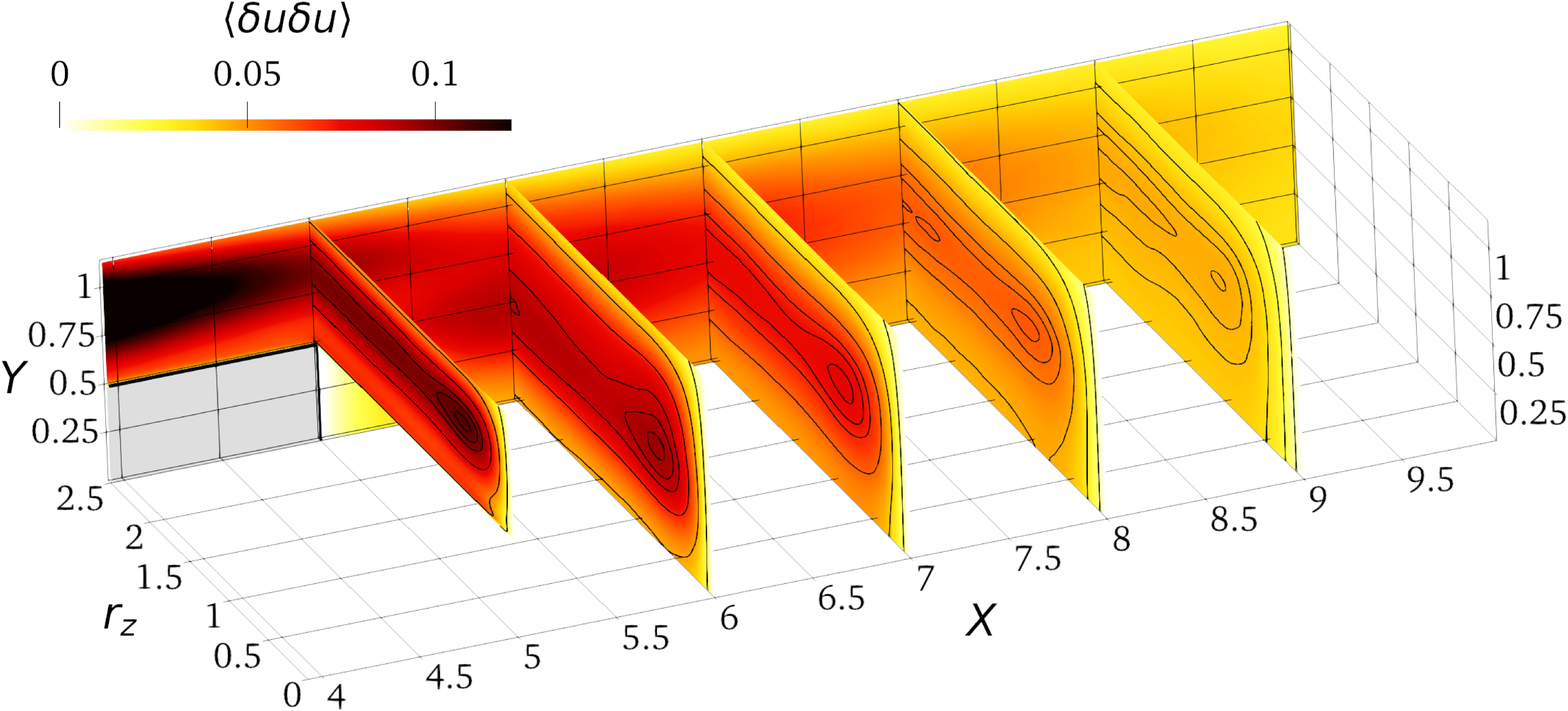}
\includegraphics[width=0.7\columnwidth]{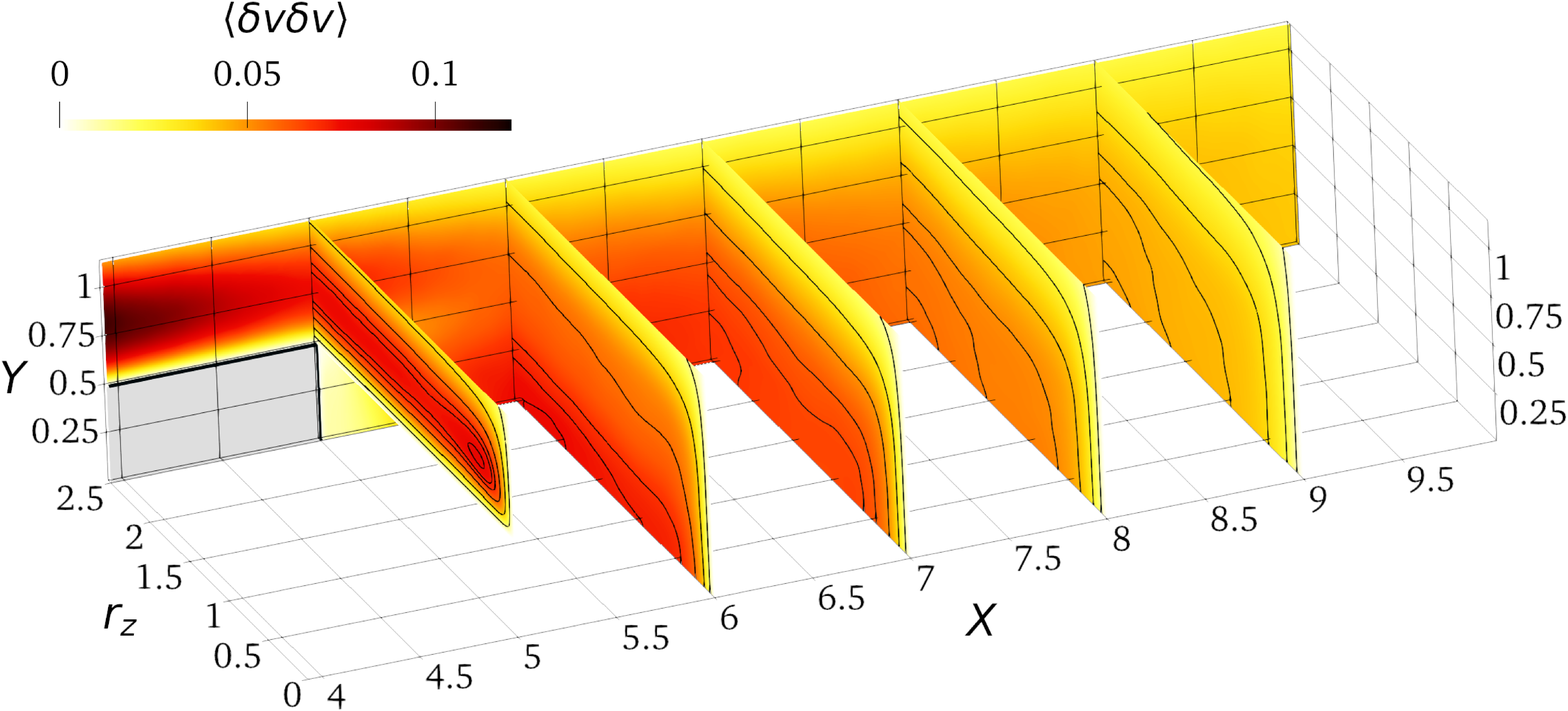}
\includegraphics[width=0.7\columnwidth]{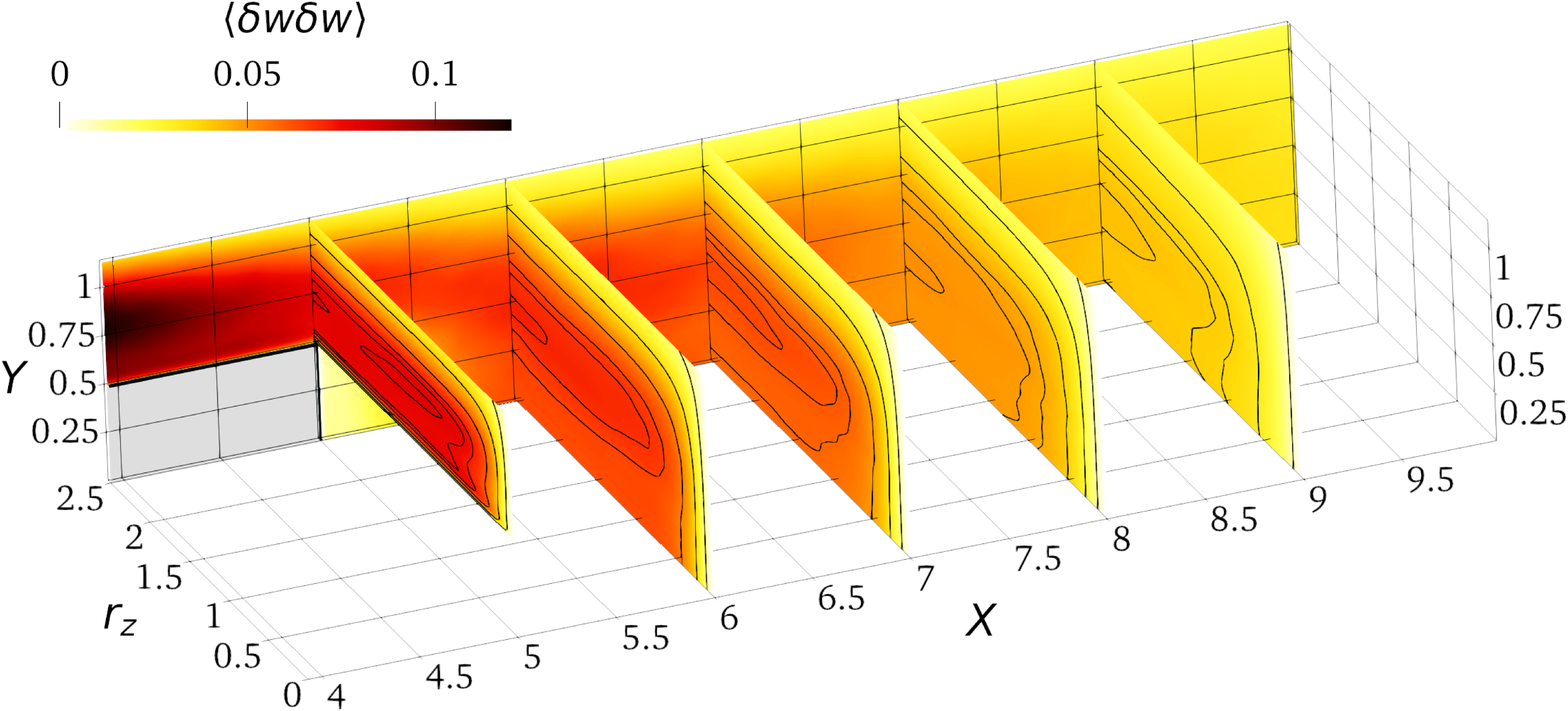}
\includegraphics[width=0.7\columnwidth]{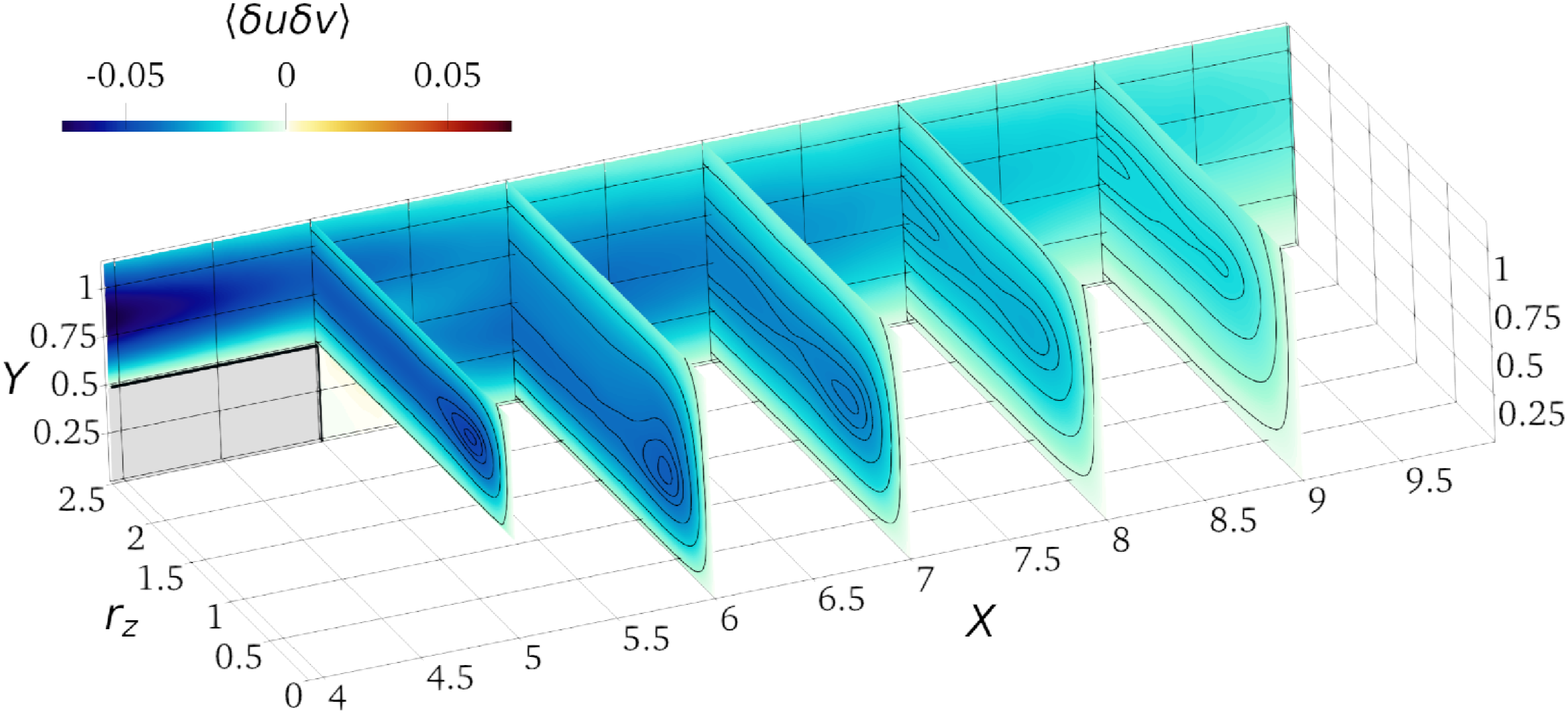}
\caption{Components of the structure function tensor plotted in the space $(X,Y,r_z)$ in the near wake.}
\label{fig:wake_dudu}
\end{figure}

Figure \ref{fig:wake_dudu} plots the three diagonal components of the structure function tensor and the off-diagonal component $\aver{\delta u \delta v}$ in the $r_x=r_y=0$ space for $4 \le X \le 10$. Structures of the turbulent boundary layer over the aft cylinder side leave a statistical footprint in the near wake, at least up to a distance of $4 D$ downstream the TE, i.e. for $X \le 9$. 

The energy associated with these structures, located slightly above the cylinder side ($Y \geq 0.5$), gradually decreases along the TE shear layer to eventually disappear at large $X$, where they are annihilated by viscous dissipation and by the reorientation and isotropisation of the pressure strain; see \S\ref{sec:wake1-pstrain}. This is particularly visible in the local peaks of $\aver{\delta u \delta u}$ and $\aver{\delta u \delta v}$ at $r_z \approx 0.5$ and $r_z \approx 2$. The advected structures follow the TE shear layer, in fact the $Y$ position of these maxima decreases from $Y \approx 0.75$ at $X \le 5$ to $Y \approx 0.5$ for larger $X$.  
%XXX can be quantitative on this? Is there a mean inclination of the shear layer? can we draw it into the figure? XXX.
Moreover, the characteristic $r_z$ scale of the streamwise-aligned vortices increases from $r_z \approx 0.5$ to $r_z \approx 1$ as they are advected downstream, due to the combined effect of viscous and pressure diffusion and of the scale-space turbulent fluxes.
%Moreover, viscous diffusion increases the characteristic $r_z$ scale of the streamwise-aligned vortices from $r_z \approx 0.5$ to $r_z \approx 1$ as they are advected downstream. 
 Unlike for the other components, the small-scale peak of $\aver{\delta v \delta v}$ associated with the streamwise-aligned vortices disappears quickly after the TE: already at $X > 6$ $\aver{\delta v \delta v}$ peaks in the core of the wake at no particular spanwise scale. As also shown by the map of $\aver{vv}$ in figure \ref{fig:uu_zoom_wake}, at these positions the large-scale vertical fluctuations produced by the spanwise-uniform von K\'arm\'an vortices dominate. Note that, as shown later, the scale information pertaining to these large vortices can be retrieved at $r_x \ne 0$ and/or $r_y \ne 0$. Last, in the near wake $\aver{\delta w \delta w}$ is distributed over $0 \le Y \le 1.4$, without an evident peak at a certain scale, indicating that as soon as the wall is removed, the $w-$structures resulting from flow impingement abruptly disappear. It will be shown \S\ref{sec:wake1-pstrain} that they are indeed destroyed by the action of the pressure strain.

%%%%%%%%%%%%%%%%%%%%%%%%%%
\subsubsection{Production}
\label{sec:wake1-prod}

\label{sec:wake-small-Production}
\begin{figure}
\centering
\includegraphics[width=0.74\columnwidth]{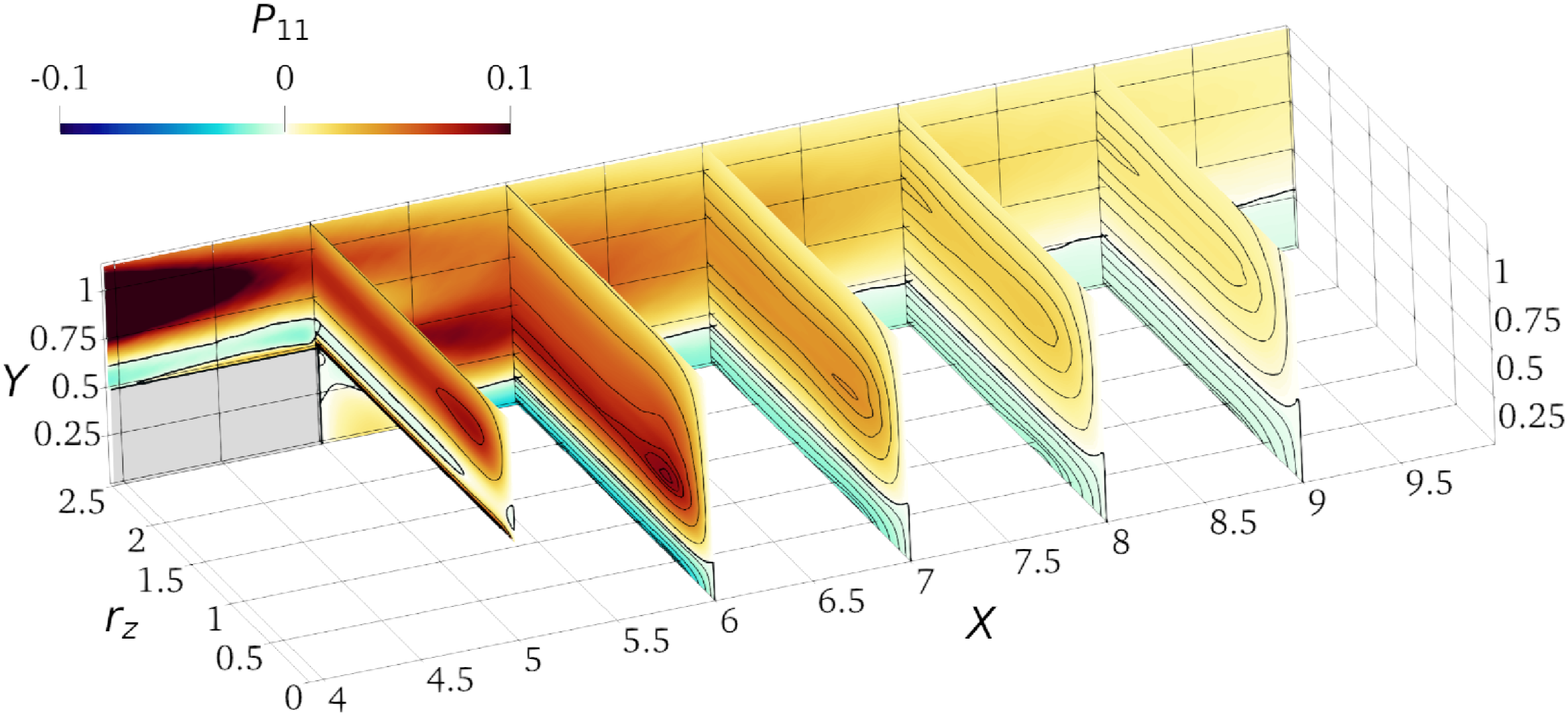}
\includegraphics[width=0.49\columnwidth]{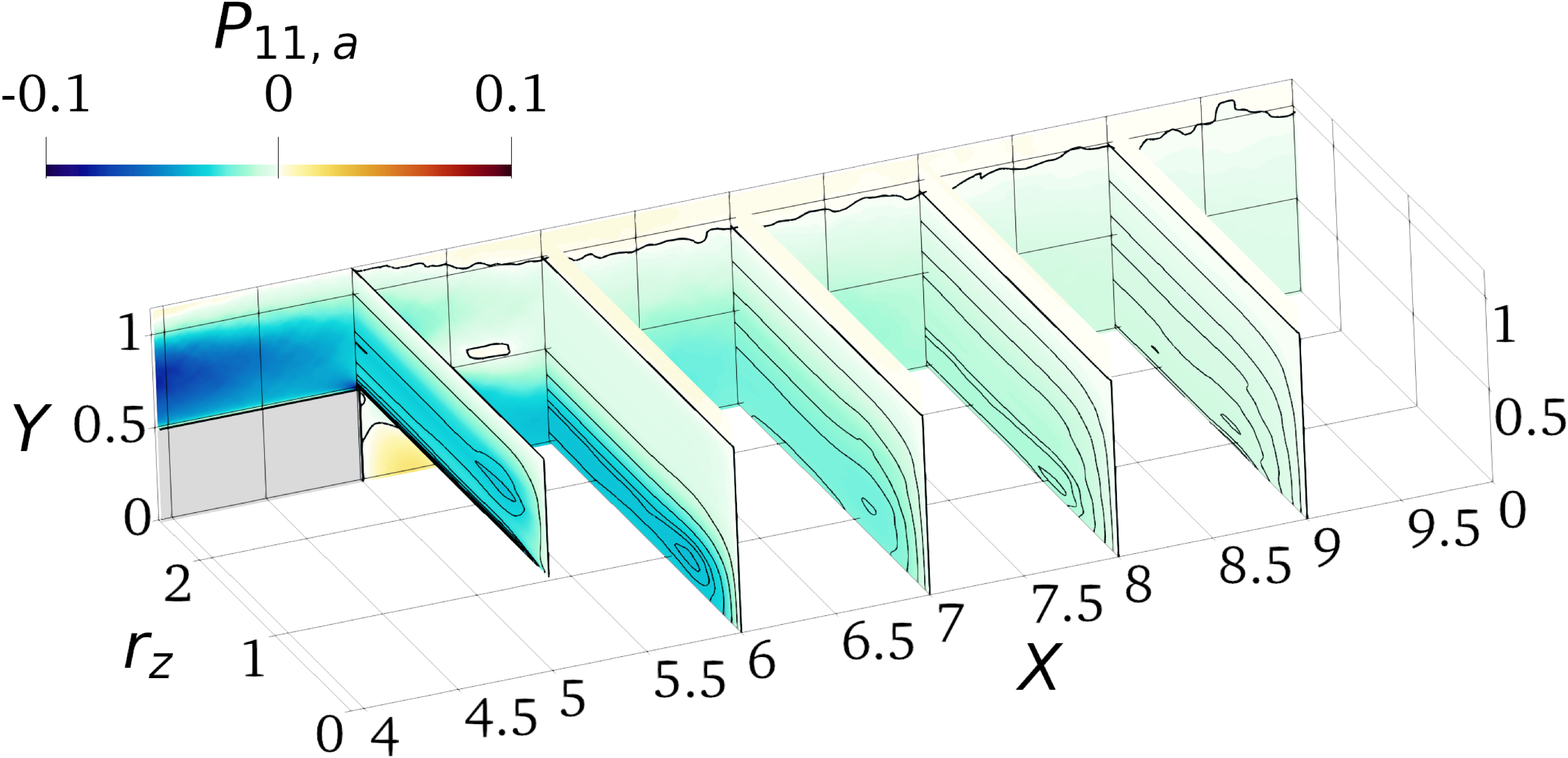}
\includegraphics[width=0.49\columnwidth]{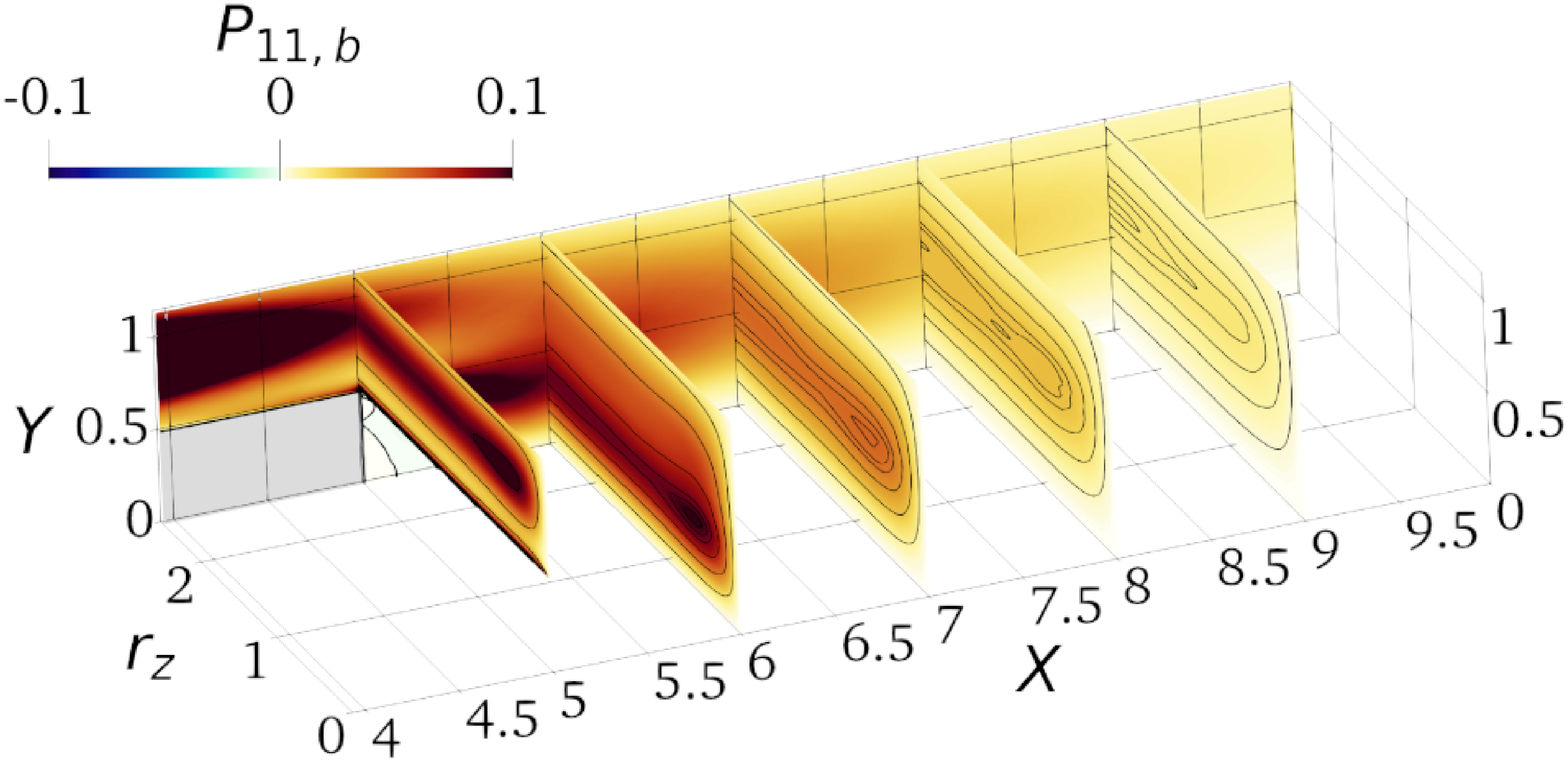}
\includegraphics[width=0.74\columnwidth]{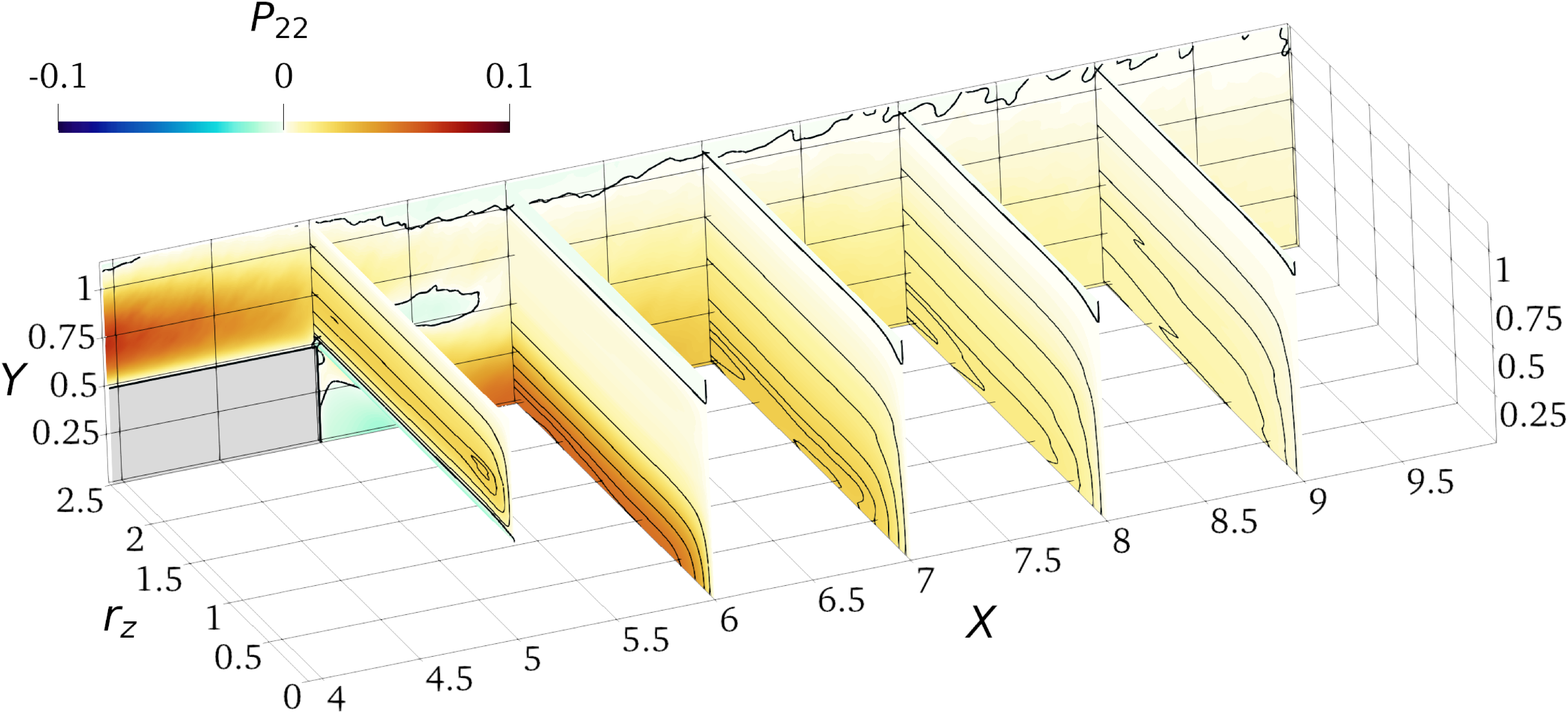}
\caption{As in figure \ref{fig:side_production}, but for $4 \le X \le 10$ and $ 0 \le Y \le 1.1$.}
\label{fig:wake_Prod}
\end{figure}

The small-scale production associated with the streamwise-aligned vortices %of the boundary layer on the side
 persists in the wake and is the main source of $u-$fluctuations up to $2 D$ downstream of the TE. However, it does not outweigh the combined effects of dissipation and redistribution, so that the streamwise-aligned structures disappear further downstream. Here the streamwise velocity fluctuations become mainly sustained by the von K\'arm\'an vortices; see \S\ref{sec:KarmanVortices} below. Figure \ref{fig:wake_Prod}, indeed, shows a local peak of $P_{11}$ at the small $r_z \approx 0.5-1$ for $X \le 7$, whereas at the largest $r_z$ for $X>7$. Similarly to what occurs over the cylinder side, $P_{11}$ is negative close to the wake centre, i.e. for $X>5.5$ and $Y \le 0.25$. However, unlike the scale-independent $P_{11}<0$ region close to the wall in the aft cylinder side (see \S\ref{sec:side_prod}), here $P_{11}$ has a local minimum at the characteristic $r_z$ scale of the streamwise-aligned vortices. The largest positive values of $P_{11}$ are in the TE shear layer, as the mean velocity gradient across it is very intense.

As in \S\ref{sec:side_prod}, $P_{11}$ is decomposed into the two terms $P_{11,a}$ and $P_{11,b}$, whose sign is determined by $\partial U / \partial x$ and $\partial U / \partial y>0$, since $\aver{\delta u \delta v}<0$ almost everywhere (see the bottom panel of figure \ref{fig:wake_dudu}). $P_{11,a}$ is negative everywhere, except within the wake vortex behind the TE. The largest negative values are observed within the TE shear layer, where $\partial U / \partial x >0$ is maximum. The scale dependency of $P_{11,a}$ derives from $\aver{\delta u \delta u}$ and shows a minimum at the $r_z$ scales of the streamwise-aligned vortices. Its vertical position evolves from $Y \approx 0.75$ immediately after the TE towards $Y=0$ at larger $X$, where $\partial U / \partial x >0$ is larger.
$P_{11,a}$ dominates within the wake vortex and in the wake centreline, but, in contrast, $P_{11,b}$ is the main contribution everywhere else. $P_{11,b}$ is positive at almost all scales and positions, except within the wake vortex due to the positive $\aver{\delta u \delta v}$ and $\partial U / \partial y$. The largest $P_{11,b}$ is found again along the TE shear layer, where the mean velocity gradients are intense. However, the downstream evolution of $P_{11,b}$ differs from that of $P_{11,a}$: it peaks at $Y \approx 0.5$ and it is zero for $Y = 0$, where $\partial U / \partial y=0$. The spanwise scales associated with local maxima of $P_{11,b}$ still identify the streamwise-aligned vortices. Like over the cylinder side, indeed, production from the streamwise-aligned structures has $P_{11,a}<0$ and $P_{11,b}>0$. However, here $P_{11,b}$ prevails only for $Y>0.25$. The positive $P_{11}$ for $Y>0.25$ is thus associated with the positive $\partial U / \partial y>0$, while the negative $P_{11}$ close to the $Y=0$ plane is the result of the mean flow acceleration $\partial U / \partial x>0$.

As over the cylinder side, the production term $P_{22}$ is almost exclusively determined by $P_{22,b}$. $P_{22}$ is positive everywhere except within the wake vortex and just above the TE shear layer. The large mean velocity gradients at the TE shear layer lead to intense $P_{22}$. Further downstream, $P_{22}$ peaks at $Y = 0$, where $\aver{\delta v \delta v}$ and $\partial V / \partial y$ are both large. For $X \approx 5$ the scale information of $P_{22}$, inherited from $\aver{\delta v \delta v}$, indicates that the vertical fluctuations are mainly sustained by the streamwise-aligned structures. For larger $X$, instead, the scale dependency of $P_{22}$ is lost as the contribution of the von K\'arm\'an vortices becomes dominant: already in the near-wake region the vertical fluctuations, unlike the streamwise ones, are mainly fed by the von K\'arm\'an-like vortices.

As a final remark, we observe that $P_{11}$ is larger than $P_{22}$. As for the cylinder side, this is consistent with the small-scale production mainly sustaining the streamwise velocity fluctuations. As shown later in \S\ref{sec:KarmanVortices}, the opposite occurs for the large-scale production associated with the von K\'arm\'an-like vortices.

%------------------------------
\subsubsection{Redistribution}
\label{sec:wake1-pstrain}

\begin{figure}
\centering
\includegraphics[width=0.74\columnwidth]{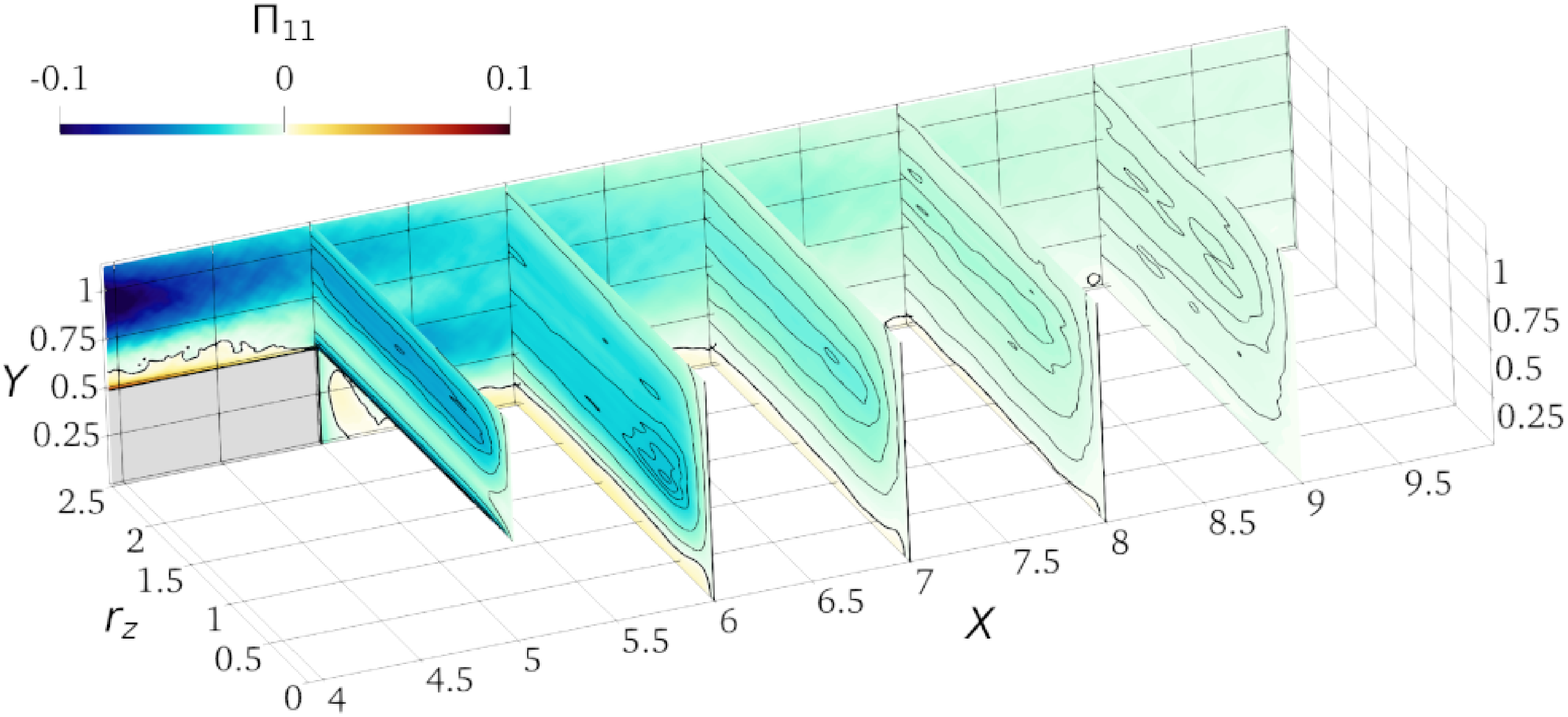}
\includegraphics[width=0.74\columnwidth]{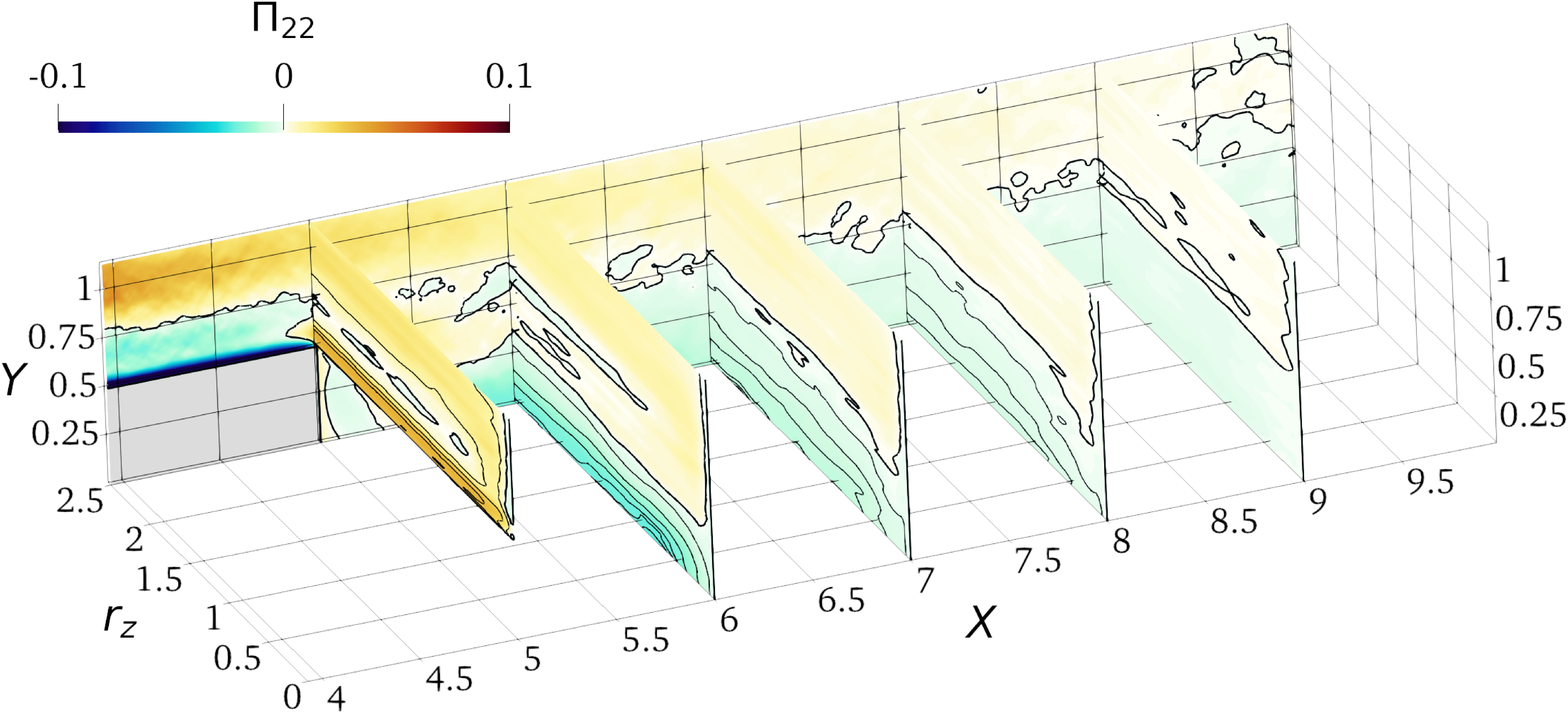}
\includegraphics[width=0.74\columnwidth]{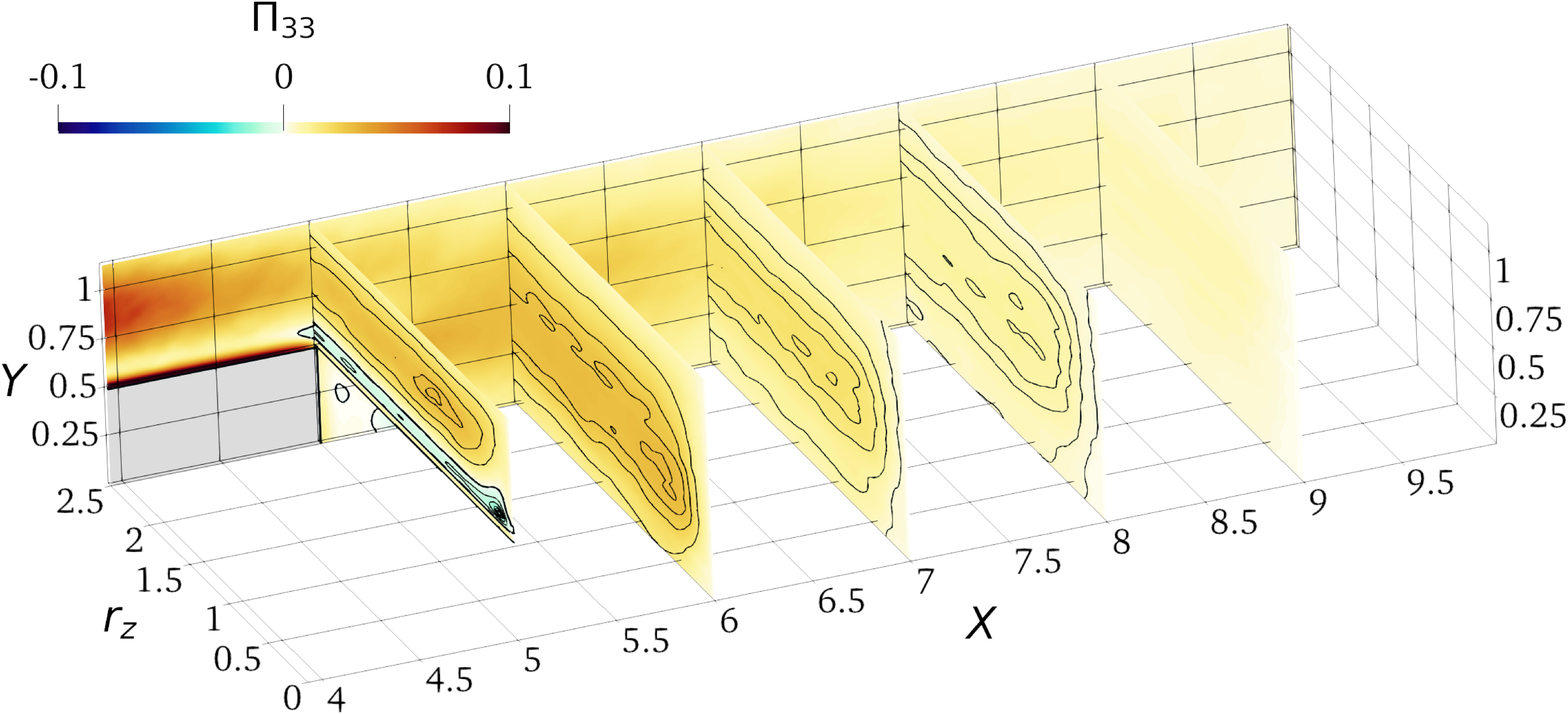}
\caption{As figure \ref{fig:side_pstrain}, but for $4 \le X \le 10$ and $ 0 \le Y \le 1.1$.}
\label{fig:wake_Pstrain}
\end{figure}

As soon as the wall vanishes and the no-penetration condition is relieved, redistribution near the wall ($Y \rightarrow 0.5$) abruptly changes from what described in \S\ref{sec:side-pstrain} for the flow over the cylinder side. Figure \ref{fig:wake_Pstrain} shows that pressure strain takes over its return-to-isotropy role and reorganises the structure of turbulence by partially redirecting the $u-$ and $w-$fluctuations towards the vertical ones. In other words, $\Pi_{22}$ becomes positive for $Y \approx 0.55$, while $\Pi_{11}$ and $\Pi_{33}$ become negative. The pressure strain transforms the $w-$structures, previously generated over the solid wall by impingement, into vertical fluctuations as soon as the wall disappears. In fact, while $\Pi_{11}$ does not show a preferential spanwise scale, $\Pi_{22}$ and $\Pi_{33}$ show their positive and negative local peaks at the small spanwise scale $r_z \approx 0.25$ that identifies the $w-$structures.

Further downstream from the TE, when the organisation of the turbulent fluctuation has changed, the pressure-strain terms near the $Y=0$ plane indicate that a fraction of the vertical energy drained from the mean flow is redistributed towards $\aver{\delta w \delta w}$ and $\aver{\delta u \delta u}$ to partially balance the negative production $P_{11}$. This is similar to the splatting observed at the side wall and, arguably, is dictated by the antisymmetric behaviour of $V$ with respect to the $Y=0$ plane. At a larger distance from the centreline, both streamwise and vertical components are fed by the mean flow, but redistribution changes with $Y$. At intermediate $Y$, say $ 0.1 < Y < 0.5$, the pressure strain partially redistributes $\aver{\delta u \delta u}$ and $\aver{\delta v \delta v}$ towards $\aver{\delta w \delta w}$. At larger $Y$, instead, the canonical scenario where  $\aver{\delta u \delta u}$ is redistributed towards both $\aver{\delta v \delta v}$ and $\aver{\delta w \delta w}$ is observed.

%%%%%%%%%%%%%%%%%%%%%%%%%%%%%%%%%%%%%%%%%
\subsection{The von K\'arm\'an vortices}
\label{sec:KarmanVortices}

The von K\'arm\'an-like spanwise vortices shed from the TE are at their early stage in the near wake, and coexist with the turbulent structures advected from the cylinder side. For their characterisation, non-zero streamwise and vertical separations are considered: the analysis is carried out first in the $r_y=r_z=0$ space at $Y=0.38$ (i.e. where the turbulent kinetic energy has a peak downstream the TE, as shown by \cite{chiarini-quadrio-2021}), and then in the $r_x=r_z=0$ space at $X=6.5$. 

\begin{figure}
\centering
\includegraphics[trim=500 0 640 0,clip,width=0.49\textwidth]{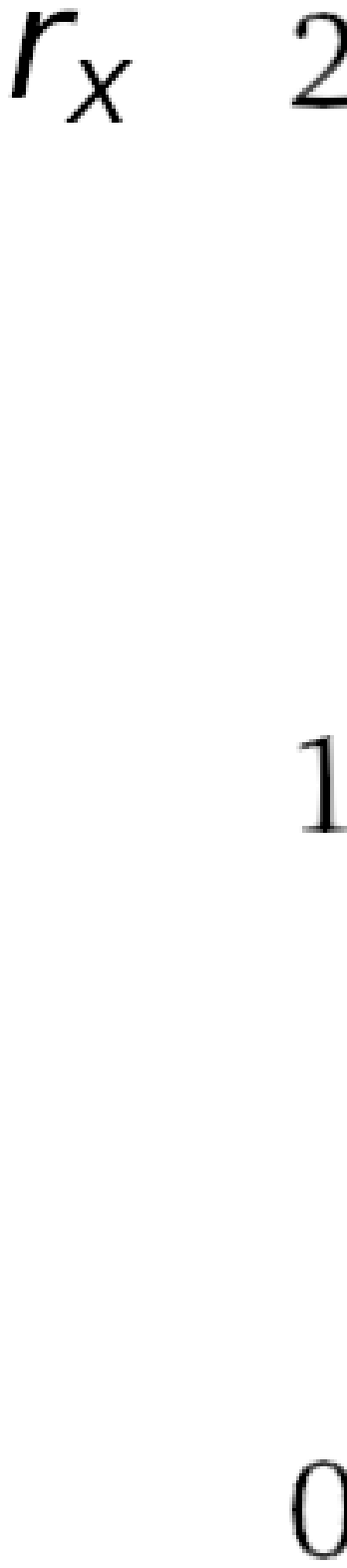}
\includegraphics[trim=500 0 640 0,clip,width=0.49\textwidth]{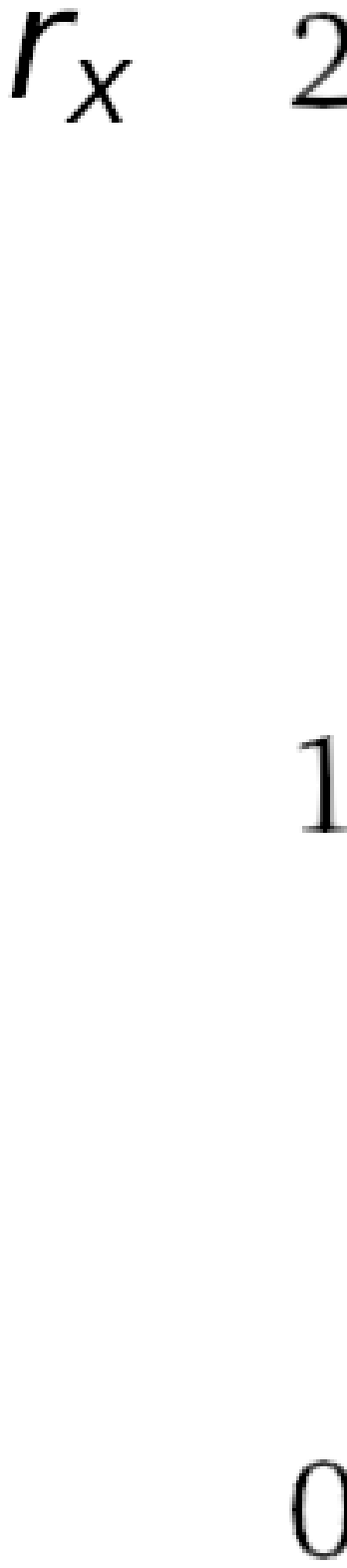}\\
\includegraphics[trim=500 0 640 0,clip,width=0.49\textwidth]{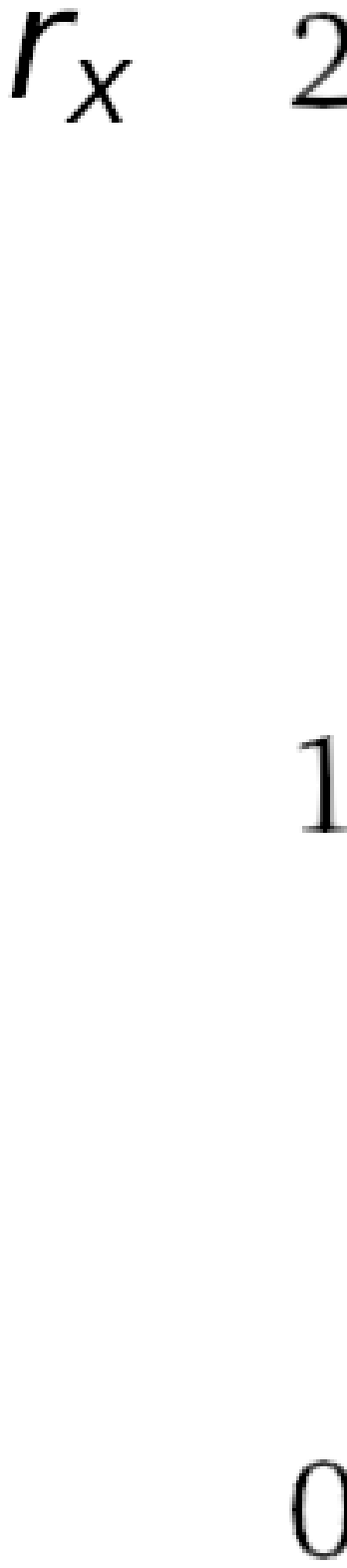}
\includegraphics[trim=500 0 640 0,clip,width=0.49\textwidth]{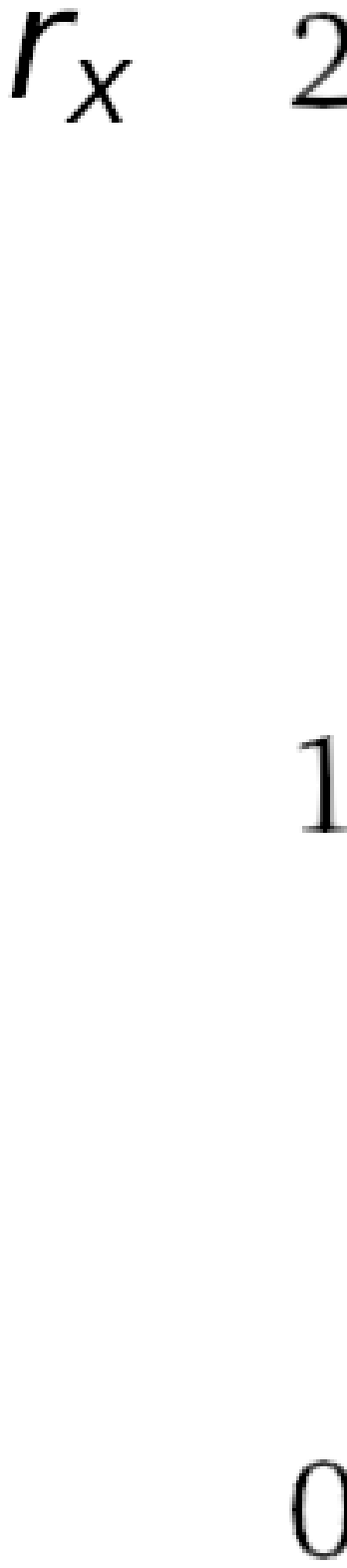} 
\caption{$(X,r_x)$ space with $r_y=r_z=0, Y=0.38$ and $5 \le X \le 9.5$.}
\label{fig:Ener_ry=0-Y=0.38}
\end{figure}

Figure \ref{fig:Ener_ry=0-Y=0.38} plots the normal components of the structure function tensor and $\aver{\delta u \delta v}$ in the $r_y=r_z=0$ space, for $5 \le X \le 9.5$. The triangular shape of the figures is due to the AGKE terms being undefined (see \S\ref{sec:methods}) for $X < 5 + r_x/2 $ and for $X > 9.5 -r_x/2 $, because of the finite size of the box where they are evaluated. Statistical trace of the spanwise rolls is found in the maps of $\aver{\delta u \delta u}$, $\aver{\delta v \delta v}$ and $\aver{\delta u \delta v}$. These structures induce a velocity field such that $R_{uu}<0$, $R_{vv}<0$ and $R_{uv}>0$ for non-zero streamwise separations, leading to peaks of the structure functions whose $r_x-$positions identify their characteristic streamwise length scale. 
The positions of these peaks are $(X,r_x)=(6.8,2.1)$ for $\aver{\delta u \delta u}$, $(X,r_x)=(8.2,2.7)$ for $\aver{\delta v \delta v}$ and $(X,r_x)=(7.19,2.5)$ for $\aver{\delta u \delta v}$. The streamwise scale $r_x = 2.5$, comparable to the cylinder cross-stream size, agrees with the visualisation of figure \ref{fig:lambda2_zoom_wake}. The streamwise size of the vortices increases downstream, arguably due to the turbulent entrainment, and the local peaks of the structure functions shift towards larger $r_x$. In contrast to the streamwise-aligned vortices, mainly organised in streamwise fluctuations (see figure \ref{fig:wake_dudu}), the large-scale von K\'arm\'an vortices are mainly organised in vertical fluctuations: in this space $\aver{\delta v \delta v}$ is larger than the other normal components of $\aver{\delta u_i \delta u_j}$. This is a typical feature of the von K\'arm\'an-like vortices and has been observed also in the wake of other bluff bodies; see for example \cite{thiesset-danaila-antonia-2014} and \cite{alvesportela-papadakis-vassilicos-2017} for the circular and square cylinders.
We mention that information about the distance between two consecutive von K\'arm\'an vortices would be available once the box is long enough to contain the pair \citep{thiesset-danaila-antonia-2014}. 

$\aver{\delta w \delta w}$ peaks close to the TE at small $r_x$, with the maximum at $(X,r_x) \approx (6.2,0.7)$. This peak cannot be related to the von K\'arm\'an-like spanwise vortices, because their characteristic scale is larger and, moreover, they do not induce spanwise velocity at their lateral sides. In fact, in the near wake $\aver{\delta w \delta w}$ is unaffected by the shed vortices, as shown for example by \cite{kiya-matsumura-1988} for the wake after a flat plate normal to the flow. An alternate explanation for the peak of $\aver{\delta w \delta w}$ rests on the observation that, at these values of $X$ and $Y$, the maps of $\aver{\delta u \delta u}$ and $\aver{\delta u \delta v}$ (figure \ref{fig:wake_dudu}) identify streamwise-aligned structures coming from the cylinder wall, with characteristic size $0.5 \le r_z \le 1$. Therefore, we conjecture that the peak of $\aver{\delta w \delta w}$ is due to these structures that, once advected in the wake, tilt around the $z$-axis to produce regions with $R_{ww}<0$ at $r_x \neq 0$. As it will be shown later in \S\ref{sec:KVpstrain}, the statistical trace of the tilting is indeed visible in the pressure-strain terms. The instantaneous snapshot in figure \ref{fig:lambda2_zoom_wake} confirms a similar tilting of the isosurfaces of $\lambda_2$, which in the near wake are inclined as they are embedded in the large-scale motion. 

\begin{figure}
\centering
\includegraphics[trim=500 0 640 0,clip,width=0.49\textwidth]{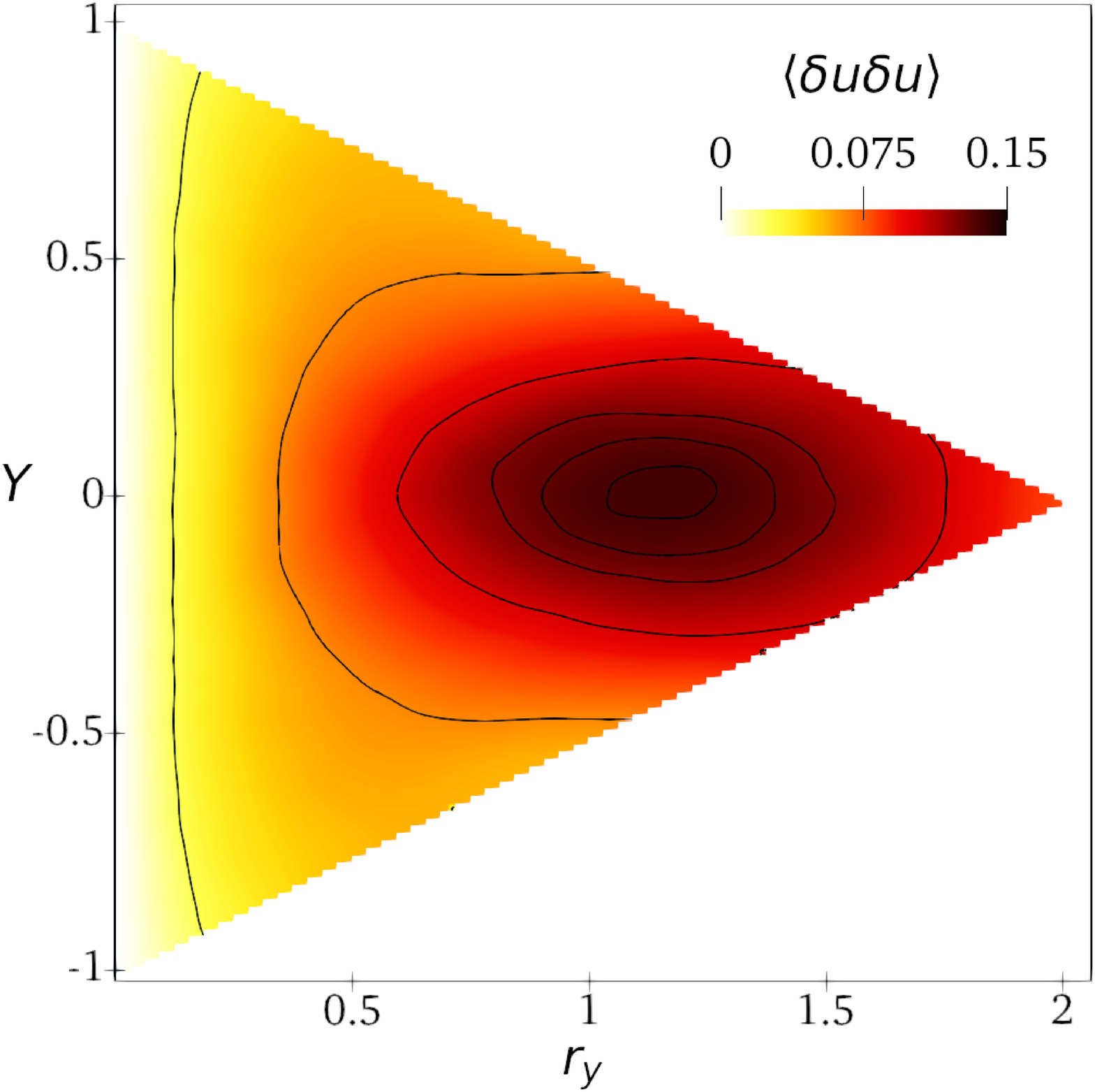}
\caption{$\aver{\delta u \delta u}$ in the $(Y,r_y)$ space with $r_x=r_z=0$, $X=6.5$ and $-1 \le Y \le 1$.}
\label{fig:Ener_rx=0-X=4} 
\end{figure}

Figure \ref{fig:Ener_rx=0-X=4} considers the $(Y,r_y)$ space (where structure functions are not defined for $Y < -1 + r_y/2$ and $Y > 1 - r_y/2$): here the von K\'arm\'an-like vortices leave a significant statistical footprint only in the map of $\aver{\delta u \delta u}$. Indeed, since $\partial V/\partial x$ is negligible compared to the other components of the $\partial U_i / \partial x_j$ tensor the $v-$fluctuations induced at their vertical sides are small, yielding only a negligible trace on the maps of of $\aver{\delta v \delta v}$ and $\aver{\delta u \delta v}$. The von K\'arm\'an-like vortices are placed symmetrically with respect to the $Y$ axis and their vertical scale is comparable with the cross-stream size of the cylinder: the peak is found at $(Y,r_y) \approx (0,1.2)$, consistently with the instantaneous visualisation of figure \ref{fig:lambda2_zoom_wake}.

%----------------------------
\subsubsection{Production}
\begin{figure}
\centering
\includegraphics[trim=500 0 640 0,clip,width=0.49\textwidth]{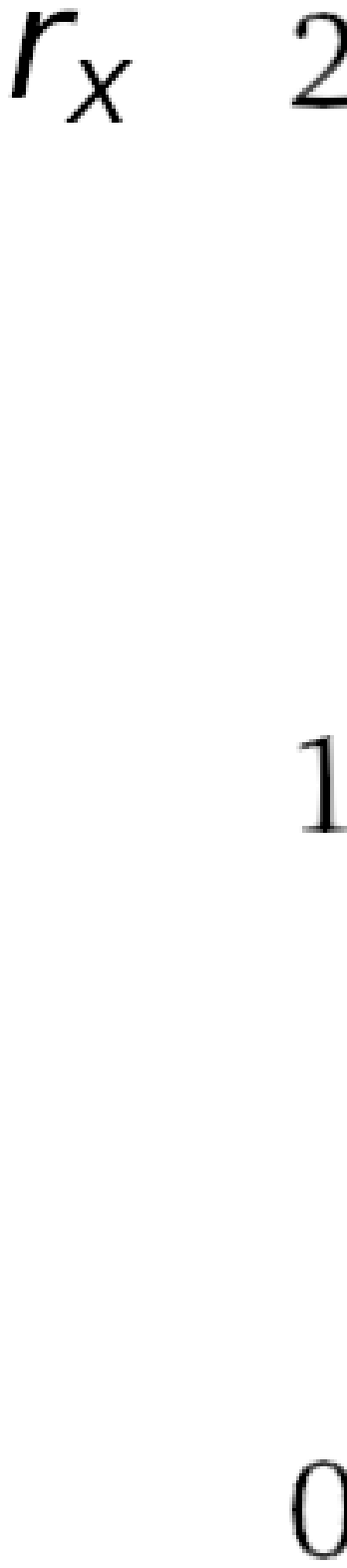}
\includegraphics[trim=500 0 640 0,clip,width=0.49\textwidth]{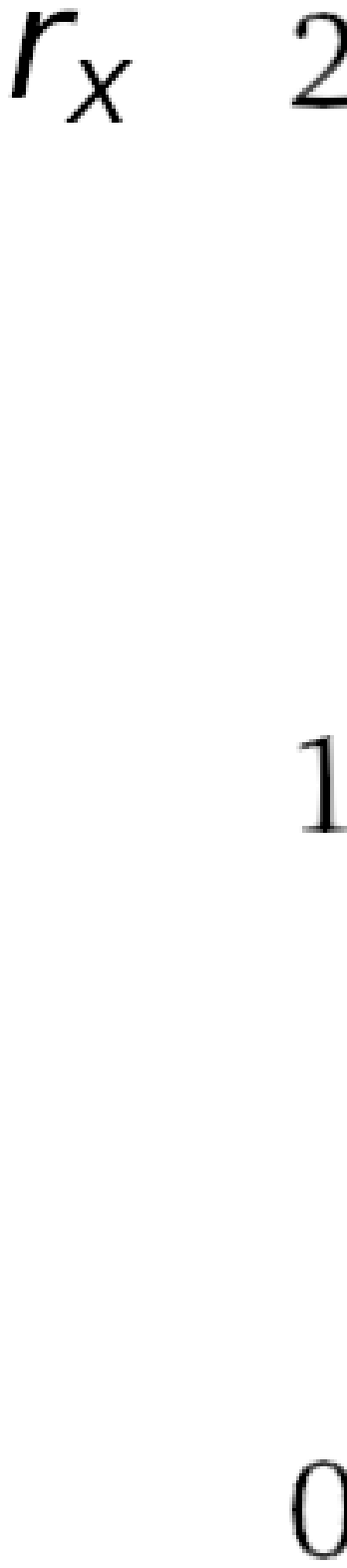}
\caption{Production terms in the $(X,r_x)$ space, $r_y=r_z=0$ and $Y=0.38$.}
\label{fig:Prod_ry=0-Y=0.38}
\end{figure}

Production in the $(X,r_x)$ space points to a second mechanism, related to the von K\'arm\'an-like vortices, acting in the near-wake region to sustain large-scale vertical and streamwise fluctuations. In figure \ref{fig:Prod_ry=0-Y=0.38}, $P_{11}$ is shown to be positive almost everywhere, except at the smallest scales $r_x \rightarrow 0$ and over the $X \approx 5 + r_x/2 $ line. Its maximum is at $(X,r_x) \approx (6.3,1.55)$, consistently with the statistical footprint of the von K\'arm\'an-like vortices. $P_{22}$, instead, is positive everywhere, with the largest values for $r_x>1$, and peaks at $(X,r_x) \approx (5.75,1.3)$, almost the same position of $P_{11}$. 
Overall, this large-scale production mechanism drains energy from the mean flow to feed both $u-$ and $v-$fluctuations at the large scales, with a preference for the former (since $P_{11}>P_{22}$). This differs from the large-scale production observed along the sides of the cylinder and related to the spanwise rolls generated by the KH instability, which is a source of streamwise fluctuations but a sink of vertical ones. Similarly to what observed in the $(X,Y,r_z$) space, $P_{11}$ is dominated by $P_{11,b}$ and $P_{22}$ by $P_{22,b}$ (not shown). Therefore, as for the small-scale motions, production is driven by the wall-normal derivatives of the mean flow. 

%-----------------------------
\subsubsection{Redistribution}
\label{sec:KVpstrain}

\begin{figure}
\centering
\includegraphics[trim=500 0 640 0,clip,width=0.49\textwidth]{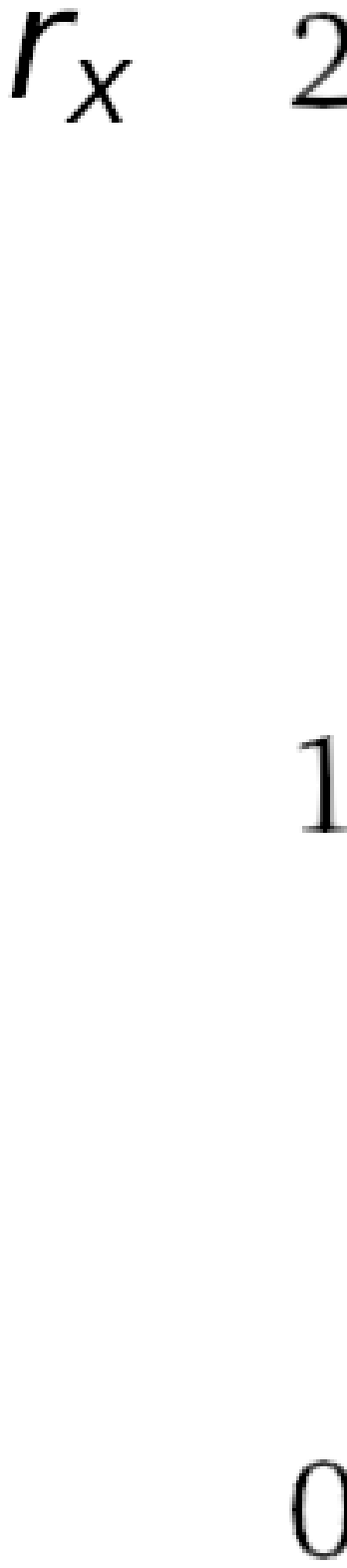}
\includegraphics[trim=500 0 640 0,clip,width=0.49\textwidth]{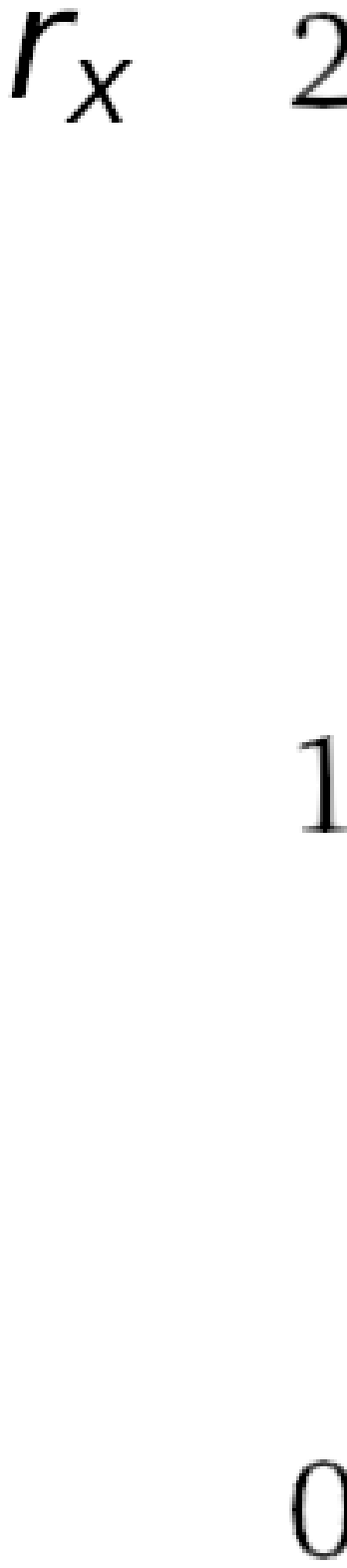}\\
\includegraphics[trim=500 0 640 0,clip,width=0.49\textwidth]{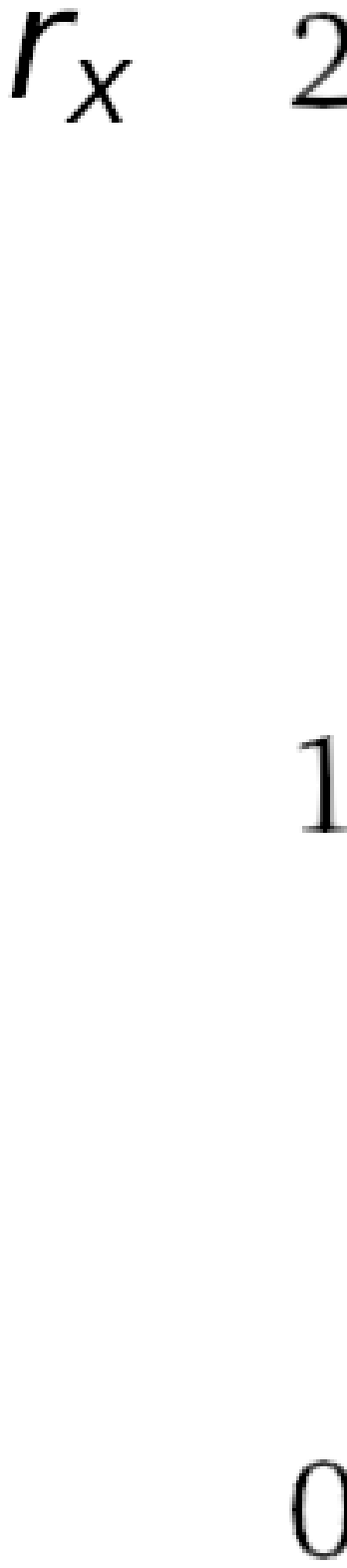}
\caption{Pressure-strain terms in the $(X,r_x)$ space with $r_y=r_z=0$ and $Y=0.38$.}
\label{fig:Pstrain_ry=0-Y=0.38}
\end{figure}

The pressure strain redistributes the $u-$fluctuations ($\Pi_{11}<0$ ) towards the other components ($\Pi_{22}>0$ and $\Pi_{33}>0$) at the largest streamwise scales associated with the von K\'arm\'an-like vortices, i.e. $r_x > 1$, as shown in figure \ref{fig:Pstrain_ry=0-Y=0.38}. Therefore, in the near-wake the large-scale vertical fluctuations, unlike the streamwise ones, are fed by both production and redistribution; this explains the larger $\aver{\delta v \delta v}$ compared to the other components observed in figure \ref{fig:Ener_ry=0-Y=0.38} and discussed in section \S\ref{sec:KarmanVortices}.

For streamwise separations $r_x<1$ associated with the small-scale structures, the character of the redistribution changes with the distance from the TE. Near the TE, the small-scale $u-$fluctuations are partially reoriented towards $v-$ and $w-$ones. For $ X > 6$, in contrast, both $u-$ and $v-$fluctuations turn into $w-$ones: $\Pi_{11}<0$, $\Pi_{22}<0$ and $\Pi_{33}>0$. This agrees with the pressure-strain term in the $r_x=r_y=0$ space for the present value of $Y$, shown in figure \ref{fig:wake_Pstrain}. Note that the positive peak of $\Pi_{33}$ occurs at scales and positions compatible with the maximum of $\aver{\delta w \delta w}$ shown in figure \ref{fig:Ener_ry=0-Y=0.38}, i.e. $(X,r_x) \approx (6,0.8)$, and that at the same scales and positions $\Pi_{11}$ features its negative peak. Therefore, as discussed above, when the streamwise-aligned vortices are advected into the wake, their tilting around their $z$-axis is accompanied by the action of the pressure strain, that partially turns their streamwise fluctuations into spanwise ones, producing regions of spanwise velocity responsible for negative $R_{ww}$ for $r_x \neq 0$.

%----------------------------------------
\subsection{Scale-space energy transfers}

\begin{figure}
\centering
\includegraphics[width=0.99\textwidth]{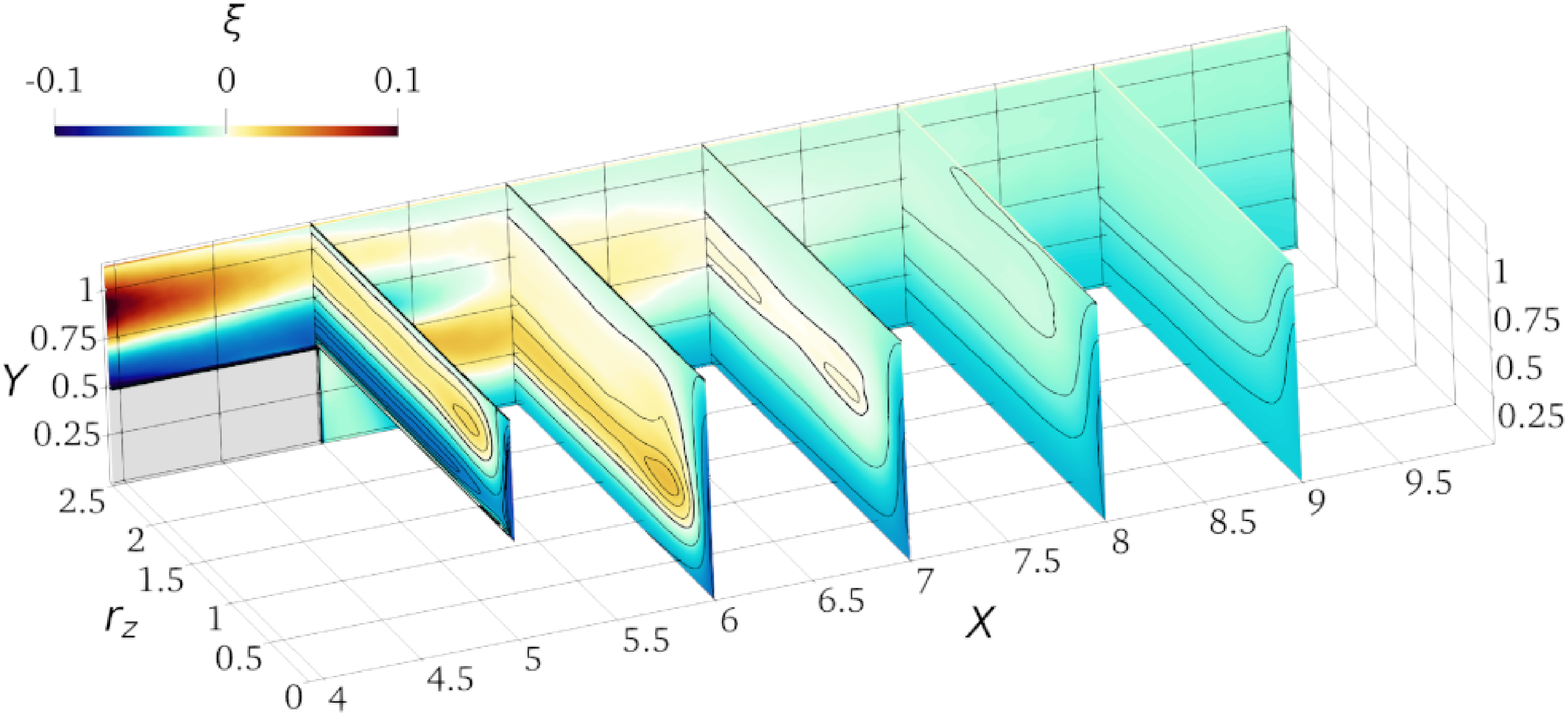}
\includegraphics[width=0.99\textwidth]{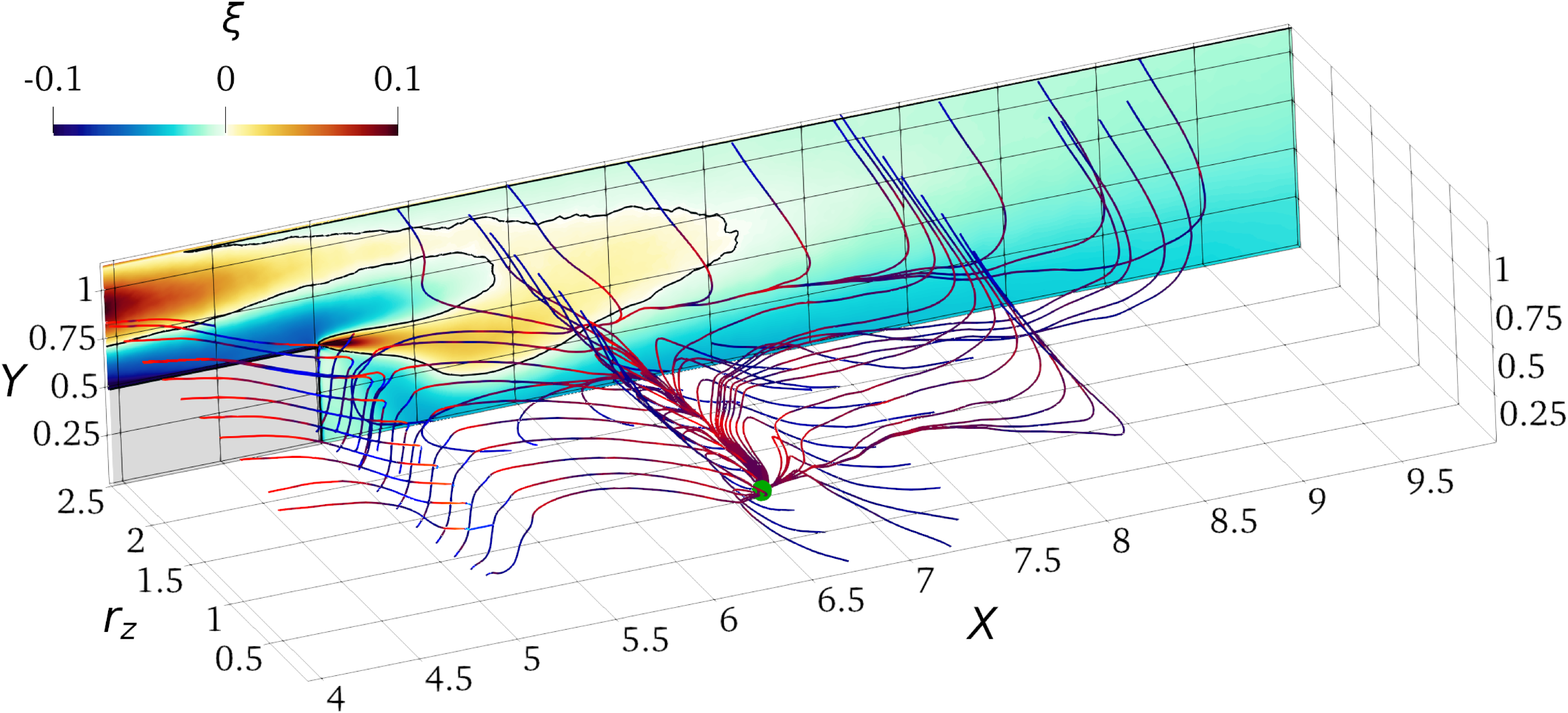}
\includegraphics[width=0.49\textwidth]{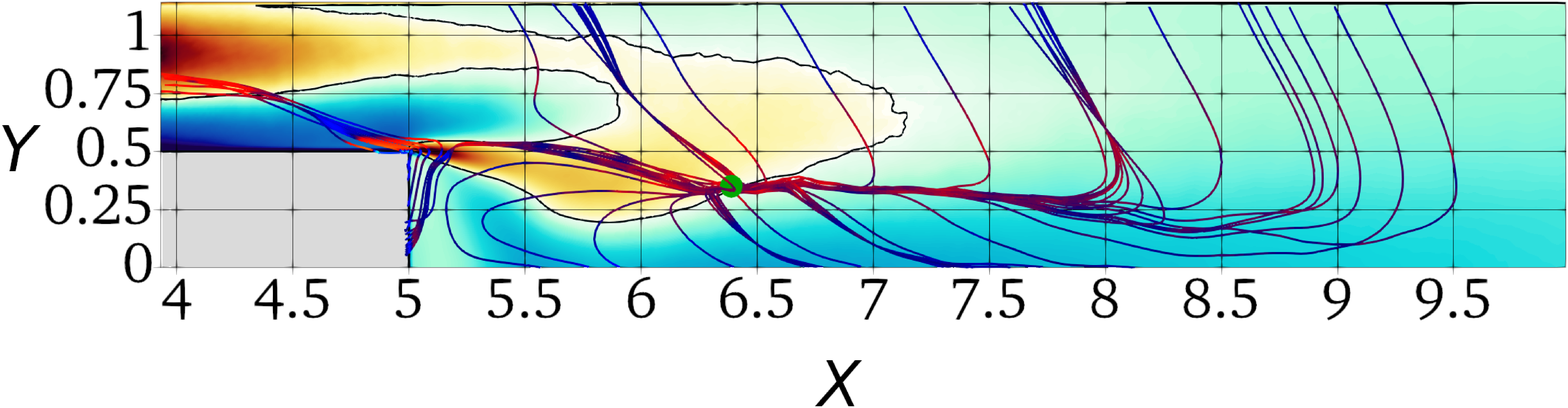}
\includegraphics[width=0.49\textwidth]{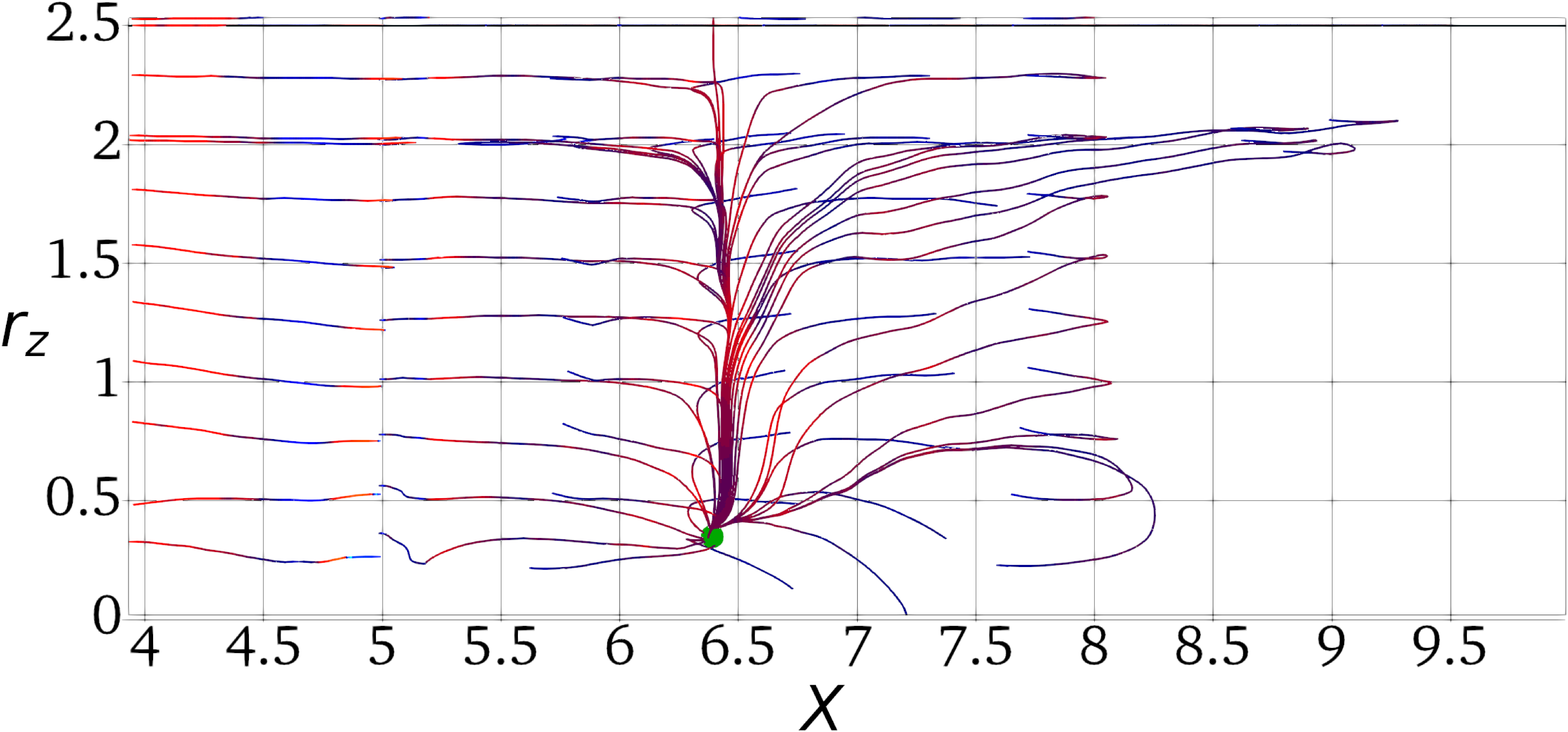}
\caption{As figure \ref{fig:side-fluxes}, but for the near-wake region $4 \le X \le 10$ and $0 \le Y \le 1$.}
\label{fig:wake-fluxes}
\end{figure}
Figure \ref{fig:wake-fluxes} depicts the scale-space energy transfers in the near wake. The source term $\xi$ shows that the positive source for $\aver{\delta q^2}$ extends from the cylinder side down to $X \approx 7$, with localised peaks at $0.5 < r_z < 1$ that are consistent with those observed in the map of $P_{11}$. The near wake has two further sinks, one within the wake vortex and the other along the wake centreline. As over the cylinder side a particular scale is not identified in these sink regions. Again at the smallest scales the dissipation dominates, yielding $\xi<0$ for $r_z \rightarrow 0$ at all $(X,Y)$ positions. The field lines coming from the cylinder side, i.e. those referred to as first family in \S\ref{side:fluxes}, are attracted by the sink within the wake vortex, and show practically no scale dependency. There are no field lines linking the cylinder side with the downstream wake, as a consequence of considering the reduced flux vector only. Indeed, when the complete flux vector is used, several flux lines connect the two regions; see for instance the single-point fluxes of $k$ \citep{chiarini-quadrio-2021b} and $\aver{u_i u_j}$ \citep{chiarini-quadrio-2021}. The implication is that the cylinder side and the near wake dynamically interact only via the energy transfers associated with the mean energy transport. 

Important field lines of the reduced flux vector in the near wake originate from a singularity point placed at $(X,Y,r_z) \approx (6.4,0.35,0.4)$ and marked with a green dot in figure \ref{fig:wake-fluxes}. They show how the excess of fluctuation energy $\aver{\delta q^2}$ produced by the small-scale production mechanism described in \S\ref{sec:wake1-prod} sustains velocity fluctuations in the sink regions of the near wake. All the lines are straight at first and keep the same $X$ and $Y$ ($\hat{\psi}_X \approx 0 $ and $\hat{\psi}_Y \approx 0$) moving towards larger $r_z$ ($\hat{\phi}_z>0$). Their colour indicates that at this stage the fluxes are locally energised. Then at a certain $r_z$ the lines suddenly bend, being attracted by the sink regions where they release energy to balance the negative $\xi$. Four different line types are identified. Lines of the first type bend upstream, descend and release energy in the wake vortex before vanishing at the TE. %Note that this occurs for all spanwise separations in the range $0 \le r_z \le 2.5$, meaning that fluctuations of different spatial extent are sustained by this small-scale production mechanism.
A second type is attracted towards the wake centreline, due to the effect of the viscous term. Note that these lines bend both upstream and downstream and do so over a range of $r_z$, thus releasing energy for a wide range of $X$ and at all spanwise scales. A third type bends upward towards larger $Y$, releasing $\aver{\delta q^2}$ in the outer sink over a wide range of $r_z$ and $X$. This transfer is representative of the turbulent entrainment in the wake. All these line types feature an inverse energy transfer in the space of scales: fluxes are energised mainly at the smallest scales and release energy at the large ones. Finally, a fourth line type goes from the singularity point towards the $r_z=0$ plane, meaning that part of the excess of $\aver{\delta q^2}$ feeds the smallest dissipative scales, highlighting in this case a more classical direct energy transfer form larger to smaller scales.

Similarly to what happens over the cylinder side, it is thus found that in the near wake the scale-energy fluxes are more complex than the classical energy cascade. The simultaneous presence of forward and reverse energy transfers challenges turbulence closures; the definition of a cross-over scale $\ell_{cross}$ between these phenomena is a sound starting point to address the issue. As expected, our data show that the spanwise cross-over scale in the near wake is larger than over the cylinder side. In particular, $\ell_{cross,z} \approx 0.1$ just after the TE and $\ell_{cross,z} \approx 0.25$ at $X=6$ (see the top panel of figure \ref{fig:wake-fluxes}), suggesting that the spanwise grid resolution for the selection of the modelling approach are less severe in the wake than over the cylinder side.

\section{Conclusions}
\label{sec:conclusions}

The present work has provided a statistical description of the large- and small-scale structures populating the flow along the side and in the near wake of a 5:1 rectangular cylinder, also known as BARC flow, at a Reynolds number based on the free stream velocity $U_\infty$ and cylinder thickness $D$ of $Re = 3000$. %, highlighting their role in the production and redistribution mechanisms of the velocity fluctuations. 
The study leverages the anisotropic generalised Kolmogorov equations, or AGKE, to provide a complete and quantitative description of the time-averaged dynamical processes behind the formation, transport and dissipation of each component of the Reynolds stress tensor, by simultaneously considering the space of scales and the physical space. 

The goal is to provide for the first time an exhaustive scale-space characterisation of the BARC flow, by focusing on the region over the cylinder side and on the near wake just after the trailing edge, where the non-equilibrium boundary layer over the cylinder side gradually develops into a free-shear flow and interacts with the large-scale motions of the von K\'arm\'an street. We have statistically described the structures in the flow, precisely identified their characteristic length scales %, revealing the existence of structures that arise due to the flow impingement over the cylinder side, so far undetected. 
and highlighted their role in the production and redistribution of the large- and small-scale velocity fluctuations. The main scale-space energy transfers, as well as implications for turbulence modelling are discussed.

Over the cylinder side, the main structures in the flow are: i) large spanwise rolls generated by the Kelvin--Helmholtz (KH) instability of the shear layer detaching from the leading edge, which are initially spanwise-invariant but later  develop a spanwise modulation; ii) hairpin-like vortices generated by the breakdown of the KH rolls; iii) small streamwise-oriented vortical structures in the aft part of the cylinder side; iv) $w-$structures in the near-wall region created by flow impingement on the cylinder surface downstream of the reattachment. The characteristic spanwise scales of the structures are $r_z \approx 2.4 D$ for the spanwise-modulated KH rolls, $r_z \approx 1.8 D$ for hairpin-like vortices and $r_z \approx 0.5 D$ for the streamwise-aligned vortices and $w-$structures. Two independent sources of velocity fluctuations have been identified: one is associated with the large KH rolls, and the other with the small streamwise vortices. The large-scale source drains energy from the mean flow to directly sustain the streamwise fluctuations, with the large-scale cross-stream fluctuations being indirectly sustained mainly by pressure-strain redistribution. In contrast, the small-scale source sustains directly both streamwise and (to a lesser extent) vertical fluctuations. Pressure strain, then, partially reorients the streamwise fluctuations into cross-stream ones, like in classical parallel wall-bounded turbulent flows. Close to the wall the flow dynamics is dictated by redistribution. Due to flow impingement, very close to the wall velocity fluctuations are organized into small-scale $w-$structures; at slightly larger wall distances they reorient into vertical fluctuations to feed the streamwise-aligned vortices.

In the near wake, small structures advected from the cylinder side coexist and interact with large von K\'arm\'an-like vortices typically shed in the wake past bluff bodies. The small-scale velocity fluctuations modify their organisation as soon as they cross the trailing edge and the wall vanishes beneath them. Indeed, the near-wall $w-$structures turn suddenly into vertical fluctuations, whereas the streamwise vortices weaken progressively until viscous dissipation and the isotropisation effect of the pressure strain annihilate them at a downstream distance from the trailing edge of approximately $4D$. Two independent sources of velocity fluctuations are identified. A small-scale source is associated with the streamwise-aligned structures transported in the near wake, and is the main contributor to $u-$fluctuations up to a downstream distance of $2 D$. A large-scale source is associated with the large wake vortices. Unlike the KH instability, here both the large-scale streamwise and vertical fluctuations are directly fed by energy drained from the mean flow, while the pressure strain partially redistributes streamwise energy towards the cross-stream components. As a result, in the near wake the fluctuating field is mainly organised in $u-$fluctuations at the small scales and in $v-$fluctuations at the large scales.

The scale-space energy transfers are far more complex than the Richardson energy cascade, with the coexistence of forward and reverse energy transfers both over the cylinder side and in the near wake. This complexity has to be considered in turbulence modelling, especially when selecting the cross-over scale $\ell_{cross}$ in Large Eddy Simulations. As a suitable candidate for the spanwise cross-over scale, we propose the minimum spanwise scale where the source term becomes positive. Our data indicate that grid resolution requirements for the selection of the modelling approach are more strict over the cylinder side, where $\ell_{cross,z} \approx 0.1 D$, than in the near wake, where the spanwise cross-over scale increases up to $\ell_{cross,z} \approx 0.25 D$.

The present study can be extended to consider higher Reynolds numbers, and/or different aspect ratios of the cylinder. The experimental work of \cite{moore-etal-2019}, for example, determined that, for smaller aspect ratios where the reattachment is intermittent or entirely absent, the main energetic scales change, owing to differences in the large-scale instabilities and in the development of the wake. 

In the present formulation, the AGKE do not discriminate between the large-scale motions due to flow instabilities and the small-scale turbulent motions, detecting scales and positions where the former force the latter, and viceversa. Work is underway to overcome this issue, by exploiting the quasi-periodic nature of the large-scale motions associated with the flow instabilities. A triple decomposition of the velocity field \citep{hussain-reynolds-1970} into mean, periodic and stochastic contributions may be used to extend the AGKE, similarly to what done by \cite{thiesset-danaila-antonia-2014} and \cite{alvesportela-papadakis-vassilicos-2020} for the GKE. This will lead to two different set of budget equations, one for the large-scale quasi-periodic motions and one for the small-scale stochastic turbulent fluctuations; their analysis will provide further insight in the multiscale spatio-temporal dynamics of the flow.

\section*{Declaration of Interests} 
The authors report no conflict of interest.

\end{document}